\documentclass[aps,prd,preprint,preprintnumbers,unsortedaddress,superscriptaddress,showpacs,nofootinbib]{revtex4-1}

\usepackage{graphicx,color}
\usepackage{relsize}
\usepackage{slashed}
\usepackage{color}
\usepackage{ulem}
\usepackage{tabu}
\usepackage{bm}

\usepackage{amsmath}
\usepackage{amssymb}
\usepackage{amsthm}
\usepackage{mathrsfs}
\usepackage{graphicx}
\usepackage{epstopdf}
\usepackage{fancyhdr}
\usepackage{array}
\usepackage[all]{xy}
\usepackage{eufrak}
\usepackage{euscript}
\usepackage{enumerate}
\usepackage{slashed}
\usepackage{hyperref}
\usepackage{setspace}
\usepackage{subfigure} % NEEDED IN ORDER TO USE SUBFIG
\usepackage{epstopdf} % COMPILE EPS (AND PDF) FIGURES USING PDFLATEX!!!
\usepackage{threeparttable}%表格内的脚注
\usepackage{booktabs}%改表格线条粗细hline, 使用\toprule, \midrule and \bottomrule
\hypersetup{pdftex,colorlinks=true,linkcolor=blue,citecolor=blue,menucolor=black,urlcolor=blue,filecolor=blue}

\newcommand{\myvec}[1]{\ensuremath{\begin{pmatrix}#1\end{pmatrix}}}

\begin{document}

%\title{Molecular $ \boldsymbol{\Omega_b }$ states}
\title{Pseudoscalar or vector meson production in non-leptonic decays of heavy hadrons}

\author{Wei-Hong Liang}
\email{liangwh@gxnu.edu.cn}
\affiliation{Department of Physics, Guangxi Normal University, Guilin 541004, China}

\author{E.~Oset}
\email{oset@ific.uv.es}
\affiliation{Departamento de F\'{i}sica Te\'{o}rica and IFIC, Centro Mixto Universidad de Valencia - CSIC,
Institutos de Investigaci\'{o}n de Paterna, Aptdo. 22085, 46071 Valencia, Spain}

\date{\today}

\begin{abstract}
We have addressed the study of non-leptonic weak decays of heavy hadrons ($\Lambda_b, \Lambda_c, B$ and $D$),
with external and internal emission to give two final hadrons,
taking into account the spin-angular momentum structure of the mesons and baryons produced.

A detailed angular momentum formulation is developed which leads to easy final formulas.
By means of them we have made predictions for a large amount of reactions,
up to a global factor, common to many of them, that we take from some particular data.
Comparing the theoretical predictions with the experimental data,
the agreement found is quite good in general and the discrepancies
should give valuable information on intrinsic form factors,
independent of the spin structure studied here.
The formulas obtained are also useful in order to evaluate meson-meson or meson-baryon loops, for instance of $B$ decays,
in which one has PP, PV, VP or VV intermediate states,
with P for pseudoscalar mesons and V for vector meson and lay the grounds for
studies of decays into three final particles.
\end{abstract}

%\pacs{Valid PACS appear here}
% PACS, the Physics and Astronomy Classification Scheme.
% Valid PACS numbers may be entered using the \verb+\pacs{#1} command.

%\keywords{Meson-baryon interaction; $\Omega_b$ states; Molecular state}

\maketitle
%\tableofcontents

%%%%%%%%%%%%%%%%%%%%%%%%%%%%%%%%%%
\section{Introduction}
\label{sec:intro}

\begin{figure}[h!]
\begin{center}
\includegraphics[scale=0.68]{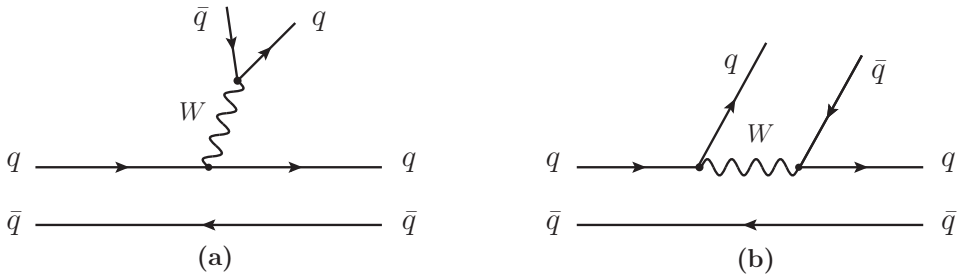}
\end{center}
\vspace{-0.65cm}
\caption{Meson decay with external emission (a) or internal emission (b).}
\label{Fig:1}
\end{figure}

Non-leptonic and semi-leptonic weak decays of heavy hadrons have
become an important source of information
on hadron dynamics \cite{cai0,thomas,Buchalla,IJMPE52,IJMPE53,IJMPE61,IJMPE62,IJMPE63,IJMPE72,IJMPE76,
IJMPE78,IJMPE79,IJMPE81,IJMPE83,IJMPE84,IJMPE89,IJMPE47,Zhao:2018zcb}.
A recent review on the subject can be seen in Ref.~\cite{review2016}.
In non-leptonic decays,
the most typical situations appear in external emission,
which we depict in Fig.~\ref{Fig:1}(a) for meson decay,
and internal emission, depicted in Fig.~\ref{Fig:1}(b)
\cite{IJMPE111,IJMPE112,IJMPE113,IJMPE114}.

In external emission, a $q\bar q$ state from the $W$ decay vertex can lead to
a pseudoscalar meson (P) or a vector meson (V),
and then the other two final $q$ and $\bar q$ states again can produce
a pseudoscalar or a vector meson.
We have thus four possibilities PP, PV, VP and VV for production.
In internal emission, a $q$ state from the first decay vertex
and a $\bar q$ from the second decay vertex merge to produce
either a pseudoscalar or a vector, and the remaining $q \bar q$ pair can again
produce a pseudoscalar or a vector.
Once again we have four possibilities, PP, PV, VP and VV for production.
Certainly there are many other decay modes, and most of them originate
from these basic structures after hadronization including
an extra $q\bar q$ pair with the quantum number of the vacuum.
Final state interaction of this pair of emerging mesons
can give rise to resonances dynamically generated
and the process provides a rich information on the nature of
such resonances \cite{review2016}.

The primary production of the PP, PV, VP, VV pairs is thus important
for the study of many other processes stemming from hadronization
of these primary meson-meson states.
There are other issues where this is important.
One of them has to do with the possible violation of universality
in $e^+e^-, \; \mu^+ \mu^-$ production in $\bar B^0 \to \gamma^* \bar K^{*0}$  decay \cite{uniex},
which is stimulating much work \cite{jorge}.
Loops involving mesons, $\bar B^0 \to D_s^- D^+$, followed by $D_s^- \to \gamma^* D_s^-$,
$D^+ D_s^- \to \bar K^{*0}$, have come to be relevant on this issue \cite{private}
and one can have them with the primary production,
$\bar B^0 \to D_s^- D^+, D_s^{*-} D^+, D_s^- D^{*+}, D_s^{*-} D^{*+}$,
with all the loops interfering among themselves.
One needs these primary amplitudes including their relative phase.

On the other hand, there is a very large amount of decays of this type measured,
and tabulated in the PDG \cite{pdg},
including $B, B_s, D, D_s$ decays, and the internal or external emission modes.
The problem arises equally in baryon decays as $\Lambda_b$ or $\Lambda_c$,
both in internal or external emission. A correlation of all these data
from a theoretical perspective is worth in itself and this is
the purpose of the present work.

The $B$ and $D$ decays into two mesons have been thoroughly studied theoretically
with different approaches,
pQCD, QCDF, SCET, BBNS factorization, light front models, and much progress has been done on the topic
\cite{Beneke,Beneke2,Rosner,Neubert3,AliLu2,Keum,Du,Chay,cdLu,Grinstein,
cai,weiwang,molina,weilu,alilu,Luwei,kuo,Gronau,Fazio,Sanda}.
One common thing to these approaches is that
different structures appearing in different reactions are identified
and conveniently parameterized in terms of parameters
(the most popular the Wilson coefficients)
that are finally obtained from experimental data.
In the present approach,
we do not evaluate these matrix elements from QCD motivated models
or elaborate quark models.
Our aim is different: we identify reactions that have the same quarks
in the initial and final states,
and the same decay topology,
and only differ by the spin rearrangements in the mesons.
We then assume the radial matrix elements to be similar in these reactions
and carry out the nontrivial Racah algebra on the weak Hamiltonian
to describe the reactions and relate them.

Another aspect of our approach is that it allows us to establish a relationship
with approaches based on heavy quark symmetry \cite{Manohar3,Mannel}
and improve upon them, in particular in the $B\to {\rm VP}$ reactions,
where the strict heavy quark symmetry gives zero for the matrix element.

Our approach leads to predictions in fair agreement with experimental data,
in particular for final states in the charm sector,
which is not so well studied.
The approach, however, leads to large discrepancies when one has pions in the final state,
which we associate with the failure of the basic assumption of
equal radial matrix elements for the same flavour quarks,
since the small pion mass leads to large momentum transfers in the reactions
with the corresponding reduction of these matrix elements.

An added value to the present work is the prediction of decay rates
for $\Lambda_b$ and $\Lambda_c$ baryons into one baryon and a meson,
which unlike $B$ decays, have not been much studied theoretically.

%%%%%%%%%%%%%%%%%%%%%%%%%%%%%%%%%%
\section{Formalism}
\label{sec:form}

We shall apply the formalism to both external emission
and internal emission and for baryon and meson decays.
We shall concentrate on the decay which are most favoured by Cabbibbo rules.

\subsection{$\boldsymbol{\Lambda_b}$ external decay}
\label{subsec:form1}

\begin{figure}[b!]
\begin{center}
\includegraphics[scale=0.72]{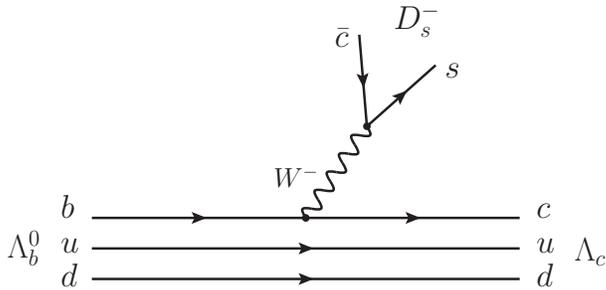}
\end{center}
\vspace{-0.5cm}
\caption{Quark description in $\Lambda_b \to D_s^- \Lambda_c$ decay with external emission.}
\label{Fig:2}
\end{figure}

We look at the process $\Lambda_b \to D_s^- \Lambda_c$
which is depicted in Fig.~\ref{Fig:2}.
Recalling the quark weak doublets
$\begin{pmatrix} \begin{smallmatrix} u\\[1.5mm]d \end{smallmatrix} \end{pmatrix}$
$\begin{pmatrix} \begin{smallmatrix} c\\[1.5mm]s \end{smallmatrix} \end{pmatrix}$
$\begin{pmatrix} \begin{smallmatrix} t\\[1.5mm]b \end{smallmatrix} \end{pmatrix}$
and that the transitions within the same column are Cabibbo favored,
the $bc$ transition is needed for the $b$ decay,
and then the $W^-$ couples to $\bar c s \equiv D_s^-$.
We shall also discuss the $D_s^{*-}$ production and the formalism is
equally valid for $\pi^-$ or $\rho^-$ production.

The next step is to realize that in $\Lambda_b^0$ the $ud$ quarks
are in isospin $I=0$ and spin $S=0$,
and they are spectators in the reaction. The final quarks produced are then
$c$ and $ud \;(I=0, S=0)$, which form the $\Lambda_c^+$.

\begin{figure}[tb!]
\begin{center}
\includegraphics[scale=0.72]{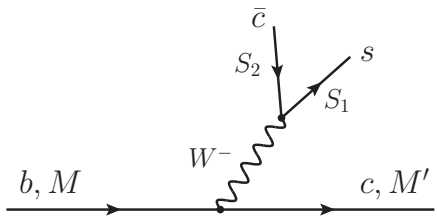}
\end{center}
\vspace{-0.5cm}
\caption{Angular momenta for the $b, c$ transition diagram of Fig.~\ref{Fig:2}.}
\label{Fig:3}
\end{figure}

Since the $ud$ quarks are spectators, we look at the matrix elements
in the weak transition for the diagram of Fig.~\ref{Fig:3}.
We shall use the fact that the $D_s^-$ and $D_s^{*-}$ have the same spatial wave functions
and only differ by the spin rearrangement, an essential input
in heavy quark spin symmetry (HQSS) \cite{Neubert,manohar}.
We also do not attempt to calculate absolute rates,
which are sensible to details of the wave function and form factors,
but just ratios.
Given the proximity of masses of $D_s^-, D_s^{*-}$,
we use again arguments of heavy quark symmetry to justify
that the spatial matrix elements will be the same
in $D_s^-$ or $D_s^{*-}$ production
and only the spin arrangements make them differ.
We advance however, that this is the only element of heavy quark symmetry that we use.
We shall see later that there are terms of type $p/m_Q$ ($m_Q$ is the mass of the heavy quark)
that are relevant in the transitions and they are kept,
while they would be neglected in calculations
making an extreme use of heavy quark symmetry.

The weak Hamiltonian is of the type $\gamma^\mu (1-\gamma_5)$ in each of the weak vertices
and then we have an amplitude
\begin{equation}%\label{eq:Lagrangian}
  t=\langle c | \gamma^\mu (1-\gamma_5) | b \rangle \;
  \langle  s | \gamma_\mu (1-\gamma_5) |c' \rangle,
\end{equation}
where $c'$ corresponds to the $\bar c$ state that forms the $D_s^-$.

In order to evaluate these matrix elements,
we choose for convenience a reference frame
where the $D_s^-$ is produced at rest.
In this frame the $\Lambda_b$ and $\Lambda_c$ have the same momentum, $p$, given by
\begin{equation}\label{eq:momentumP}
  p=\frac{\lambda^{1/2} (M^2_{\Lambda_b}, M^2_{D_s}, M^2_{\Lambda_c})}{2\, M_{D_s}}.
\end{equation}

Furthermore, neglecting the internal momentum of the quark versus $p$,
which is of the order of $5000$ MeV, we can write
\begin{equation}\label{eq:versus}
  \frac{p_c^\mu}{m_c}=\frac{p_{\Lambda_c}^\mu}{M_{\Lambda_c}}, ~~~~~~~~
  \frac{p_b^\mu}{m_b}=\frac{p_{\Lambda_b}^\mu}{M_{\Lambda_b}},
\end{equation}
since these ratios are related to just the velocity of $\Lambda_c$ or $\Lambda_b$.

In that frame the quarks of  $D_s^-$ will be at rest and we take the usual spinors
\begin{equation}\label{eq:spinor}
u_r=\mathcal{A} \myvec{\chi_r \\[2mm]
             \mathcal{B} \; (\vec{\sigma}\cdot \vec{p}\,) \; \chi_r},
~~~~~~~ \chi_1=\myvec{1\\0}, ~~~~~ \chi_2=\myvec{0\\1},
\end{equation}
with
\begin{equation}\label{eq:ABFactor}
  \mathcal{A}= \left( \frac{E_p +m}{2\, m} \right)^{1/2},~~~~~\mathcal{B}= \frac{1}{E_p +m},
\end{equation}
where $p, m$ and $E_p$ are the momentum, mass and energy of the quark.
The spinors $v_r$ for the antiparticles in the Itzykson-Zuber convention \cite{itzy} are,
\begin{equation}
  v_r=\mathcal{A} \myvec{\mathcal{B} \; (\vec{\sigma}\cdot \vec{p}\,) \; \chi_r \\[2mm]
      \chi_r}.
\end{equation}
We take the Dirac representation for the $\gamma^\mu$ matrices,
\begin{equation}\label{eq:gammaMatrix}
  \gamma^0= \myvec{I & 0 \\[1mm]   0 & -I}, ~~~~
  \gamma^i= \myvec{0 & \sigma^i \\[1mm]   -\sigma^i & 0}, ~~~~
  \gamma_5= \myvec{0 & I \\[1mm]   I & 0}.
\end{equation}
By using Eq.~\eqref{eq:versus}, we can rewrite
\begin{equation}\label{eq:ABFactor2}
 \mathcal{A}= \left( \frac{\frac{E_\Lambda}{M_\Lambda}+1}{2} \right)^{1/2},~~~~~
 \mathcal{B}_Q \; p_Q = \mathcal{B} \cdot p, ~~~~~
 \mathcal{B}= \frac{1}{M_\Lambda (1+ \frac{E_\Lambda}{M_\Lambda})},
\end{equation}
where $M_\Lambda, E_\Lambda$ refer to the mass and energy of
the $\Lambda_b$ or $\Lambda_c$ in the $D_s^-$ rest frame,
and $\mathcal{B}_Q, p_Q$, the $\mathcal{B}$ factor of Eq.~\eqref{eq:ABFactor}
and the $b$ or $c$ quark momentum, respectively.

We also note that in the spinor and $\gamma^\mu$ convention that we use we have $\gamma_5 u_r =v_r$,
such that
\begin{align}\label{eq:gammaSign}
(\gamma^\mu -\gamma^\mu \, \gamma_5)\; | v_r \rangle
& = (\gamma^\mu -\gamma^\mu \, \gamma_5)\; \gamma_5 \;| u_r \rangle \nonumber\\
& = -(\gamma^\mu -\gamma^\mu \, \gamma_5)\; | u_r \rangle,
\end{align}
and we can just use spinors corresponding to particles instead of antiparticles when these occur,
just changing a global sign, since we only have one antiparticle, $\bar c$.

The next consideration is that we must combine the spins of $S_1, S_2$
to form a pseudoscalar or a vector and then
we must implement the particle-hole conjugation.
For this let us recall that a state with angular momentum
and the third component $j, m$ behaves as a hole
(antiparticle in this case) of $j, -m$ with a phase
\begin{equation}\label{eq:phase}
  | j m \rangle \to (-1)^{j-m} \; \langle j, -m |.
\end{equation}
Since $(-1)^{2m}= -1$ for the quarks, we incorporate the sign of Eq.~\eqref{eq:gammaSign}
and the phase of Eq.~\eqref{eq:phase} considering the spinor of spin $S_2$
in Fig.~\ref{Fig:3} as a state that combines with $S_1$ with spin third component $-S_2$
and phase $(-1)^{1/2+S_2}$,
\begin{equation}
  | S_2 \rangle \to \langle -S_2 | \; (-1)^{1/2+S_2}.
\end{equation}
Then $\langle 1/2 ,\, S_1 | \; \langle 1/2,\, -S_2 | \, (-1)^{1/2+S_2}$ will combine
to give total spin $j=0,1$ for pseudoscalar or vector production.

The next step is to realize that the state $| 1/2 ,\, S_1 \rangle \; | 1/2,\, -S_2 \rangle $,
which will form the $D_s^-$,
is at rest and the $\gamma^\mu, \gamma^\mu \gamma_5$ matrices reduce to
$\gamma^0 \equiv 1$, $\gamma^i \gamma_5 \equiv \sigma^i$ in the bispinor $\chi_r$ space,
such that we are led to evaluate the matrix element
\begin{equation}\label{eq:element1}
  \langle S_1 | S_2 \rangle \; \langle M' | \gamma^0 -\gamma^0 \gamma_5  |M \rangle
  + \langle S_1 | \sigma^i  | S_2  \rangle  \;
  \langle M' | \gamma^i -\gamma^i \gamma_5 | M \rangle,
\end{equation}
where the $\langle M' | \cdots | M\rangle $ matrix elements
are evaluated in the $D_s^-$ rest frame.
This is done in Appendix \ref{App:A}.

The width for $\Lambda_b \to D_s^- \Lambda_c$ or $D_s^{*-} \Lambda_c$ is given by
\begin{equation}\label{eq:width1}
  \Gamma = \frac{1}{2 \pi} \frac{M_{\Lambda_c}}{M_{\Lambda_b}} \;
  \overline{\sum}\sum |t|^2 \;P_{D_s^{(*)-}},
\end{equation}
where, as shown in Appendix \ref{App:A}, $\overline{\sum}\sum |t|^2$ is given by
%\begin{spacing}{2.0}
\begin{equation}\label{eq:Sumt2}
  \overline{\sum}\sum |t|^2 =
    \left\{
    \begin{array}{ll}
     2(\mathcal{A}\mathcal{A}')^2 \,\big[\,
     (1+\mathcal{B}\mathcal{B}' \,\vec{p}^{\;2})^2+(\mathcal{B}+\mathcal{B}')^2 \,\vec{p}^{\; 2}
     \,\big],& {\rm for~} j=0; \\[2.5mm]
    2(\mathcal{A}\mathcal{A}')^2 \,\big[\,
     3+3(\mathcal{B}^2+\mathcal{B}'^{\,2}) \,\vec{p}^{\;2} -4\mathcal{B}\mathcal{B}' \,\vec{p}^{\;2} +3 (\mathcal{B}\mathcal{B}')^2 \,\vec{p}^{\;4}
     \,\big],& {\rm for~} j=1, \\
    \end{array}
   \right.
\end{equation}
% \end{spacing}
with $j=0$ for $D_s^-$ production and $j=1$ for $D_s^{*-}$ production.
The momentum $P_{D_s^{(*)-}}$ is the $D_s^-$ or $D_s^{*-}$ momentum in the decay
of the $\Lambda_b$ in its rest frame,
\begin{equation}
  P_{D_s^{(*)}}=
  \frac{\lambda^{1/2}(M^2_{\Lambda_b}, M^2_{\Lambda_c}, M^2_{D_s^{(*)}})}{2\, M_{\Lambda_b}},
\end{equation}
and $\mathcal{A}$ and $\mathcal{B}$ in Eq.~\eqref{eq:Sumt2} are given by Eq.~\eqref{eq:ABFactor2},
the prime magnitudes for $\Lambda_c$ and those without prime for $\Lambda_b$.
We recall that $p$ in Eq.~\eqref{eq:Sumt2} is the momentum of $\Lambda_b$  or
$\Lambda_c$ in the rest frame of $D_s^{(*)-}$ given in Eq.~\eqref{eq:momentumP}.

We observe that in the absence of the $p$ terms,
the production of the vector state has a strength a factor
of three bigger than the pseudoscalar one.
However, the $\mathcal{B}$ terms are relevant and are responsible for diversions from this ratio,
as we shall see in the Results section.

\subsection{External emission in $\boldsymbol{B}$ decays}
\label{subsec:form2}

\begin{figure}[b!]
\begin{center}
\includegraphics[scale=0.7]{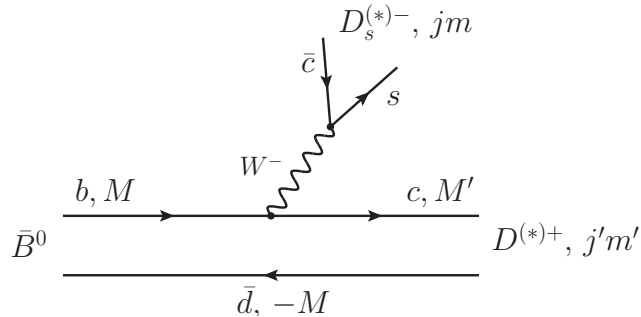}
\end{center}
\vspace{-0.65cm}
\caption{External emission for $\bar B^0 \to D_s^{(*)-} D^{(*)+}$  decay.}
\label{Fig:4}
\end{figure}

We will be looking at the decays $\bar B^0 \to D_s^- D^+$,
$D_s^- D^{*+}$, $D_s^{*-} D^+$, $D_s^{*-} D^{*+}$.
The quark diagram for the transitions is shown in Fig.~\ref{Fig:4},
where $jm$ denotes the spin and its third component of the meson $D_s^{(*)-}$
that comes from the $W^-$ conversion into $\bar c s$,
and $j' m'$ denotes the spin and its third component of the meson $D^{(*)+}$
that comes from the combination of $c \bar d$.
We must couple the $b\bar d$ quarks to spin zero,
$\bar c s$ to $j\,m$ and $c\bar d$ to $j' \, m'$.
This is done explicitly in Appendix \ref{App:B}.
The results that we obtain there are summarized as follows.

According to Appendix \ref{App:B}, we obtain
\begin{alignat}{2}
{\rm A)}~~~    & j=0,\; j'=0:   \nonumber\\ %[2cm]
               & \overline{\sum}\sum |t|^2 =(\mathcal{A}\mathcal{A}')^2 \cdot 2\,
               (1+\mathcal{B}\mathcal{B}' \, \vec{p}^{\, 2})^2; \label{eq:A}\\
\vspace{3cm}
{\rm B)}~~~    &  j=0,\; j'=1:   \nonumber\\
               & \overline{\sum}\sum |t|^2 =(\mathcal{A}\mathcal{A}')^2 \cdot 2\,
               (\mathcal{B}+\mathcal{B}')^2 \, \vec{p}^{\, 2}; \label{eq:B}\\
\vspace{3cm}
{\rm C)}~~~    &  j=1,\; j'=0:   \nonumber\\
               & \overline{\sum}\sum |t|^2 =(\mathcal{A}\mathcal{A}')^2 \cdot 2\,
               (\mathcal{B}+\mathcal{B}')^2 \, \vec{p}^{\, 2}; \label{eq:C}\\
\vspace{3cm}
{\rm D)}~~~    &  j=1,\; j'=1:   \nonumber\\
               & \overline{\sum}\sum |t|^2
               =(\mathcal{A}\mathcal{A}')^2 \big[\,6+4\mathcal{B}^2\,\vec{p}^{\; 2} +4 \mathcal{B}'^2 \,\vec{p}^{\; 2}
               -12\mathcal{B}\mathcal{B}' \,\vec{p}^{\; 2} +6 (\mathcal{B}\mathcal{B}')^2 \, \vec{p}^{\; 4}\,\big]; \label{eq:D}
\end{alignat}
with $\mathcal{A},\mathcal{B},\mathcal{A}',\mathcal{B}',p$ given by
\begin{equation}\label{eq:ABFactor3}
\begin{aligned}
& \mathcal{A}= \left( \frac{\frac{E_B}{M_B}+1}{2} \right)^{1/2},
& \qquad
& \mathcal{A}'= \left( \frac{\frac{E_{D^+}}{M_{D^+}}+1}{2} \right)^{1/2},   \\ %[2cm]
\vspace{3cm}
& \mathcal{B}p= \frac{p}{M_B (1+ \frac{E_B}{M_B})},
& \qquad
&   \mathcal{B}'p= \frac{p}{M_{D^+} (1+ \frac{E_{D^+}}{M_{D^+}})},
\end{aligned}
\end{equation}
and we must change $D^+$ to $D^{*+}$ in case of $D^{*+}$ production.
The momentum $p$, as discussed before,
is the momentum of $B$ or $D^+ (D^{*+})$ in the rest frame
of the $D_s^- (D_s^{*-})$, given by
\begin{equation}
  p= \frac{\lambda^{1/2}(M_B^2,\, M^2_{D_s^{(*)}}, \, M^2_{D^{(*)+}})}{2\, M_{D_s^{(*)}}},
\end{equation}
where by $D_s^{(*)}$ we indicate either $D_s^-$ or $D_s^{*-}$ and the same for $D^{(*)+}$.
The energies in Eq.~\eqref{eq:ABFactor3} are $E_i=\sqrt{M_i^2 +p^2}$.
In this case, following the normalization convention of
the meson fields in Mandl and Shaw \cite{mandl},
the width is given by
\begin{equation}\label{eq:widthM}
  \Gamma=\frac{1}{8\pi}\, \frac{1}{M^2_B}\; \overline{\sum}\sum |t|^2 \;P_{D_s^{(*)}},
\end{equation}
with $P_{D_s^{(*)}}$ the $D_s^{(*)-}$ momentum in the $\bar B^0$ rest frame,
\begin{equation}
  P_{D_s^{(*)}}= \frac{\lambda^{1/2}(M_B^2,\, M^2_{D_s^{(*)}}, \, M^2_{D^{(*)}})}{2\, M_B}.
\end{equation}

We can see that the cases B) and C),
corresponding to $D_s^- D^{*+}$ and $D_s^{*-} D^+$ productions,
have the same strength, which is in very good agreement with experiment \cite{pdg},
as we shall see in the Results section.

\section{Internal emission}
\label{sec:Internal}

We study now another topology of the weak decay process, the internal emission,
and again we differentiate the case of baryon decay from the one of meson decay.

\subsection{$\boldsymbol{\Lambda_b}$ decay in internal emission}
\label{subsec:IntB}

\begin{figure}[t]
\begin{center}
\includegraphics[scale=0.7]{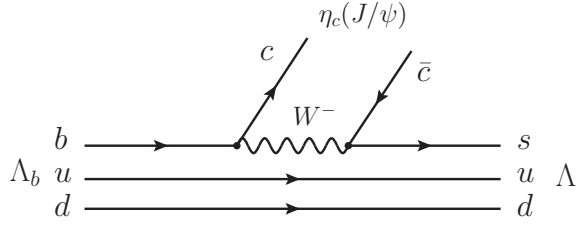}
%\hspace{0.9cm}
%\includegraphics[scale=0.7]{Fig1-b.eps}
\end{center}
\vspace{-0.65cm}
\caption{Quark description of $\Lambda_b \to \eta_c (J/\psi) \Lambda$ decay by internal emission.}
\label{Fig:5}
\end{figure}

\begin{figure}[t]
\begin{center}
\includegraphics[scale=0.7]{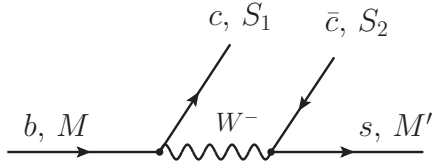}
\end{center}
\vspace{-0.65cm}
\caption{Description of spin third components of the quarks
in the interaction line of Fig.~\ref{Fig:5}.}
\label{Fig:6}
\end{figure}

We look now at the process depicted in Fig.~\ref{Fig:5}
for the decay $\Lambda_b \to \eta_c (J/\psi) \Lambda$.
Once again we look at the most favored Cabibbo-allowed process.
The $b$ quark converts to a $c$ quark and the $W^-$ produced produces a $\bar c s$ pair.
The $s$ quark combines with the $ud$ quarks from $\Lambda_b$,
which are in $I=0, S=0$ and act as spectators.
The $b$ quark and the $s$ quark provide the spin of
$\Lambda_b$ and $\Lambda$ respectively.
Hence, the whole process is studied by looking at
the upper line in the diagram of Fig.~\ref{Fig:6},
which shows the spin components of the particles.
In this case we also take the $\eta_c$ or $J/\psi$ in the rest frame,
where the $\Lambda_b$ and $\Lambda$ have the same momentum $p$,
given by Eq.~\eqref{eq:momentumP} substituting $D_s (D_s^*)$
and $\Lambda_c$ by $\eta_c (J/\psi)$ and $\Lambda$.
Then the $c$ and $\bar c$ have four-spinors at rest
while $b$ and $s$ have four-spinors for moving particles.
Then we must evaluate the operator ($c'$ standing for the $\bar c$ state)
\begin{align}\label{eq:element2}
& ~~~\langle \bar u_s |\gamma^0-\gamma^0 \gamma_5| u_{c'} \rangle \;
\langle \bar u_c | \gamma^0 -\gamma^0 \gamma_5  |u_b \rangle
  - \langle \bar u_s | \gamma^i -\gamma^i \gamma_5 | u_{c'}  \rangle  \;
  \langle \bar u_c | \gamma^i -\gamma^i \gamma_5 | u_b \rangle \nonumber\\
&=\mathcal{A}\mathcal{A}' \left[ (\chi_s, \; -\chi_s \, \vec{\sigma} \cdot \vec{p} \, \mathcal{B}')
 \Big(
 \begin{array}{cc}
 I & -I\\[-2mm]
 I & -I\\
 \end{array}
 \Big)
 \Big(
 \begin{array}{c}
 \chi_{c'}\\[-1.5mm]
 0\\
 \end{array}
 \Big)
 (\chi_c,\, 0)
 \Big(
 \begin{array}{cc}
 I & -I\\[-2mm]
 I & -I\\
 \end{array}
 \Big)
 \Big(
 \begin{array}{c}
 \chi_{b}\\[-1mm]
 \mathcal{B} \, \vec \sigma \cdot \vec p \, \chi_b\\
 \end{array}
 \Big) \right. \nonumber\\%[0.5mm]
 & \qquad ~
- \left. (\chi_s, \; -\chi_s \, \vec{\sigma} \cdot \vec{p} \, \mathcal{B}')
\Big(
 \begin{array}{cc}
 -\sigma^i & \sigma^i\\[-2mm]
 -\sigma^i & \sigma^i\\
 \end{array}
 \Big)
 \Big(
 \begin{array}{c}
 \chi_{c'}\\[-1.5mm]
 0\\
 \end{array}
 \Big)
 (\chi_c, \; 0)
 \Big(
 \begin{array}{cc}
 -\sigma^i & \sigma^i\\[-2mm]
 -\sigma^i & \sigma^i\\
 \end{array}
 \Big)
 \Big(
 \begin{array}{c}
 \chi_{b}\\[-1mm]
 \mathcal{B} \, \vec \sigma \cdot \vec p \, \chi_b\\
 \end{array}
 \Big)
 \right],
\end{align}
where, $\mathcal{A}, \mathcal{A}', \mathcal{B}, \mathcal{B}'$ are the coefficients of Eq.~\eqref{eq:ABFactor2}
for the $\Lambda_b$ and $\Lambda$ respectively.

The matrix element of Eq.~\eqref{eq:element2} can be written as
\begin{equation}\label{eq:Amplitude3}
  t=\mathcal{A}\mathcal{A}'\;\big[t_1 + t_2 + t_3 + t_4 + t_5 + t_6 + t_7 + t_8 \big],
\end{equation}
with
\begin{eqnarray}\label{eq:ti26}
  t_1 &=& \langle M' | S_2 \rangle \; \langle S_1 | M \rangle, \nonumber\\[2mm]
  t_2 &=& -\mathcal{B}' \, \langle M' |\vec \sigma \cdot \vec p \,| S_2 \rangle \;
  \langle S_1 | M \rangle, \nonumber\\[2mm]
  t_3 &=& -\mathcal{B} \, \langle M' | S_2 \rangle \;
  \langle S_1 | \vec \sigma \cdot \vec p\, | M \rangle, \nonumber\\[2mm]
  t_4 &=& \mathcal{B}\mathcal{B}' \, \langle M' |\vec \sigma \cdot \vec p \,| S_2 \rangle \;
  \langle S_1 |\vec \sigma \cdot \vec p \,| M \rangle,\\[2mm]
  t_5 &=& -\langle M' |\sigma^i| S_2 \rangle \;
  \langle S_1 |\sigma^i | M \rangle, \nonumber\\[2mm]
  t_6 &=& \mathcal{B}'\, \langle M' |\vec \sigma \cdot \vec p\,\sigma^i| S_2 \rangle \;
  \langle S_1 |\sigma^i | M \rangle, \nonumber\\[2mm]
  t_7 &=& \mathcal{B}\, \langle M' |\sigma^i| S_2 \rangle \;
  \langle S_1 |\sigma^i \, \vec \sigma \cdot \vec p \,| M \rangle, \nonumber \\[2mm]
  t_8 &=& -\mathcal{B}\mathcal{B}'\, \langle M' |\vec \sigma \cdot \vec p \,\sigma^i| S_2 \rangle \;
  \langle S_1 |\sigma^i \,\vec \sigma \cdot \vec p \,| M \rangle. \nonumber
\end{eqnarray}

In Appendix \ref{App:C}, we evaluate explicitly these terms.
We write here the final results which are relevant to compare
$\Lambda_b \to \eta_c \Lambda$ and $\Lambda_b \to J/\psi \Lambda$,
which correspond to $j=0$ and $j=1$ respectively.
\begin{equation}\label{eq:Sumtt2}
  \overline{\sum}\sum |t|^2 =
    \left\{
    \begin{array}{ll}
     2(\mathcal{A}\mathcal{A}')^2 \,\big[\,
     (1+\mathcal{B}\mathcal{B}' \,\vec{p}^{\;2})^2+(\mathcal{B}+\mathcal{B}')^2 \,\vec{p}^{\; 2}\,
     \big],& {\rm for~} j=0; \\[2mm]
    2(\mathcal{A}\mathcal{A}')^2 \,\big[\,
     3+3(\mathcal{B}^2+\mathcal{B}'^2) \,\vec{p}^{\;2} -4\mathcal{B}\mathcal{B}' \,\vec{p}^{\;2} +3 (\mathcal{B}\mathcal{B}')^2 \,\vec{p}^{\;4}\,
     \big],& {\rm for~} j=1, \\
    \end{array}
   \right.
\end{equation}

Note that in the absence of the $p$-dependent terms,
the ratio of production of $j=1$ to $j=0$ is a factor 3,
apart from phase space.
This is what was obtained in Ref.~\cite{xieliang}
with the strict application of heavy quark spin symmetry.
The $p$-dependent terms, however, change this ratio, as we shall see.

It is also remarkable that this result is the same one obtained for external emission
(see Eqs.~\eqref{eq:Sumt2} and \eqref{eq:Sumtt2}),
even if the original matrix elements are different in both cases.

\subsection{Internal emission for meson decays}
\label{subsec:InterM}

\begin{figure}[t]
\begin{center}
\includegraphics[scale=0.7]{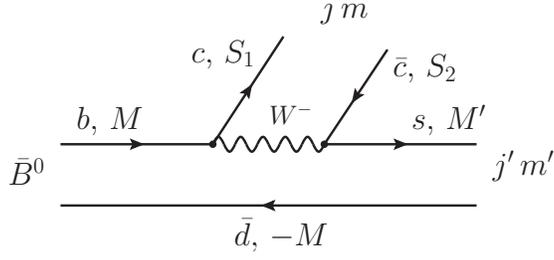}
\end{center}
\vspace{-0.6cm}
\caption{Quark description of $\bar B^0 \to \eta_c (J/\psi) \bar K^0
(\bar K^{*0})$ decay with labels for the spin components.}
\label{Fig:7}
\end{figure}

Now we look at the diagram of Fig.~\ref{Fig:7}.
In the former section we have coupled the $c\bar c$ to $jm$.
Here we must couple in addition $b\bar d$ to $00$ and $s\bar d$ to $j' m'$.

Once again we take the different terms and project over $00$ for the $\bar B^0$
and $j' m'$ for the final $\bar K^0$.
%The terms to consider are $t'_i$ of Eqs.(A.24.1???) and (A.24.2???) for $j=0$ and $j=1$ respectively.
Details of the calculations are shown in Appendix \ref{App:D}.

The results obtained are the following.
$\overline{\sum}\sum |t|^2 $ is given by
\begin{equation}\label{eq:Sumtt4}
  \overline{\sum}\sum |t|^2 =
    \left\{
    \begin{array}{ll}
     2\,(\mathcal{A}\mathcal{A}')^2 \;  (1+\mathcal{B}\mathcal{B}' \, \vec{p}^{\; 2})^2,& {\rm for~} j=0,\, j'=0; \\[2mm]
    2\,(\mathcal{A}\mathcal{A}')^2 \;  (\mathcal{B}+\mathcal{B}')^2 \; \vec{p}^{\; 2},& {\rm for~} j=0, \,j'=1; \\[2mm]
    2\,(\mathcal{A}\mathcal{A}')^2 \; (\mathcal{B}+\mathcal{B}')^2 \; \vec{p}^{\; 2}, & {\rm for~} j=1,\, j'=0; \\[2mm]
    2\,(\mathcal{A}\mathcal{A}')^2 \,
    \big[\,3+2\mathcal{B}^2\,\vec{p}^{\; 2} +2 \mathcal{B}'^2 \,\vec{p}^{\; 2}
               -6\mathcal{B}\mathcal{B}' \,\vec{p}^{\; 2} +3 (\mathcal{B}\mathcal{B}')^2 \, \vec{p}^{\; 4}\,\big],
               & {\rm for~} j=1,\, j'=1; \\
    \end{array}
   \right.
\end{equation}
These results are the same as those obtained for external emission
in Eqs.~\eqref{eq:A}-\eqref{eq:D}
even if the original matrix elements are quite different.

%%%%%%%%%%%%%%%%%%%%%%%%%%%%%%%%%%
\section{Results}
\label{sec:res}

\subsection{$\boldsymbol{\Lambda_b, \Lambda_c}$ decays in external emission}
\label{subsec:resA}
We apply the above formulas to $\Lambda_b$ and $\Lambda_c$ decays,
both with the Cobibbo most favored as well as with Cabibbo-suppressed modes.
The decay width in the Mandl-Shaw normalization is given by Eq.~\eqref{eq:width1}
which we reproduce here as generic for all the decays studied here,
\begin{equation}\label{eq:width5}
  \Gamma = \frac{1}{2 \pi} \frac{M_{\Lambda_f}}{M_{\Lambda_i}} \; \overline{\sum}\sum |t|^2 \;P_f,
\end{equation}
where $\Lambda_i, \Lambda_f$ refer to the initial $\Lambda$ ($\Lambda_b$ or $\Lambda_c$)
and the final $\Lambda$ ($\Lambda_c$ or $\Lambda$) in $\Lambda_b \to D_s^{(*)-} \Lambda_c$
or $\Lambda_c \to \pi (\rho) \Lambda$,
and $P_f$ is the momentum of the final $m_f$ meson,
\begin{equation}
  P_f=
  \frac{\lambda^{1/2}(M^2_{\Lambda_i}, M^2_{\Lambda_f}, M^2_{m_f})}{2\, M_{\Lambda_i}}.
\end{equation}

We apply it to
\begin{enumerate}[1)]
\setlength{\itemsep}{1.5pt}
\setlength{\parsep}{0.5pt}
\setlength{\parskip}{0.2pt}
\item $\Lambda_b \to D_s^{-} \Lambda_c$, \;$D_s^{*-} \Lambda_c$;
\item $\Lambda_b \to \pi^- \Lambda_c$, \;$\rho^- \Lambda_c$;
\item $\Lambda_b \to D^- \Lambda_c$, \;$D^{*-} \Lambda_c$; ~ (Cabibbo suppressed)
\item $\Lambda_b \to K^- \Lambda_c$, \;$K^{*-} \Lambda_c$; ~ (Cabibbo suppressed)
\item $\Lambda_c \to \pi^+ \Lambda$, \;$\rho^+ \Lambda$;
\item $\Lambda_c \to K^+ \Lambda$, \;$K^{*+} \Lambda$.~ (Cabibbo suppressed)
\end{enumerate}
The Cabibbo-suppressed rate can be calculated using the same formulas,
but multiplying by
\begin{equation}\label{eq:FC}
  {\rm FC} = \left(\frac{\sin \theta_C}{\cos \theta_C}\right)^2, ~~~~~~ \sin \theta_C =0.22534.
\end{equation}
We can then obtain the widths for all these decays using the same global constant,
related to the spatial matrix elements of the quark wave functions,
which we do not evaluate, but assume equal for all cases.
While this is very good when dealing with $D_s^{-}$ or $D_s^{*-}$ production,
it is not so much for the other cases,
so we should keep this in mind when comparing with data.

In Table \ref{tab:tab1} we give the results.
We separate the cases of $\Lambda_b$ and $\Lambda_c$ decays
since they involve different Cabibbo-Kobayashi-Maskawa matrix elements
and the spatial wave functions can also be different.
We also make a different block for the decays of $\Lambda_b$ in the
light sector for the same reasons.
The experimental data in this table and the following ones are taken from averages of the PDG \cite{pdg}.
\begin{table}[h]
\renewcommand\arraystretch{1.5}
\caption{Branching ratios for $\Lambda_b, \Lambda_c$ decays in external emission. }
\centering
\begin{tabular}{l  c  c }
\toprule[1.0pt]\toprule[1.0pt]
%\hline\hline
{Decay process~} ~& BR (Theo.)  &  BR (Exp.)\\
\hline
 $\Lambda_b \to D_s^{-} \Lambda_c$ & ~(fit to the Exp.)~  & ~$(1.10 \pm 0.10) \times 10^{-2}$~\\
\hline
$\Lambda_b \to D_s^{*-} \Lambda_c$ & $(1.35 \pm 0.12) \times 10^{-2}$   &  \\
\hline
 $\Lambda_b \to D^{-} \Lambda_c$ {\footnotesize (Cabibbo suppressed)}
 & $(6.89\pm 0.62) \times 10^{-4}$  & $(4.6 \pm 0.6) \times 10^{-4}$\\
\hline
 $\Lambda_b \to D^{*-} \Lambda_c$ {\footnotesize(Cabibbo suppressed)}
 & $(8.19\pm 0.74) \times 10^{-4}$  & \\
 \midrule[1.0pt]\midrule[1.0pt]
%\hline\hline
$\Lambda_b \to \pi^- \Lambda_c$ &  $(8.60\pm 0.72)\times 10^{-2}$ & $(4.9 \pm 0.4) \times 10^{-3}$ \\
\hline
$\Lambda_b \to \rho^- \Lambda_c$ &  $(2.79\pm 0.23)\times 10^{-3}$ &  \\
\hline
$\Lambda_b \to K^- \Lambda_c$ {\footnotesize(Cabibbo suppressed)}
 &  ~(fit to the Exp.)~  &  ~$(3.59 \pm 0.30)\times 10^{-4}~$ \\
\hline
$\Lambda_b \to K^{*-} \Lambda_c$  {\footnotesize(Cabibbo suppressed)}
&  $(1.12 \pm 0.09)\times 10^{-4}$  &   \\
\midrule[1.0pt]\midrule[1.0pt]
%\hline\hline
$\Lambda_c \to \pi^+ \Lambda$ & $0.17 \pm 0.03$ & $(1.30 \pm 0.07) \times 10^{-2}$ \\
\hline
$\Lambda_c \to \rho^+ \Lambda$ &  $(5.0\pm 1.0)\times 10^{-3}$ & $< 6\%$ \\
\hline
$\Lambda_c \to K^+ \Lambda$ {\footnotesize(Cabibbo suppressed)}
 &  (fit to the Exp.)  &  $(6.1 \pm 1.2)\times 10^{-4}$ \\
\hline
$\Lambda_c \to K^{*+} \Lambda$ {\footnotesize(Cabibbo suppressed)}
 &  $(1.8\pm 0.4)\times 10^{-4}$  &   \\
%\hline\hline
\bottomrule[1.0pt]\bottomrule[1.0pt]
\end{tabular}
\label{tab:tab1}
\end{table}

In the first block we show the results for $\Lambda_b \to \Lambda_c D_s^- (D_s^{*-}, D^-, D^{*-})$,
where the latest two modes are Cabibbo-suppressed
(we do not count for this purpose the $b \to c$ transition
which is common to all these decay modes.).
The theoretical errors contain only the relative errors of the experimental
datum used for the fit (in some cases later on where the experimental numbers
have a different $+$ or $-$ error,
we take the bigger relative error for simplicity in the results.).

We fit the $\Lambda_b \to D_s^- \Lambda_c$ decay rate
and make predictions for the other decay modes,
we can only compare with $\Lambda_b \to D^- \Lambda_c$ which is Cabibbo-suppressed.
We find results which are barely compatible within uncertainties.
We should stress that on top of the factors found by us,
one should implement extra form factors
which stem from the spatial matrix elements involving the wave functions of the quarks.
These depend on momentum transfers and hence the masses.
Our position is that, given the small mass differences between $D_s^-, D_s^{*-}, D^-$
and $D^{*-}$, these intrinsic form factors should not differ much.
In any case, the differences found between our theory and the experimental results
could serve to quantify ratios of form factors in those decays.
Yet, in the present case,
we can only conclude that they are very similar for these decays.

Another comment worth making is that Eqs.~\eqref{eq:Sumt2},
in a strict use of heavy quark symmetry, neglecting terms
of type $p/m_Q$,
would give a rate three times bigger for the decay into a vector than for
the related pseudoscalar.
Yet, the theoretical ratio obtained is $1.23$, indicating the relevant role played
by the $p/m_Q$ terms ($\mathcal{B}p$, $\mathcal{B}'p$ terms) in the decay rates.

In the second block we fit the rate for $\Lambda_b \to K^- \Lambda_c$ to
the experimental datum and observe one result
which is common to all the results that follow: the rate for $\Lambda_b \to \pi^- \Lambda_c$
is grossly overestimated. Two reasons can be given for it.
First, the $\pi^-$ has been considered as a $q\bar q$ state, but being a light Goldstone boson,
its structure should be more complex.
Second, the light $\pi^-$ mass will have as a consequence that the intrinsic momentum transfers
and will be much larger and consequently the form factors much smaller.
This is one of the cases where the discrepancies found here can be used to determine empirically
the intrinsic form factors of the reaction.

In the last block, where we fit $\Lambda_c \to K^+ \Lambda$, we observe again the overestimation
of the $\Lambda_c \to \pi^+ \Lambda$ mode. The prediction for $\Lambda_c \to \rho^+ \Lambda$
is consistent with the experimental upper bound.

\begin{figure}[b]
\begin{center}
\includegraphics[scale=0.7]{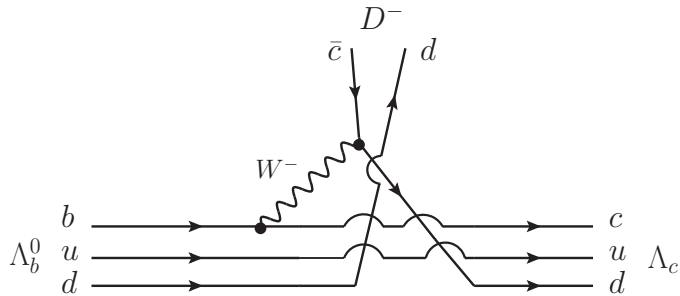}
\end{center}
\vspace{-0.6cm}
\caption{Mechanism with a different topology in the $\Lambda_b \to D^- \Lambda_c$ decay.}
\label{Fig:new}
\end{figure}

At this point it is mandatory to make one more observation concerning some of
the decays in table \ref{tab:tab1}.
We take the $\Lambda_b \to D^- \Lambda_c$ reaction for the discussion
and the arguments can be applied also to
the $\Lambda_b \to D^{*-} \Lambda_c$, $\Lambda_b \to \pi^- \Lambda_c$,
$\Lambda_b \to \rho^- \Lambda_c$,
$\Lambda_c \to\ \pi^+ \Lambda$ and $\Lambda_c \to \rho^+ \Lambda$.
For these decays there is an alternative decay topology to the external emission
involving transfer diagrams,
which for the $\Lambda_b \to D^- \Lambda_c$ case we depict in Fig.~\ref{Fig:new}.
The new mechanism is also Cabibbo-suppressed in the upper $W^-$ vertex, as the mechanism in external emission of table \ref{tab:tab1}
(see Fig.~\ref{Fig:2} replacing $\bar c s$ by $\bar c d$). But there is a very distinct feature in the new topology:
A $d$ quark from the $\Lambda_b$ is transferred to the $D^-$ meson and the $d$ quark
originating from $W^- \to \bar c d$ is trapped by the final $\Lambda_c$ state.
These transfer reactions occur in nuclear reactions and also reactions using quark degrees of freedom.
Normally these mechanisms involve large momentum transfers and they are highly penalized.
Reduction factors of three orders of magnitude or more are common in nuclear reactions involving transfers of nucleons from the projectile to the target \cite{Dillig}.
In quark models of hadron interaction \cite{Vijande} such mechanisms are taken into account by means of the ``rearrangement" diagrams which are also drastically reduced compared to the direct diagrams \cite{david,Ortega}.
In our case, to have an idea of the momentum transfers involved, let us go to the frame where the meson produced is at rest, the momentum of $\Lambda_b$ and $\Lambda_c$ is $5670~{\rm MeV}/c$ (see Eq.~\eqref{eq:momentumP}).
Then in the mechanism of Fig.~\ref{Fig:new} we have to make a large momentum transfer to bring the quark $d$ of the $\Lambda_b$ to the $D^-$ at rest, and similarly a large momentum transfer to bring the quark $d$ produced at rest in the $W^- \to \bar c d$ vertex to the fast moving $\Lambda_c$ in that frame.
The matrix elements accounting for such mechanisms involve form factors with large momentum transfers that make these mechanisms extremely small.

 It is interesting to compare our results with those of Ref.~\cite{Zhao:2018zcb}, where using a quark-diquark picture and light front dynamics the baryonic decay rates of Table I have also been evaluated. It is not possible to compare absolute values because they have been fitted to different observables, but some of the ratios can be compared. Our emphasis has been in relating the vector or pseudoscalar decay modes. In this sense the ratio between the branching ratios for $\Lambda_c \to K^+ \Lambda$ and $\Lambda_c \to K^{*+} \Lambda$ is $3.39$ in our case versus $0.53$ in Ref.~\cite{Zhao:2018zcb}.  The ratios between two vector channels are more similar, in this sense the ratio between the branching ratios for $\Lambda_c \to \rho^+ \Lambda$ and $\Lambda_c \to K^* \Lambda$ is $27.8$ in our case versus $21.5$  in Ref.~\cite{Zhao:2018zcb}.  In the case of $\Lambda_b$ decays the ratio of rates between $\Lambda_b \to K^- \Lambda_c$ and $\Lambda_b \to K^{*-} \Lambda_c$ is $3.21$ in our case versus $0.55$ in Ref.~\cite{Zhao:2018zcb}. However, the ratio of rates between $\Lambda_b \to D_s^- \Lambda_c$ and $\Lambda_b \to D_s^{*-} \Lambda_c$ are more similar, $0.81$ in our case versus $0.67$ in Ref.~\cite{Zhao:2018zcb}.  It is clear that the models are providing different results and this makes the measurement of the missing rates more urgent to keep learning about the theoretical aspects of these reactions.

\subsection{$\boldsymbol{\Lambda_b, \Lambda_c}$ decays in internal emission}
\label{subsec:resB}

In this case we have
\begin{enumerate}[1)]
\setlength{\itemsep}{1.5pt}
\setlength{\parsep}{0.5pt}
\setlength{\parskip}{0.2pt}
\item $\Lambda_b \to J/\psi \Lambda$, \;$\eta_c \Lambda$;
\item $\Lambda_b \to D^0 \Lambda$, \;$D^{*0} \Lambda$.~ (Cabibbo suppressed)
\end{enumerate}
There are no decays of this type for $\Lambda_c^+$ with $\Lambda$ in the final state.

\begin{table}[h]
\renewcommand\arraystretch{1.5}
\caption{Branching ratios for $\Lambda_b$ decays in internal emission.}
\centering
\begin{tabular}{l  c  c }
\toprule[1.0pt]\toprule[1.0pt]
%\hline\hline
{Decay process~} ~& ${\rm BR}(b \to \Lambda_b) \times$ BR (Theo.)  & ~~~ ${\rm BR}(b \to \Lambda_b) \times$ BR (Exp.)\\
\hline
 $\Lambda_b \to J/\psi \Lambda$ & {(fit to the Exp.)}  & $(5.8 \pm 0.8) \times 10^{-5}$\\
\hline
$\Lambda_b \to \eta_c \Lambda$ & {$(3.9\pm 0.5) \times 10^{-5}$}   &  \\
\hline
 $\Lambda_b \to D^0 \Lambda$ {\footnotesize (Cabibbo suppressed)}
 & {$(8.9\pm 1.2) \times 10^{-6}$}  &\\
\hline
 $\Lambda_b \to D^{*0} \Lambda$ {\footnotesize (Cabibbo suppressed)}
 & {$(9.5\pm 1.3) \times 10^{-6}$}  & \\
\bottomrule[1.0pt]\bottomrule[1.0pt]
%\hline\hline
\end{tabular}
\label{tab:tab2}
\end{table}

In Table \ref{tab:tab2} we make predictions for three decay modes,
fitting $\Lambda_b \to J/\psi \Lambda$
for which there are experimental data.
Once again we see that the ratio of rate for  $\Lambda_b \to J/\psi \Lambda$
and $\Lambda_b \to \eta_c \Lambda$ is only $1.49$ instead of the factor three that one would
obtain with strict heavy quark symmetry.
Once again, the $\mathcal{B}p, \mathcal{B}'p$ terms are responsible for this difference.

\subsection{$\boldsymbol{B}$ decays in external emission}
\label{subsec:resC}

Let us see
\begin{enumerate}[1)]
\setlength{\itemsep}{1.5pt}
\setlength{\parsep}{0.5pt}
\setlength{\parskip}{0.2pt}
\item $\bar B^0 \to D_s^- D^+ $, \;$D_s^{*-} D^+$, \; $D_s^- D^{*+}$, \; $D_s^{*-} D^{*+}$;
\item $\bar B^0 \to \pi^- D^+ $, \;$\rho^- D^+$, \;
      $\pi^- D^{*+}$, \; $\rho^- D^{*+}$; %~ (Cabibbo suppressed)
\item $B^- \to D_s^- D^0 $, \;$D_s^{*-} D^0$, \;
      $D_s^- D^{*0}$, \; $D_s^{*-} D^{*0}$;
\item $B^- \to \pi^- D^0 $, \;$\rho^- D^0$, \; $\pi^- D^{*0}$, \;
      $\rho^- D^{*0}$; % ~ (Cabibbo suppressed)
\item $\bar B_s^0 \to D_s^- D_s^+ $, \;$D_s^{*-} D_s^+$, \;
      $D_s^- D_s^{*+}$, \; $D_s^{*-} D_s^{*+}$;
\item $\bar B_s^0 \to \pi^- D_s^+ $, \;$\rho^- D_s^+$, \;
      $\pi^- D_s^{*+}$, \; $\rho^- D_s^{*+}$; %~ (Cabibbo suppressed)
\item $ B_c^- \to D_s^- \, \eta_c $, \;$D_s^{*-} \, \eta_c$, \;
      $D_s^- \, J/\psi$, \; $D_s^{*-} \,J/\psi$;
\item $ B_c^- \to \pi^- \, \eta_c $, \;$\rho^- \, \eta_c$, \;
      $\pi^- \, J/\psi$, \; $\rho^- \,J/\psi$.
\end{enumerate}

Because we have only mesons, the width in the Mandl-Shaw normalization is given by
\begin{equation}\label{eq:widthM8}
  \Gamma = \frac{1}{8\pi}\; \frac{1}{M^2_{B_i}} \; \overline{\sum} \sum |t|^2 \; P_f,
\end{equation}
with
\begin{equation}\label{eq:pf}
  P_f = \frac{\lambda^{1/2} (M^2_{B_i},\, M^2_1, \, M^2_2)}{2\, M_{B_i}},
\end{equation}
where $M_1, M_2$ are the masses of the final mesons.

The results for $B$ decays in external emission are shown in Tables \ref{tab:tab3}, \ref{tab:tab4}, \ref{tab:tab5} and \ref{tab:tab6}.
\begin{table}[h!]
\renewcommand\arraystretch{1.5}
\caption{Branching ratios for $\bar B^0$ decays in external emission.}
\centering
\begin{tabular}{l  c  c }
\toprule[1.0pt]\toprule[1.0pt]
%\hline\hline
{Decay process~} ~& BR (Theo.)  &  BR (Exp.)\\
\hline
 $\bar B^0 \to D_s^{-} D^+$ &  $(1.31 \pm 0.10) \times 10^{-2}$  & $(7.2 \pm 0.8) \times 10^{-3}$\\
\hline
$\bar B^0 \to D_s^{*-} D^+$ & $(7.25 \pm 0.58)\times 10^{-3}$   & $(7.4 \pm 1.6)\times 10^{-3}$ \\
\hline
 $\bar B^0 \to D_s^{-} D^{*+}$ & $(7.68 \pm 0.61) \times 10^{-3}$  & $(8.0 \pm 1.1) \times 10^{-3}$\\
\hline
 $\bar B^0 \to D_s^{*-} D^{*+}$ & ~(fit to the Exp.)~  & ~~$ (1.77 \pm 0.14)\times 10^{-2}$ \\
 \midrule[1.0pt]\midrule[1.0pt]
%\hline\hline
$\bar B^0 \to \pi^- D^+$ %{\footnotesize (Cabibbo suppressed)}
&  $(26.5\pm 4.4)\%$ & $(2.52 \pm 0.13) \times 10^{-3}$ \\
\hline
$\bar B^0 \to \rho^- D^+$ %{\footnotesize (Cabibbo suppressed)}
& (fit to the Exp.) & $(7.9\pm 1.3)\times 10^{-3}$ \\
\hline
$\bar B^0 \to \pi^- D^{*+}$ %{\footnotesize (Cabibbo suppressed)}
&  $(23.0\pm 3.8)\%$  &  ~~$(2.74 \pm 0.13)\times 10^{-3}$ \\
\hline
$\bar B^0 \to \rho^- D^{*+}$  %{\footnotesize (Cabibbo suppressed)}
& $(8.1\pm 1.3)\times 10^{-3}$  &  $(2.2^{+1.8}_{-2.7})\times 10^{-3}$ \\
%\hline\hline
\bottomrule[1.0pt]\bottomrule[1.0pt]
\end{tabular}
\label{tab:tab3}
\end{table}

In Table \ref{tab:tab3} we distinguish again the heavy sector from the light sector.
In the heavy sector, fitting $\bar B^0 \to D_s^{*-} D^{*+}$ to the experiment,
we obtain results in good agreement with experiment.
In the case of $\bar B^0 \to D_s^{-} D^{+}$, there is an overestimate of about $50\%$
counting the extremes of the error bands,
which indicates again the reduction effect
that the form factor would have by going from the masses
of $D_s^{*-} D^{*+}$ to the lighter ones of $D_s^{-} D^{+}$.
It is interesting to remark that, within same masses,
the rates for PV and VP decay modes ($D_s^{*-} D^{+}$ and $D_s^{-} D^{*+}$) are the same.
We see this in experiment within errors.
Even more, the ratios of the centres of the results are $1.08$ for experiment and $1.06$ for the theory.
More interesting is to realize that these two modes are proportional to $(\mathcal{B}p+\mathcal{B}'p)^2$
(see Eqs.~\eqref{eq:B}, \eqref{eq:C}) and would be strictly zero in the heavy quark counting.
Let us also stress that in this counting the rate for VV decay to PP decay would be a factor of three.
Experimentally it is $2.45$, indicating the more moderate role of the $\mathcal{B}p, \mathcal{B}'p$ terms in this case.

In the light sector we see again the gross overestimate of the rates
in the modes with a pion in the final state.
More surprising is the discrepancy of a factor $1.7$,
counting the extremes of the errors, for the $\bar B^0 \to \rho^- D^{*+}$ decay,
although given the large experimental errors speculation
is not appropriate at present time.
Concerning the two pionic modes, it is still rewarding to see that
the ratio of the rates of the two pionic decay modes is in good agreement with experiment.
This should be the case
if the discrepancies of the absolute rates are due to the intrinsic form factor,
because this should be very similar for $\pi^- D^+$ or $\pi^- D^{*+}$.

\begin{table}[h!]
\renewcommand\arraystretch{1.5}
\caption{Branching ratios for $B^-$ decays in external emission.}
\centering
\begin{tabular}{l  c  c }
\toprule[1.0pt] \toprule[1.0pt]
%\hline\hline
{Decay process~} ~& BR (Theo.)  &  BR (Exp.)\\
\hline
 $B^- \to D_s^{-} D^0$ & $(1.27\pm 0.18) \times 10^{-2}$  & $(9.0 \pm 0.9) \times 10^{-3}$\\
\hline
$B^- \to D_s^{*-} D^0$ & $(7.03\pm 0.99) \times 10^{-3}$   & $(7.6 \pm 1.6)\times 10^{-3}$ \\
\hline
 $B^- \to D_s^{-} D^{*0}$ & $(7.43 \pm 1.04) \times 10^{-3}$  & $(8.2 \pm 1.7) \times 10^{-3}$\\
\hline
 $B^- \to D_s^{*-} D^{*0}$ & ~(fit to the Exp.)~  &~$ (1.71 \pm 0.24)\times 10^{-2}$~ \\
\midrule[1.0pt]\midrule[1.0pt]
%\hline\hline
$B^- \to \pi^- D^0$ %{\footnotesize (Cabibbo suppressed)}
& $(44.9\pm 6.0)\%$ & $(4.80 \pm 0.15) \times 10^{-3}$ \\
\hline
$B^- \to \rho^- D^0$ %{\footnotesize (Cabibbo suppressed)}
&  (fit to the Exp.) & $(1.34\pm 0.18)\times 10^{-2}$ \\
\hline
$B^- \to \pi^- D^{*0}$ %{\footnotesize (Cabibbo suppressed)}
&  $(38.9\pm 5.2)\%$  &  ~$(5.18 \pm 0.26)\times 10^{-3}~$ \\
\hline
$B^- \to \rho^- D^{*0}$ %{\footnotesize (Cabibbo suppressed)}
 &  $(1.37\pm 0.18)\times 10^{-2}$  &  $(9.8\pm 1.7)\times 10^{-3}$ \\
\bottomrule[1.0pt]\bottomrule[1.0pt]
%\hline\hline
\end{tabular}
\label{tab:tab4}
\end{table}

The results in Table \ref{tab:tab4} are related to those in Table \ref{tab:tab3},
only the spectator $\bar d$ quark is substituted by a $\bar u$.
Once again, in the heavy sector we find a good agreement with experiment.
There is still a small overestimate of the rate for $B^- \to D_s^{-} D^0$
like in the related former case of $\bar B^0 \to D_s^{-} D^{+}$,
but counting extremes in the errors the discrepancy is only of $10\%$.

In the light sector we find again the discrepancy in the modes with a final pion.
Interesting is the rate for $B^- \to \rho^- D^{*0}$,
where counting the extreme of the errors the discrepancy is only of $10\%$,
unlike the larger discrepancy in $\bar B^0 \to \rho^- D^{*+}$ that we discussed before.
The ratio of the experimental rates for $B^- \to D_s^{-} D^{*0}$ and $B^- \to D_s^{*-} D^{0}$
is $1.08$ versus $1.06$ for the theory, and the ratio of the experimental rates of
$B^- \to D_s^{*-} D^{*0}$ to $B^- \to D_s^{-} D^{0}$ is $1.9$
 while in the counterpart of $\bar B^0$ decay it is $2.45$.
 However, the ratios can be made compatible playing with uncertainties.

 In the light sector we find again the large rate for the pionic decay modes
  and the $B^- \to \rho^- D^{*0}$ rates are compatible within uncertainties.
  The ratio of the two pionic decay rates is roughly compatible with experiment within errors.

\begin{table}[h!]
\renewcommand\arraystretch{1.5}
\caption{Branching ratios for $\bar B_s^0$ decays in external emission.}
\centering
\begin{tabular}{l  c  c }
\toprule[1.0pt]\toprule[1.0pt]
%\hline\hline
{Decay process~} ~& BR (Theo.)  &  BR (Exp.)\\
\hline
 $\bar B_s^0 \to D_s^{-} D_s^+$ &  $(1.06\pm 0.14) \times 10^{-2}$  & $(4.4 \pm 0.5) \times 10^{-3}$\\
\hline
$\bar B_s^0 \to D_s^{*-} D_s^+ +D_s^- D_s^{*+}$~~~ & $(1.08\pm 0.14) \times 10^{-2}$   & $(1.37 \pm 0.16)\times 10^{-2}$ \\
\hline
 $\bar B_s^0 \to D_s^{*-} D_s^{*+}$ & ~(fit to the Exp.)~  & ~$ (1.43 \pm 0.19)\times 10^{-2}$~ \\
\midrule[1.0pt]\midrule[1.0pt]
%\hline\hline
$\bar B_s^0 \to \pi^- D_s^+$ %{\footnotesize (Cabibbo suppressed)}
&  $(23.1\pm 4.7)\%$ & $(3.00 \pm 0.23) \times 10^{-3}$ \\
\hline
$\bar B_s^0 \to \rho^- D_s^+$ %{\footnotesize (Cabibbo suppressed)}
& (fit to the Exp.) & $(6.9\pm 1.4)\times 10^{-3}$ \\
\hline
$\bar B_s^0 \to \pi^- D_s^{*+}$ %{\footnotesize (Cabibbo suppressed)}
&  $(20.1\pm 4.1)\%$  &  ~$(2.0 \pm 0.5)\times 10^{-3}~$ \\
\hline
$\bar B_s^0 \to \rho^- D_s^{*+}$  %{\footnotesize (Cabibbo suppressed)}
& $(7.1\pm 1.4)\times 10^{-3}$  &  $(9.6 \pm 2.1)\times 10^{-3}$ \\
\bottomrule[1.0pt]\bottomrule[1.0pt]
%\hline\hline
\end{tabular}
\label{tab:tab5}
\end{table}

\begin{figure}[t]
\begin{center}
\includegraphics[scale=0.7]{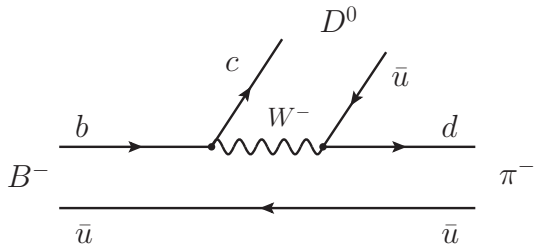}
\end{center}
\vspace{-0.6cm}
\caption{Internal emission mechanism for $B^- \to \pi^- D^0$ decay.}
\label{Fig:new2}
\end{figure}

At this point it is important to have a look at the results of
tables \ref{tab:tab3} and \ref{tab:tab4} from a different perspective.
As we have mentioned, the difference between tables \ref{tab:tab3} and \ref{tab:tab4} is that
we have replaced the spectator $\bar d$ quark by a $\bar u$.
In this sense, within the pure mechanism for external emission, we should expect the same rates,
up to a minor effect of the difference of masses in the phase space.
This is actually the case in tables \ref{tab:tab3} and \ref{tab:tab4} in the first block,
both theoretically and experimentally.
However, in the second block the experimental numbers are about double in table \ref{tab:tab4} than in table \ref{tab:tab3}.
This requires an explantation.
Indeed, while the first block of decays proceeds through external emission,
the second block can also proceed via internal emission,
as shown in Fig.~\ref{Fig:new2} for $B^- \to \pi^- D^0$.
Internal emission is color suppressed and is expected to be reduced by
about a factor of three relative to external emission
and thus about one order of magnitude in the rate.
This is the general rule experimentally,
but in processes where the two mechanisms are possible we expect an interference,
and assuming constructive interference we would have a factor $(1+\frac{1}{3})^2 \simeq 1.8$
in the rates of the second block in table \ref{tab:tab4} versus the counterpart in table \ref{tab:tab3}.
This is actually the case and has been studied in Refs.~\cite{bassel1,bassel2,bassel3,neubert2,thomas,Browder}.
Two amplitudes $a_1, a_2$ are considered to account for the effective charge current
(i.e. the external emission in our case)
and effective neutral current (i.e. the internal emission in our case).
The amplitudes $a_1$ and $a_2$ are fitted to experiment for
the $\bar B^0 \to \pi^- D^+, \rho^- D^+, \pi^- D^{*+}$
and $\rho^- D^{*+}$, and both the relative magnitude and the sign are obtained for $a_1$ and $a_2$,
with $a_2 / a_1 =0.25 \pm 0.07 \pm 0.06$ \cite{neubert2,thomas}.
In addition to some coefficient close to $1$ in the $a_2 / a_1$ term, this reproduces the experiment in a picture qualitatively similar as the one exposed above based on the color counting and constructive interference.
Different values for $|a_2 / a_1|$ are obtained in Ref.~\cite{cdLu},
with a relatively small phase in the ratio $a_2 / a_1$,
but qualitatively similar.

Assuming similar relative contributions of the internal emission mechanism in all cases,
the ratios of rates in the second block of table \ref{tab:tab4} would still make sense,
exception made of the pionic production mode for the reasons exposed along the work.
Yet, the ratio between the related $\pi^- D^0$ and $\pi^- D^{*0}$ modes should also be meaningful
and we see that it is in good agreement with experiment within errors.

The rates in Table \ref{tab:tab5} are also related to those in the former two tables.
By fixing the rate for $\bar B_s^0 \to D_s^{*-} D_s^{*+}$, the rates for
$\bar B_s^0 \to D_s^{*-} D_s^{+} + D_s^{-} D_s^{*+}$ are compatible within errors,
and the one for $\bar B_s^0 \to D_s^{-} D_s^+$ is a bit overestimated even counting errors.

In the light sector we find again a gross overestimate for the pionic decay modes but the results for
$\bar B_s^0 \to \rho^- D_s^{*+}$ obtained fixing the rate of $\bar B_s^0 \to \rho^- D_s^+$
are compatible with experiment.
Once again, the ratio of the two pionic decay modes is compatible with experiment.
Actually, the fact that the ratio of decays to $\pi {\rm P}$ and $\pi {\rm V}$ are compatible
with experiment in spite of the very different expressions of $\overline \sum \sum |t|^2$
comes to reinforce our statement that it is the intrinsic form factor
(independent on whether we have P or V in the final state)
which is responsible for the overcounting of the rates in the theory.

\begin{table}[t]
\renewcommand\arraystretch{1.5}
\caption{Rates of branching ratios for $B_c^-$ decays in external emission.}
\centering
\begin{tabular}{l  c  c }
\specialrule{0em}{2pt}{1pt}
\toprule[1.0pt]\toprule[1.0pt]
%\hline\hline
{~~~~~~~~~~~~Rate~~~~~~~~~~~~} ~& ~~~Theo.~~~  &  ~~~Exp.~~~\\
\hline
 $\Gamma_{B_c^- \to D_s^{-} \eta_c}/\Gamma_{ B_c^- \to D_s^{-} J/\psi}$ & $1.76$   & \\
\specialrule{0em}{2pt}{2pt}
\hline
$\Gamma_{B_c^- \to D_s^{*-} \eta_c}/\Gamma_{ B_c^- \to D_s^{-} J/\psi}$ & $0.90$   & \\
\specialrule{0em}{2pt}{2pt}
\hline
 $\Gamma_{B_c^- \to D_s^{*-} J/\psi}/\Gamma_{ B_c^- \to D_s^{-} J/\psi}$ & $2.47$   & $2.5 \pm 0.5$\\
 \specialrule{0em}{3pt}{3pt}
\midrule[1.0pt]\midrule[1.0pt]
%\hline\hline
$\Gamma_{B_c^- \to D_s^{-} J/\psi}/\Gamma_{ B_c^- \to \pi^{-} J/\psi}$ &  $0.05$  & $3.1 \pm 0.5$ \\
\specialrule{0em}{2pt}{2pt}
\hline
$\Gamma_{ B_c^- \to D_s^{*-} J/\psi}/\Gamma_{ B_c^- \to \pi^{-} J/\psi}$
& $0.11$   & $10.4 \pm 3.1 \pm 1.6$\\
\specialrule{0em}{2pt}{2pt}
\hline
$\Gamma_{ B_c^- \to \pi^- \eta_c}/\Gamma_{ B_c^- \to \pi^{-} J/\psi}$
& $1.11$   & \\
 \specialrule{0em}{3pt}{3pt}
\midrule[1.0pt]\midrule[1.0pt]
%\hline\hline
$\Gamma_{ B_c^- \to D_s^{-} \eta_c}/\Gamma_{ B_c^- \to \rho^{-} J/\psi}$ &  $2.22$  & \\
 \specialrule{0em}{2pt}{2pt}
\hline
$\Gamma_{ B_c^- \to D_s^{*-} \eta_c}/\Gamma_{ B_c^- \to \rho^{-} J/\psi}$ & $1.13$   & \\
\specialrule{0em}{2pt}{2pt}
\hline
$\Gamma_{ B_c^- \to D_s^{-} J/\psi}/\Gamma_{ B_c^- \to \rho^{-} J/\psi}$ &  $1.26$  & \\
\specialrule{0em}{2pt}{2pt}
\hline
$\Gamma_{ B_c^- \to D_s^{*-} J/\psi}/\Gamma_{ B_c^- \to \rho^{-} J/\psi}$ & $3.11$   & \\
\specialrule{0em}{2pt}{2pt}
\hline
$\Gamma_{ B_c^- \to \pi^{-} \eta_c}/\Gamma_{ B_c^- \to \rho^{-} J/\psi}$ &  $31.52$  & \\
\specialrule{0em}{2pt}{2pt}
\hline
$\Gamma_{ B_c^- \to \rho^{-} \eta_c}/\Gamma_{ B_c^- \to \rho^{-} J/\psi}$ & $0.94$   & \\
\specialrule{0em}{2pt}{2pt}
\hline
$\Gamma_{ B_c^- \to \pi^{-} J/\psi}/\Gamma_{ B_c^- \to \rho^{-} J/\psi}$ & $28.30$   & \\
\specialrule{0em}{2pt}{2pt}
\bottomrule[1.0pt]\bottomrule[1.0pt]
%\hline\hline
\end{tabular}
\label{tab:tab6}
\end{table}

In Table \ref{tab:tab6} we show predicted ratios for several decay modes of $B^-_c$.
Since we expect the rates for decay modes with pions in the final state to be grossly overcounted,
it is not surprising to see that ratios of heavy decay modes to
those with pions in the final state are rather small compared with experiment.
Yet, the only ratio that we can compare involving heavy decay modes (line three of the table)
is in very good agreement with experiment.

\subsection{$\boldsymbol{D}$ decays in external emission}
\label{subsec:resD}

We look at
\begin{enumerate}[1)]
\setlength{\itemsep}{1.5pt}
\setlength{\parsep}{0.5pt}
\setlength{\parskip}{0.2pt}
\item $D^0 \to \pi^+ K^- $, \;$\rho^+ K^-$, \; $\pi^+ K^{*-}$, \; $\rho^+ K^{*-}$;
\item $D^0 \to \pi^+ \pi^- $, \;$\rho^+ \pi^-$, \; $\pi^+ \rho^-$,\;
      $\rho^+ \rho^-$; ~ (Cabibbo suppressed)
\item $D^+ \to \pi^+ \bar K^0 $, \;$\rho^+ \bar K^0$, \;
      $\pi^+ \bar K^{*0}$, \; $\rho^+ \bar K^{*0}$;
\item $D^+ \to \pi^+ \pi^0 $, \;$\rho^+ \pi^0$, \;
      $\pi^+ \rho^0$, \; $\rho^+ \rho^0$;  ~ (Cabibbo suppressed)
\item $D_s^+ \to \pi^+ \eta $, \;$\rho^+ \eta$, \; $\pi^+ \phi$, \; $\rho^+ \phi$;
\item $D_s^+ \to \pi^+ \eta' $, \;$\rho^+ \eta'$;
\item $D_s^+ \to \pi^+ K^0 $, \;$\rho^+ K^0$, \;
      $\pi^+ K^{*0}$, \; $\rho^+  K^{*0}$.
      ~ (Cabibbo suppressed)
\end{enumerate}

To evaluate the rates with $\pi^0, \rho^0, \eta, \eta'$ production,
we must look at the $d\bar d$ for $\pi^0, \rho^0$, and $s \bar s$ for $\eta, \eta'$.
We have with the $\eta, \eta'$ mixing of Ref.~\cite{Bramon}
\begin{alignat*}{2}
& \rho^0 = \frac{1}{\sqrt{2}}\, (u\bar u - d\bar d),
& \qquad
& \pi^0 =\frac{1}{\sqrt{2}}\, (u\bar u - d\bar d),  \\[2mm]
& \eta = \frac{1}{\sqrt{3}}\, (u\bar u + d\bar d - s\bar s),
& \qquad
& \eta' = \frac{1}{\sqrt{6}}\, (u\bar u + d\bar d +2 s\bar s).
\end{alignat*}
Thus in the case of $\pi^0, \rho^0$ production,
we must multiply by $\frac{1}{2}$ the standard formula,
in the case of $\eta$ production we must multiply by $\frac{1}{3}$
and in the case of $\eta'$ production by $\frac{2}{3}$.

The results for $D$ decays in external emission are shown in Tables \ref{tab:tab7}, \ref{tab:tab8} and \ref{tab:tab9}.

\begin{table}[tb]
\renewcommand\arraystretch{1.5}
\caption{Branching ratios for $D^0$ decays in external emission.}
\centering
\begin{threeparttable}
\begin{tabular}{l  c  c }
\toprule[1.0pt]\toprule[1.0pt]
%\hline\hline
{Decay process~} ~& BR (Theo.)  &  BR (Exp.)\\
\hline
 $D^0 \to \pi^{+} K^-$ &  $(18.86\pm 2.62)\%$  & $(3.89 \pm 0.04)\%$\\
\hline
 $D^0 \to \pi^{+} K^{*-}$ & ~(fit to the Exp.)~  & ~~$(5.83^{+0.81}_{-0.56})\%$ \tnote{1}
 %\footnotemark[1]
 ~~\\
\hline
$D^0 \to \pi^+ \pi^- $ {\footnotesize (Cabibbo suppressed)}
& $(4.36\pm 0.61)\%$ & ~~$(1.407 \pm 0.025) \times 10^{-3}$ \\
\hline
$D^0 \to \pi^+ \rho^-$ {\footnotesize (Cabibbo suppressed)}~~
& $(4.45\pm 0.62) \times 10^{-3}$   &  ~$(5.08 \pm 0.25)\times 10^{-3}~$ \\
\midrule[1.0pt]\midrule[1.0pt]
%\hline\hline
$D^0 \to \rho^+ K^-$ & ~(fit to the Exp.)~  & $(11.1 \pm 0.7)\%$ \\
\hline
 $D^0 \to \rho^+ K^{*-}$ & $(8.2\pm 0.5) \times 10^{-2}$  & \\
\hline
$D^0 \to \rho^+ \pi^-$ {\footnotesize (Cabibbo suppressed)}
& $(3.2\pm 0.2) \times 10^{-2}$ & $(10.0\pm 0.4)\times 10^{-3}$ \\
\hline
$D^0 \to \rho^+\rho^-$ {\footnotesize (Cabibbo suppressed)}
 & $(6.5\pm 0.4) \times 10^{-3}$   &   \\
\bottomrule[1.0pt]\bottomrule[1.0pt]
%\hline\hline
\end{tabular}
   \begin{tablenotes}
        \footnotesize
        \item[1]This datum is obtained by averaging
            $D^0 \to K^{*-} \pi^+; \; K^{*-} \to K^- \pi^0$
            and $D^0 \to K^{*-} \pi^+; \; K^{*-} \to \bar K^0_S \pi^-$.
   \end{tablenotes}
\end{threeparttable}
\label{tab:tab7}
\end{table}

In Table \ref{tab:tab7} we show results for $D^0$ decays. We separate the sectors with $\pi$ or $\rho$
in the final state because of the different masses.
In the pionic sector we find that the $D^0 \to \pi^+ \pi^- $ mode, with two pions in the final state
is overcounted, but the $D^0 \to \pi^+ \rho^-$ mode comes out fine when the $D^0 \to \pi^{+} K^{*-}$
is fitted to experiment.
The $D^0 \to \pi^{+} K^-$ mode is also overcounted.

In the $\rho$ sector, once again the $\rho^+ \pi^-$ mode is overcounted, following the general trend.

\begin{table}[t!]
\renewcommand\arraystretch{1.5}
\caption{Branching ratios for $D^+$ decays in external emission.}
\centering
\begin{tabular}{l  c  c }
\toprule[1.0pt]\toprule[1.0pt]
%\hline\hline
{Decay process~} ~& BR (Theo.)  &  BR (Exp.)\\
\hline
 $D^+ \to \pi^{+} \bar K^0$ &  $(4.84\pm 0.55)\%$  & $(2.94\pm 0.16)\%$\\
\hline
 $D^+ \to \pi^{+} \bar K^{*0}$ & ~(fit to the Exp.)~ & ~~$(1.50\pm 0.17)\%$\\
\hline
$D^+ \to \pi^+ \pi^0 $ {\footnotesize (Cabibbo suppressed)}~~
& $(5.86\pm 0.66)\times 10^{-3}$ &~ $(1.17\pm 0.06)\times 10^{-3}$ \\
\hline
$D^+ \to \pi^+ \rho^0$ {\footnotesize (Cabibbo suppressed)}
&  $(5.79\pm 0.66)\times 10^{-4}$  &  ~$(8.0 \pm 1.4)\times 10^{-4}~$ \\
\midrule[1.0pt]\midrule[1.0pt]
%\hline\hline
$D^+ \to \rho^+ \bar K^0$ & ~(fit to the Exp.)~ & $(11.8^{+1.2}_{-0.8})\%$ \\
\hline
 $D^+ \to \rho^+ \bar K^{*0}$ & $(8.8\pm 0.9)\%$  & \\
\hline
$D^+ \to \rho^+ \pi^0$ {\footnotesize (Cabibbo suppressed)}
& $(1.8\pm 0.2)\%$ &  \\
\hline
$D^+ \to \rho^+\rho^0$ {\footnotesize (Cabibbo suppressed)}
 &  $(3.5\pm 0.4)\times 10^{-3}$  &   \\
\bottomrule[1.0pt]\bottomrule[1.0pt]
%\hline\hline
\end{tabular}
\label{tab:tab8}
\end{table}

The $D^+$ decay modes of Table \ref{tab:tab8} are closely related to
those of the $D^0$ decay discussed before.
In the pionic decay sector we fit $D^+ \to \pi^{+} \bar K^{*0}$
and then the $D^+ \to \pi^+ \rho^0$ rate is basically compatible within errors with experiment,
and the $D^+ \to \pi^{+} \bar K^0$ mode is a bit overcounted.
The $D^+ \to \pi^+ \pi^0 $ mode, with two pions in the final state,
is also overcounted following the general trend.
In the $\rho$ sector we do not have experimental rates to compare and we leave there the predictions.

\begin{table}[bt]
\renewcommand\arraystretch{1.5}
\caption{Branching ratios for $D_s^+$ decays in external emission.}
\centering
\begin{tabular}{l  c  c }
\toprule[1.0pt]\toprule[1.0pt]
%\hline\hline
{Decay process~} ~& BR (Theo.)  &  BR (Exp.)\\
\hline
 $D_s^+ \to \pi^{+} \eta$ & ~(fit to the Exp.)~   & $(1.70 \pm 0.09)\%$\\
\hline
 $D_s^+ \to \pi^{+} \eta'$ & $(1.09\pm 0.06) \%$   & $(3.94 \pm 0.25)\%$\\
\hline
$D_s^+ \to \pi^+ \phi $
& $(1.33\pm 0.07)\%$ & $(4.5 \pm 0.4) \%$ \\
\hline
 $D_s^+ \to \pi^{+} K^{0}$ {\footnotesize (Cabibbo suppressed)}
& $(3.14\pm 0.17)\times 10^{-3}$  &~~ $(2.44\pm 0.12)\times 10^{-3}$\\
\hline
$D_s^+ \to \pi^+ K^{*0}$ {\footnotesize (Cabibbo suppressed)}~~
& $(1.04\pm 0.06)\times 10^{-3}$   &  ~~$(2.13 \pm 0.36)\times 10^{-3}$ \\
\midrule[1.0pt]\midrule[1.0pt]
%\hline\hline
$D_s^+ \to \rho^+ \eta$ &  $(4.7\pm 1.3)\%$ & $(8.9 \pm 0.8)\%$ \\
\hline
$D_s^+ \to \rho^+ \eta'$ & $(1.4 \pm 0.4)\%$ & $(5.8\pm 1.5)\%$ \\
\hline
 $D_s^+ \to \rho^+ \phi$ & ~(fit to the Exp.)~  & $(8.4^{+1.9}_{-2.3})\%$\\
\hline
$D_s^+ \to \rho^+ K^0$ {\footnotesize (Cabibbo suppressed)}
 & $(9.1\pm 2.5)\times 10^{-3}$   &   \\
\hline
$D_s^+ \to \rho^+ K^{*0}$ {\footnotesize (Cabibbo suppressed)}
 & $(6.9\pm 1.9)\times 10^{-3}$   &   \\
\bottomrule[1.0pt]\bottomrule[1.0pt]
%\hline\hline
\end{tabular}
\label{tab:tab9}
\end{table}

In Table \ref{tab:tab9} we see results for $D_s^+$ decay.
There we have fitted the $D_s^+ \to \pi^{+} \eta$ mode
and the rates for $D_s^+ \to \pi^+ \eta'$, $D_s^+ \to \pi^+ \phi$
modes are a bit smaller than those of experiment.
Following the general trend it is better to assume that
the $D_s^+ \to \pi^{+} \eta$ rate, with smaller masses,
would be a bit overcounted and the rates for the bigger mass modes
would be in better agreement with experiment.
The Cabibbo-suppressed modes of $D_s^+ \to \pi^{+} K^{0}$ and $D_s^+ \to \pi^+ K^{*0}$
would be in fair agreement with experiment within errors.

In the $\rho$ sector, if we fit $D_s^+ \to \rho^+ \phi$, the $D_s^+ \to \rho^+ \eta$ rate is a bit small
compared with experiment and the $D_s^+ \to \rho^+ \eta'$ rate smaller by more than a factor of two,
counting errors.

\subsection{$\boldsymbol{B}$ decays in internal emission}
\label{subsec:resE}

We look at the cases
\begin{enumerate}[1)]
\setlength{\itemsep}{1.5pt}
\setlength{\parsep}{0.5pt}
\setlength{\parskip}{0.2pt}
\item $\bar B^0 \to \eta_c \bar K^0 $, \;$J/\psi \bar K^0$, \;
      $\eta_c \bar K^{*0} $, \; $J/\psi \bar K^{*0}$ \,;
\item $\bar B^0 \to \psi (2S) \bar K^0 $, \; $\psi (2S) \bar K^{*0}$ \,;
\item $\bar B^0 \to D^0 \pi^0 $, \;$D^{*0} \pi^0$, \; $D^0 \rho^0$, \;
      $D^{*0} \rho^0$ \,;%~ (Cabibbo suppressed)
\item $B^- \to \eta_c K^- $, \;$J/\psi K^-$, \; $\eta_c K^{*-} $, \; $J/\psi K^{*-}$ \,;
\item $B^- \to \psi (2S) K^- $, \; $\psi (2S) K^{*-}$ \,;
\item $B^- \to D^0 \pi^- $, \;$D^{*0} \pi^-$, \; $D^0 \rho^-$, \;
      $D^{*0} \rho^-$ \,;%~ (Cabibbo suppressed)
\item $\bar B_s \to \eta_c \eta $, \;$J/\psi \eta$, \; $\eta_c \phi $, \; $J/\psi \phi$ \,;
\item $\bar B_s \to \eta_c \eta' $, \;$J/\psi \eta'$ \,;
\item $\bar B_s \to D^0 K^0 $, \;$D^{*0} K^0$, \; $D^0 K^{*0}$, \; $D^{*0} K^{*0}$ \,;
      %~ (Cabibbo suppressed)
\item $B_c^- \to \eta_c D_s^- $, \;$J/\psi D_s^-$, \;
      $\eta_c D_s^{*-} $, \; $J/\psi D_s^{*-}$ \,;
\item $B_c^- \to D^0 D^- $, \;$D^{*0} D^-$, \; $D^0 D^{*-} $, \;
      $D^{*0} D^{*-}$ \,. %~ (Cabibbo suppressed)
\end{enumerate}

Note again that in the case of $\pi^0, \rho^0$ production
we must multiply the standard formula by $\frac{1}{2}$
and in the case of $\eta, \eta'$ production
by $\frac{1}{3}$ and $\frac{2}{3}$ respectively,
considering the $d\bar d$ and $s \bar s$ content of these mesons.
Note also that the case 2) is unrelated to the other,
because we have a different radial wave function,
but we can calculate the ratio of the two,
and relate to the rates of case 5).

Tables \ref{tab:tab10}, \ref{tab:tab11} and \ref{tab:tab12}
show the branching ratios for $\bar B^0, B^-$ and $\bar B_s$ decays in internal emission,
and Table \ref{tab:tab13}
shows the rates of branching ratios for $\bar B_c^-$ decays in internal emission.

\begin{table}[bt]
\renewcommand\arraystretch{1.5}
\caption{Branching ratios for $\bar B^0$ decays in internal emission.}
\centering
\begin{tabular}{l  c  c }
\toprule[1.0pt]\toprule[1.0pt]
%\hline\hline
{Decay process~} ~& BR (Theo.)  &  BR (Exp.)\\
\hline
 $\bar B^0 \to \eta_c \bar K^0$ & $(1.23\pm 0.05)\times 10^{-3}$   & $(8.0 \pm 1.2)\times 10^{-4}$\\
\hline
 $\bar B^0 \to J/\psi \bar K^0$ & ~~(fit to the Exp.)~~   & ~$(8.73 \pm 0.32)\times 10^{-4}$~\\
\hline
$\bar B^0 \to \eta_c \bar K^{*0} $ & $(4.53\pm 0.17)\times 10^{-4}$ & $(6.3 \pm 0.9)\times 10^{-4}$ \\
\hline
 $\bar B^0 \to J/\psi \bar K^{*0}$ & $(1.31\pm 0.05)\times 10^{-3}$  & ~~$(1.28\pm 0.05)\times 10^{-3}$\\
\midrule[1.0pt]\midrule[1.0pt]
%\hline\hline
$\bar B^0 \to \psi(2S) \bar K^{0}$ & $(2.9\pm 0.2)\times 10^{-4}$   &  $(5.8 \pm 0.5)\times 10^{-4}$ \\
\hline
$\bar B^0 \to \psi(2S) \bar K^{*0}$ & ~~(fit to the Exp.)~~  & $(5.9 \pm 0.4)\times 10^{-4}$ \\
\midrule[1.0pt]\midrule[1.0pt]
%\hline\hline
$\bar B^0 \to D^0 \pi^0$ %{\footnotesize (Cabibbo suppressed)}
& $(2.11\pm 0.14)\times 10^{-3}$ & ~~$(2.63\pm 0.14)\times 10^{-4}$ \\
\hline
 $\bar B^0 \to D^{*0} \pi^0$ %{\footnotesize (Cabibbo suppressed)}
 & $(1.71\pm 0.11)\times 10^{-3}$  & $(2.2\pm 0.6)\times 10^{-4}$\\
\hline
$\bar B^0 \to D^0 \rho^0 $ %{\footnotesize (Cabibbo suppressed)}
 & ~~(fit to the Exp.)~~  & ~~$(3.21\pm 0.21)\times 10^{-4}$  \\
\hline
$\bar B^0 \to D^{*0} \rho^0 $ %{\footnotesize (Cabibbo suppressed)}
 & $(4.63 \pm 0.30) \times 10^{-4}$   & $< 5.1 \times 10^{-4}$  \\
\bottomrule[1.0pt]\bottomrule[1.0pt]
%\hline\hline
\end{tabular}
\label{tab:tab10}
\end{table}

In Table \ref{tab:tab10} we show results for $\bar B^0$ decays with internal emission.
In the $\eta_c, J/\psi$ decay sector,
the rates obtained are in quite good agreement with experiment,
with the  $\bar B^0 \to \eta_c \bar K^0$ rate a bit overcounted,
following the trend for all PP decays,
due to the smaller masses involved,
which would produce larger momentum transfers,
and, thus, reduced intrinsic form factors.

The modes with $\psi(2S)$ in the final state can be considered in just rough agreement.
In the $D^0, D^{*0}$ decay sector, once again the pionic modes are grossly overcounted
and the predictions for $\bar B^0 \to D^{*0} \rho^0 $ are compatible with the upper experimental bound.

\begin{table}[t!]
\renewcommand\arraystretch{1.5}
\caption{Branching ratios for $B^-$ decays in internal emission.}
\centering
\begin{tabular}{l  c  c }
\toprule[1.0pt]\toprule[1.0pt]
%\hline\hline
{Decay process~} ~& BR (Theo.)  &  BR (Exp.)\\
\hline
 $B^- \to \eta_c K^-$ & $(1.45 \pm 0.04)\times 10^{-3}$   & $(9.6 \pm 1.1)\times 10^{-4}$\\
\hline
 $B^- \to J/\psi K^-$ & ~~(fit to the Exp.)~~   & ~$(1.026 \pm 0.031)\times 10^{-3}$~\\
\hline
$B^- \to \eta_c K^{*-} $ & $(5.32 \pm 0.16)\times 10^{-4}$ & $(1.0^{+0.5}_{-0.4})\times 10^{-3}$ \\
\hline
 $B^- \to J/\psi K^{*-}$ & $(1.53 \pm 0.05)\times 10^{-3}$  & $(1.43\pm 0.08)\times 10^{-3}$\\
\midrule[1.0pt]\midrule[1.0pt]
%\hline\hline
$B^- \to \psi(2S) K^{-}$ & $(3.4\pm 0.7)\times 10^{-4}$   &  $(6.26 \pm 0.24)\times 10^{-4}$ \\
\hline
$B^- \to \psi(2S) K^{*-}$ & ~~(fit to the Exp.)~~  & $(6.7 \pm 1.4)\times 10^{-4}$ \\
%\midrule[1.0pt]\midrule[1.0pt]
%%\hline\hline
%$B^- \to D^0 \pi^-$ %{\footnotesize (Cabibbo suppressed)}
%& $(8.52\pm 1.14)\times 10^{-2}$ & $(4.80\pm 0.15)\times 10^{-3}$ \\
%\hline
% $B^- \to D^{*0} \pi^-$% {\footnotesize (Cabibbo suppressed)}
% & $(6.91 \pm 0.93) \times 10^{-2}$  & $(5.18\pm 0.26)\times 10^{-3}$\\
%\hline
%$B^- \to D^0 \rho^- $% {\footnotesize (Cabibbo suppressed)}
% & ~~(fit to the Exp.)~~   & $(1.34\pm 0.18)\%$  \\
%\hline
%$B^- \to D^{*0} \rho^- $ %{\footnotesize (Cabibbo suppressed)}
% & $(1.93 \pm 0.26) \times 10^{-2}$   & $(9.8 \pm 1.7) \times 10^{-3}$  \\
\bottomrule[1.0pt]\bottomrule[1.0pt]
%\hline\hline
\end{tabular}
\label{tab:tab11}
\end{table}

In Table \ref{tab:tab11} we show results for $B^-$ decays in internal emission.
The results are related to the former ones
since we have just changed a $\bar d$ spectator quark by a $\bar u$ quark.
The results are similar to those of $\bar B^0$ decays,
with a bit of overcounting for the  $B^- \to \eta_c K^-$ rate.
The other rates involving $\eta_c$ or $J/\psi$ are basically compatible with experiment
within errors.
%In the $D^0, D^{*0}$ decay sector, the modes with a pion in the final state are overcounted,
%as usual, and the $D^{*0} \rho^-$ mode also a bit overcounted.
We omit to present results in Table \ref{tab:tab11} for the $B^- \to D^0 \pi^-, D^{*0} \pi^-, D^0 \rho^-, D^{*0} \rho^-$ decays.
Indeed, these modes proceed both via internal but also external emission and there is a constructive interference between these modes.
We discussed this issue when referring to Table \ref{tab:tab4} in subsection \ref{subsec:resC}.
The external emission mode, color favored, is dominant,
but the color suppressed internal emission mode,
through interference increases the decay width of these modes by about a factor of two.

\begin{table}[h!]
\renewcommand\arraystretch{1.5}
\caption{Branching ratios for $\bar B_s$ decays in internal emission.}
\centering
\begin{tabular}{l  c  c }
\toprule[1.0pt]\toprule[1.0pt]
%\hline\hline
{Decay process~} ~& BR (Theo.)  &  BR (Exp.)\\
\hline
 $\bar B_s \to \eta_c \eta$ & $(3.63\pm 0.27)\times 10^{-4}$   & \\
\hline
 $\bar B_s \to J/\psi \eta$ &  $(2.55\pm 0.19)\times 10^{-4}$  & ~$(4.0 \pm 0.7)\times 10^{-4}$~\\
\hline
$\bar B_s \to \eta_c \phi $ & $(3.64\pm 0.27)\times 10^{-4}$ &  \\
\hline
 $\bar B_s \to J/\psi \phi$ & ~~(fit to the Exp.)~~  & ~~$(1.08\pm 0.08)\times 10^{-3}$\\
\hline
$\bar B_s \to \eta_c \eta'$ & $(3.88\pm 0.29)\times 10^{-4}$   &   \\
\hline%\hline
$\bar B_s \to J/\psi \eta'$ & $(2.24\pm 0.17)\times 10^{-4}$  & $(3.3 \pm 0.4)\times 10^{-4}$ \\
\midrule[1.0pt]\midrule[1.0pt]
%\hline\hline
$\bar B_s \to D^0 K^0$ & ~~(fit to the Exp.)~~ & $(4.3\pm 0.9)\times 10^{-4}$ \\
\hline
 $\bar B_s \to D^{*0} K^0$  & $(3.4\pm 0.7)\times 10^{-4}$  & $(2.8\pm 1.1)\times 10^{-4}$\\
\hline
$\bar B_s \to D^0 K^{*0} $% {\footnotesize (Cabibbo suppressed)}
 & $(2.1 \pm 0.4)\times 10^{-4}$   & $(4.4\pm 0.6)\times 10^{-4}$  \\
\hline
$\bar B_s \to D^{*0} K^{*0} $ %{\footnotesize (Cabibbo suppressed)}
 & $(3.0\pm 0.6) \times 10^{-4}$   &   \\
\bottomrule[1.0pt]\bottomrule[1.0pt]
%\hline\hline
\end{tabular}
\label{tab:tab12}
\end{table}

In Table \ref{tab:tab12} we show results for $\bar B_s$ decays in internal emission.
In the $J/\psi, \eta_c$ sector the results obtained are fair compared with the experiment.
Those in the $D^0, D^{*0}$ decay sector are also fair.

\begin{table}[h!]
\renewcommand\arraystretch{1.5}
\caption{Rates of branching ratios for $B_c^-$ decays in internal emission.}
\centering
\begin{tabular}{l  c  c }
\toprule[1.0pt]\toprule[1.0pt]
%\hline\hline
{~~~~~~~~~~~~Rate~~~~~~~~~~~~} ~& ~~~Theo.~~~  &  ~~~Exp.~~~\\
\hline
 $\Gamma_{B_c^- \to \eta_c D_s^{-}}/\Gamma_{B_c^- \to J/\psi D_s^{-} }$ & $2.03$   & \\
\hline
$\Gamma_{B_c^- \to \eta_c D_s^{*-} }/\Gamma_{ B_c^- \to J/\psi D_s^{-}}$ & $0.98$   & \\
\hline
 $\Gamma_{B_c^- \to J/\psi D_s^{*-} }/\Gamma_{B_c^- \to J/\psi D_s^{-}}$ & $3.44$   & $2.5 \pm 0.5$\\
\specialrule{0em}{2pt}{2pt}
\midrule[1.0pt]\midrule[1.0pt]
%\hline\hline
$\Gamma_{B_c^- \to D^{0} D^-}/\Gamma_{B_c^- \to D^{*0} D^-}$ &  $1.42$  & \\
\hline
$\Gamma_{B_c^- \to D^{0} D^{*-}}/\Gamma_{B_c^- \to D^{*0} D^-}$ &  $1.07$  & \\
\hline
$\Gamma_{B_c^- \to D^{*0} D^{*-}}/\Gamma_{B_c^- \to D^{*0} D^-}$ &  $1.61$  & \\
\bottomrule[1.0pt]\bottomrule[1.0pt]
%\hline\hline
\end{tabular}
\label{tab:tab13}
\end{table}

In Table \ref{tab:tab13} we show results of rates involving $B^-_c$ decays.
The only one measured is in fair agreement with experiment.

At this point we would like to make some discussion about the momentum transfer involved in the reactions
and the repercussion in form factors.
In Table \ref{tab:tabnew} we show some reactions involving pions in the final state and the reaction used to normalize the data,
together with the overcounting factor in the $\pi$ decay mode.
The momentum transfer from one hadron to another is calculated in the rest frame of the decaying particle.

\begin{table}[tbh]
\renewcommand\arraystretch{1.5}
\caption{The momentum transfer ($q$) in reactions involving pions in the final state
and the approximated $\pi$ overcounting factor (OCF).}
\centering
\begin{tabular}{l  c  l  c}
\toprule[1.0pt]\toprule[1.0pt]
%\hline\hline
%{Reaction} ~& Momentum transfer [MeV/$c$]  &  fitted to & $\pi$ overcounting factor (approximated) \\
{~~Reaction} ~& ~~$q$ [MeV/$c$]~~  &  ~~~fitted to & ~~OCF~~ \\
\hline
 $\Lambda_b \to \pi^- \Lambda_c$ & $2342$   & $\Lambda_b \to K^- \Lambda_c$  &  $15$ \\
\hline
$\Lambda_b \to K^- \Lambda_c$ & $2314$   &  & \\
\hline
 $\Lambda_c \to \pi^+ \Lambda$ & $864$   & $\Lambda_c \to K^+ \Lambda$  & $15$\\
\hline
%\hline\hline
$\Lambda_c \to K^+ \Lambda$ &  $781$  & &\\
\specialrule{0em}{2pt}{2pt}
\midrule[1.0pt]\midrule[1.0pt]
$\bar B^0 \to \pi^- D^+$ &  $2306$   & $\bar B^0 \to \rho^- D^+$ & $100$\\
\hline
$\bar B^0 \to \rho^- D^+$ &  $2235$  & &\\
\hline
$B^- \to \pi^- D^0$ &  $2308$  & $B^- \to \rho^- D^0$ & $100$\\
\hline
$B^- \to \rho^- D^0$ & $2237$ &  & \\
\specialrule{0em}{2pt}{2pt}
\midrule[1.0pt]\midrule[1.0pt]
$\bar B_s^0 \to \pi^- D_s^+$ &  $2320$ & $\bar B_s^0 \to \rho^- D_s^+$ & $100$\\
$\bar B_s^0 \to \rho^- D_s^+$  & $2248$ &  &\\
\specialrule{0em}{2pt}{2pt}
\midrule[1.0pt]\midrule[1.0pt]
$D^0 \to \pi^+ K^-$ &  $861$  & $D^0 \to \pi^+ K^{*-}$ & $5$\\
\hline
$D^0 \to \pi^+ K^{*-}$ &  $711$  & &\\
\hline
$D^+ \to \pi^+ \bar K^0$ &  $863$  & $D^+ \to \pi^+ \bar K^{*0}$ & $2$\\
\hline
$D^+ \to \pi^+ \bar K^{*0}$ & $712$ &  & \\
\specialrule{0em}{2pt}{2pt}
\midrule[1.0pt]\midrule[1.0pt]
$\bar B^0 \to \pi^0 D^0$ &  $2308$  & $\bar B^0 \to \rho^0 D^0$ & $8$\\
\hline
$\bar B^0 \to \rho^0 D^0$ &  $2237$  & &\\
\specialrule{0em}{2pt}{2pt}
\midrule[1.0pt]\midrule[1.0pt]
$B^- \to D_s^- D^0$ &  $1814$  &  & \\
$B^- \to D_s^{*-} D^0$  & $1734$ &  &\\
\bottomrule[1.0pt]\bottomrule[1.0pt]
%\hline\hline
\end{tabular}
\label{tab:tabnew}
\end{table}

In the second and third blocks, where the $\pi$ overcounting factor is of the order of $100$,
we see that the momentum transfer is very large and the difference between momenta in the $\pi^- D^+$
and $\rho^- D^+$ decay modes of of about $70\; {\rm MeV}/c$.
This seems to indicate that we are at the tail of the form factor where it decreases rapidly
and a difference of $70\; {\rm MeV}/c$ can make such a difference.
On the contrary in the first block, where the overcounting factor is of the order of $15$,
the difference of momenta between $\Lambda_b \to \pi^- \Lambda_c$ and $\Lambda_b \to K^- \Lambda_c$
is only about $30\; {\rm MeV}/c$ which makes the changes in the form factor less drastic.
For the case of $\Lambda_c \to \pi^+ \Lambda$ and $\Lambda_c \to K^+ \Lambda$ the difference
of momenta is of the order of $83\; {\rm MeV}/c$,
larger than before, but the total momentum transfers are substantially smaller,
so we are in a region where the form factor do not fall down so fast.

In the fourth block the momenta are similar as in the $\Lambda_c$ decay modes.
The difference of momenta between $D^0 \to \pi^+ K^-$ and $D^0 \to \pi^+ K^{*-}$
is $150\, {\rm MeV}/c$ and the total momentum transfer is much smaller than in the $\bar B$ decay modes,
so we obtain an overcounting factor of about $5$, much smaller than in the $\bar B$ decays.
The overcounting factor is about a factor of two for
the $D^+ \to \pi^+ \bar K^0$ and $D^+ \to \pi^+ \bar K^{*0}$,
but counting errors in the rates,
the difference between these two cases is not very large,
and the important thing is that qualitatively
we can understand the reason for these different overcounting factors.

In the fifth block for the $\bar B^0 \to \pi^0 D^0$ and $\bar B^0 \to \rho^0 D^0$ decay modes,
we find a surprise, since the difference of momenta between them is of the order of $70\; {\rm MeV}/c$,
like that in the second block and we should also expect an overcounting factor of the order of $100$.
Yet, the overcounting factor is only of the order of $8$.
The difference between these decays is that $\bar B^0 \to \pi^- D^+$ proceeds via external emission
and $\bar B^0 \to D^0 \pi^0$ proceeds via internal emission.
We find a plausible explanation for this: In external emission the momentum transfer, $q$,
is carried by a single $W$ (see Fig.~\ref{Fig:4}),
while in internal emission this momentum can be shared by two $Wq\bar q$ transitions (see Fig.~\ref{Fig:7}).
It is well known in nuclear physics, applying Glauber theory,
that in such cases,
the optimal rate appears when the momentum transferred is equally shared in the two scattering points \cite{glauber,oset}.
Then, assuming a simple form factor $e^{- \alpha^2 q^2}$,
typical of quark models,
we would have $e^{- \alpha^2 (q/2)^2}\, e^{- \alpha^2 (q/2)^2} = e^{- \alpha^2 q^2/2}$
in internal emission versus $e^{- \alpha^2 q^2}$ in external emission.
So, the effect of form factors should be more drastic in external emission.

Finally in the sixth block we show for reference the momentum transfers in the
$B^- \to D_s^- D^0$ and $B^- \to D_s^{*-} D^0$.
We see that the momenta are smaller than in the pionic modes studied before,
and this should make the predictions in that sector more reliable.

\subsection{$\boldsymbol{D}$ decays in internal emission}
\label{subsec:resF}

We have the following cases.
\begin{enumerate}[1)]
\setlength{\itemsep}{1.5pt}
\setlength{\parsep}{0.5pt}
\setlength{\parskip}{0.2pt}
\item $D^0 \to \bar K^0 \pi^0 $, \;$\bar K^{*0} \pi^0$, \;
$\bar K^0 \rho^0 $, \; $\bar K^{*0} \rho^0$ \,;
\item $D^0 \to \pi^0 \pi^0 $, \;$\rho^0 \pi^0$, \;
$\pi^0 \rho^0 $, \; $\rho^0 \rho^0$ \,;~ (Cabibbo suppressed)
\item $D^+ \to \bar K^0 \pi^+ $, \;$\bar K^{*0} \pi^+$, \;
$\bar K^0 \rho^+ $, \; $\bar K^{*0} \rho^+$ \,;
\item $D^+ \to \pi^0 \pi^+ $, \;$\rho^0 \pi^+$, \; $\pi^0 \rho^+ $, \;
$\rho^0 \rho^+$ \,;~ (Cabibbo suppressed)
\item $D_s^+ \to \bar K^0 K^+ $, \;$\bar K^{*0} K^+$, \;
$\bar K^0 K^{*+} $, \; $\bar K^{*0} K^{*+}$ \,;
\item $D_s^+ \to \pi^0 K^+ $, \;$\rho^0 K^+$, \;
$\pi^0 K^{*+} $, \; $\rho^0 K^{*+}$ \,.~ (Cabibbo suppressed)
\end{enumerate}

We should note that the decay modes for $D^+$ are the same ones as in external emission of Table \ref{tab:tab8}
and there can be a mixing.
The amplitudes with external emission are bigger, since the mode is color favored,
and this mode will dominate.
However, as we saw in the discussion concerning the $\bar B^0 \to \pi^- D^+$ and $B^- \to \pi^- D^0$ decays,
the interference of the two mechanisms can lead to larger decay rates than with the external emission alone.
However, in the present case by comparing the $D^+ \to \pi^+ \bar K^0$ and $D^0 \to \pi^+ K^-$,
if the pattern of interference was like in $B \to \pi D$ decays
we should expect a bigger rate for $D^+ \to \pi^- \bar K^0$, which proceeds via the two mechanisms.
Yet, experimentally the rate for $D^0 \to \pi^+ K^-$ is bigger than for $D^+ \to \pi^+ K^0$
and the same happens with the $D^+ \to \pi^+ \rho^0$ versus $D^0 \to \pi^+ \rho^-$,
where the second rate is bigger than the first, even if we multiply by a factor $2$
the $D^+ \to \pi^+ \rho^0$ rate to account for the reduction factor of $1/2$ mentioned above.
It is clear that the pattern of interference is different for $D$ mesons and $B$ mesons.
Yet, as in the case of Tables \ref{tab:tab3} and \ref{tab:tab4},
the ratio of rates in Table \ref{tab:tab8} should be fair.

For $D^0$ and $D_s$ decays, the modes obtained here are different than for external emission,
but they can be reached by strong interaction rescattering.
By looking at Tables \ref{tab:tab3} and \ref{tab:tab10}
for $\bar B^0$ decay in external and internal emission,
we see that the former have one order of magnitude bigger rates than the latter.
It is then quite likely that the internal emission modes are easier reached by external emission and strong rescattering, and, thus, predictions made from the internal emission formulas would be misleading.
We thus refrain from showing results for these modes.

One might argue that the $B$ decay modes from internal emission could also
be obtained by external emission followed by strong interaction rescattering.
Yet, this is far more unlikely than in $D$ decays.
Indeed, take $\bar B^0 \to D_s^- D^+$ in external emission and $B^0 \to \eta_c \bar K^0$ in internal emission.
$D_s^- D^+$ and $\eta_c \bar K^0$ are coupled channels,
but transition from one to the other requires the exchange of a $D_s^*$ vector meson
in the extended hidden gauge approach
and is penalized by the large mass in the $D_s^*$ propagator \cite{ramos,feijoo,omegac}.
On the contrary, if we take $D^0 \to \pi^+ K^-$ from external emission
and $D^0 \to \pi^0 \bar K^0$ from internal emission,
the $\pi^+ K^- \to \pi^0 \bar K^0$ transition requires the exchange of a $\rho$
and gives rise to the standard chiral potential.
The $B$ decay modes by internal emission are thus, genuine modes,
with expected small interference from external emission followed by rescattering.

\section{Conclusions}
\label{sec:conc}

We have made a study of the properties of internal and external emission
in the weak decay of heavy hadrons from the point of view of the spin-angular momentum
structure, differentiating among the vector and pseudoscalar decay modes.
The rest of the structure is given by intrinsic form factors
related to the spatial wave functions of the quark states,
which do not differentiate the spin of the mesons formed.
In this sense, for similar masses of the decay products,
like $\eta_c, J/\psi$ or $D, D^*$,
we can obtain rates of decays up to a global factor.
Yet, we are not using heavy quark symmetry,
and actually we show that the $B \to {\rm PV}, {\rm VP}$ decay modes
are proportional to $(\mathcal{B}p)^2$ or $(\mathcal{B}'p)^2$,
which are terms of type $(\frac{p}{m_Q})^2$, with $m_Q$ the heavy quark mass,
and would be neglected in a strict heavy quark symmetry counting.
We show that these modes have a similar strength than the
$B\to {\rm PP, \; VV}$ modes which survive in the heavy quark limit,
and these predictions are corroborated by experiment.

The derivation of the final formulas requires a good deal of angular momentum algebra
that we have written in the appendices.
Yet, the final formulas are rather easy and we can show
that $\overline \sum \sum |t|^2$ is formally
the same for internal and external emission.

We applied the formulas to correlate a large amount of data in $\Lambda_b, \Lambda_c$ decays
and $B$ or $D$ decays that involve more than $100$ reactions.
We have taken a given datum for a certain decay rate and
then have made predictions for the related reactions.
The agreement in general is quite good and the discrepancies are systematic.
The most remarkable one is that decay modes involving pions
in the final states are overcounted in our approach.
We gave an explanation for that, because the small mass of
the pion leads to larger momentum transfers that reduce the intrinsic form factors related to
the spatial wave functions of the quarks involved,
which are independent of the spin rearrangements,
since all the quarks are in their ground state and only spin rearrangement
differentiates the pseudoscalar from a vector, say $\eta_c, J/\psi$ or $D, D^*$.

The results obtained go beyond the evaluation of ratios and the predictions
made for decay rates. We have evaluated the amplitudes for $B \to {\rm PP, \; PV, \; VP, \; VV}$
with the momentum and spin structure and proper relative phase,
and thus, this information is valuable
if one wishes to evaluate loops that contain these intermediate channels,
as one would like to do in studies related to the possible lack of universality.

The discrepancies for the case of pion production modes can be used to find information on the
intrinsic form factors involved in the reactions beyond the spin-angular momentum
structure that we have studied in detail.

The formulas obtained are ready to compare with future measurements
that would involve polarizations of the vector mesons produced.

In most cases we have made predictions for rates that have not been yet measured.
The results obtained here can be compared with future measurements.
The rates obtained can also be used in analyses that require estimates of
some rates to induce other rates.

Finally the formalism deduced here also lays the grounds for further studies in which one can
have internal or external emission,
as we have done, and in the final state we hadronize creating
a $q\bar q$ pair with the quantum numbers of the vacuum,
which together with the primary $q' \bar q'$ pair formed leads to two mesons.
In this case one would address decay processes with three particles
in the final state and help correlate a larger amount of decay modes already observed.

\begin{acknowledgments}

This work is partly supported by the National Natural Science Foundation
of China (NSFC) under Grant Nos. 11565007, 11747307 and 11647309.
This work is also partly supported by
the Spanish Ministerio de Economia y Competitividad
and European FEDER funds under the contract number
FIS2011-28853-C02-01, FIS2011-28853-C02-02,
FIS2014-57026-REDT, FIS2014-51948-C2-1-P, and FIS2014-51948-C2-2-P,
and the Generalitat Valenciana in the program Prometeo II-2014/068.

\end{acknowledgments}

\begin{appendix}
\section{External emission $\boldsymbol{\Lambda_b \to D_s^- \,(D_s^{*-})\, \Lambda_c}$ decay }
\label{App:A}

As was shown in Eq.~\eqref{eq:element1}, we must evaluate the matrix element
\begin{equation}\label{eq:App:element1}
 t= \langle S_1 | S_2 \rangle \;
 \langle M' | \gamma^0 -\gamma^0 \gamma_5  |M \rangle
  + \langle S_1 | \sigma^i  | S_2  \rangle  \;
  \langle M' | \gamma^i -\gamma^i \gamma_5 | M \rangle.
\end{equation}
Using the spinors of Eqs.~\eqref{eq:spinor}, \eqref{eq:ABFactor2}
and the $\gamma^\mu$ matrices of Eq.~\eqref{eq:gammaMatrix}, and the property
\begin{equation}\label{eq:App:sigma}
  \sigma_i \, \sigma_j = \delta_{ij}+i \, \epsilon_{ijk} \,\sigma_k,
\end{equation}
we get the result \footnote{We use $\sigma_i \equiv \sigma^i$, $i=1, 2, 3$ along the derivation,
where $\sigma_i$ are the Pauli matrices.}
\begin{equation}\label{eq:App:amplitude}
  t= \mathcal{A}\mathcal{A}'\, [t_1 + t_2 + t_3 +t_4 +t_5+t_6],
\end{equation}
where
\begin{eqnarray}
  t_1 &=& (1+ \mathcal{B}\mathcal{B}'\, \vec{p}^{\,2})\; \langle S_1 | S_2 \rangle \;
  \langle M' | M \rangle, \nonumber\\[2mm]
  t_2 &=& -(\mathcal{B}+\mathcal{B}') \; \langle S_1 | S_2 \rangle \;
  \langle M' |\vec \sigma \cdot \vec p \,| M \rangle, \nonumber\\[2mm]
  t_3 &=& (\mathcal{B}+\mathcal{B}') \; \langle S_1 |\vec \sigma \cdot \vec p \,| S_2 \rangle \;
          \langle M' | M \rangle, \nonumber\\[2mm]
  t_4 &=& (-1+ \mathcal{B}\mathcal{B}'\, \vec{p}^{\,2})\; \langle S_1 | \sigma^i | S_2 \rangle \;
           \langle M' |\sigma^i | M \rangle, \\[2mm]
  t_5 &=& -2\mathcal{B}\mathcal{B}' \; \langle S_1 |\vec \sigma \cdot \vec p \,| S_2 \rangle \;
          \langle M' |\vec \sigma \cdot \vec p \,| M \rangle, \nonumber\\[2mm]
  t_6 &=& i(\mathcal{B}-\mathcal{B}')\; \epsilon_{ijk} \; p^j \;\langle S_1 | \sigma^i | S_2 \rangle \;
           \langle M' |\sigma^k | M \rangle, \nonumber
\end{eqnarray}
with $\mathcal{A}, \mathcal{A}'$ or $\mathcal{B}, \mathcal{B}'$ coming from Eq.~\eqref{eq:ABFactor2}
for $\Lambda_b$ and $\Lambda_c$ respectively.

We proceed now to evaluate the terms $t_i$ of the former equation.
We use angular momentum algebra following Rose convention and formalism \cite{rose}.

\subsection{Term $\boldsymbol{t_1}$}
We combine now $S_1, \, -S_2$ with the phase $(-1)^{1/2+S_2}$ to give angular momentum $jm$,
with $j=0, 1$,
\begin{equation}\label{eq:App:t1}
  t_1 = (1+ \mathcal{B}\mathcal{B}' \, \vec{p}^{\,2})\; \delta_{S_1 \, S_2} \; \delta_{MM'},
\end{equation}
and combinating to angular momentum $j,m$ we get
\begin{eqnarray}
  t_1 &\to& (1+ \mathcal{B}\mathcal{B}' \, \vec{p}^{\,2})\; \sum_{S_1} \,
  \mathcal{C}(\frac{1}{2}\, \frac{1}{2}\, j; \, S_1, -S_2, m)\;
  (-1)^{1/2+S_2}\, \delta_{S_1 \, S_2} \; \delta_{MM'} \nonumber \\[2mm]
   &=& (1+ \mathcal{B}\mathcal{B}' \, \vec{p}^{\,2})\; \sum_{S_1} \,
        \mathcal{C}(\frac{1}{2}\, \frac{1}{2}\, j; \, S_1, -S_1, 0)\;
        (-1)^{1/2+S_1}\,  \delta_{MM'}\; \delta_{m0}.
\end{eqnarray}
Using explicitly the Clebsch-Gordan coefficients (CGC), we find the sum in this last equation
zero for $j=1$ and
\begin{equation}\label{eq:App:t1:2}
 t_1 = -\sqrt{2}\, (1+ \mathcal{B}\mathcal{B}' \, \vec{p}^{\,2})\;  \delta_{MM'}\; \delta_{m0} \; \delta_{j0}
\end{equation}
for $j=0$.

\subsection{Term $\boldsymbol{t_2}$}
We write
\begin{equation}\label{eq:App:sigp}
  \vec \sigma \cdot \vec p = \sum_\mu (-1)^\mu\; \sigma_{-\mu} \; p_\mu,
\end{equation}
with $\sigma_\mu$ in spherical basis ($-\frac{1}{\sqrt{2}}(\sigma_x +i \sigma_y)$,
$\frac{1}{\sqrt{2}}(\sigma_x -i \sigma_y)$, $\sigma_z$), and
\begin{equation*}
    p_\mu =  \sqrt{\frac{4\pi}{3}} \; p \; Y_{1\mu},
\end{equation*}
in terms of the spherical harmonics.

Then
\begin{equation*}
  t_2 = -(\mathcal{B}+\mathcal{B}') \; \delta_{S_1 \, S_2} \; \sum_\mu (-1)^\mu\; p_\mu \,
  \langle M' |\sigma_{-\mu} | M \rangle.
\end{equation*}
Using the Wigner-Eckart theorem
\begin{equation}\label{eq:Wingner}
  \langle M' |\sigma_{-\mu} | M \rangle =\sqrt{3}\;
  \mathcal{C}(\frac{1}{2}\, 1\, \frac{1}{2}; \, M, -\mu, M'),
\end{equation}
which implies $M-\mu=M'$, one has
\begin{equation}\label{eq:App:t2}
  t_2 = -(\mathcal{B}+\mathcal{B}') \; \delta_{S_1 \, S_2} \; (-1)^{M-M'}\; p_{M-M'}\; \sqrt{3}\;\,
  \mathcal{C}(\frac{1}{2}\, 1\, \frac{1}{2}; \, M, M'-M, M').
\end{equation}
Combining now to angular momentum $jm$, we find
\begin{eqnarray}
 t_2 &\to&  \sum_{S_1} \, -(\mathcal{B}+\mathcal{B}') \; \delta_{S_1 \, S_2} \;
 (-1)^{M-M'}\; (-1)^{1/2+S_2}\; \nonumber \\
 &&\times \sqrt{3}\;\,
 \mathcal{C}(\frac{1}{2}\, 1\, \frac{1}{2}; \, M, M'-M, M')\;
 \mathcal{C}(\frac{1}{2}\, \frac{1}{2}\, j; \, S_1, -S_2, m)\;   p_{M-M'},
\end{eqnarray}
which implies $m=0$, and summing explicitly over $S_1$ as in the case of $t_1$, we get
\begin{equation}\label{eq:App:t2:2}
  t_2 = \sqrt{2} (\mathcal{B}+\mathcal{B}') \; \delta_{j \, 0} \; \delta_{m \, 0} \; (-1)^{M-M'}\;
  \sqrt{3}\; \mathcal{C}(\frac{1}{2}\, 1\, \frac{1}{2}; \, M, M'-M, M')\;   p_{M-M'},
\end{equation}
which can also be written, using Eqs.~\eqref{eq:App:sigp}, \eqref{eq:Wingner}, as
\begin{equation}\label{eq:App:t2:3}
  t_2 = \sqrt{2} (\mathcal{B}+\mathcal{B}') \; \delta_{j \, 0} \; \delta_{m \, 0} \;
  \langle M' |\vec \sigma \cdot \vec p \,| M\rangle.
\end{equation}

\subsection{Term $\boldsymbol{t_3}$}
\begin{eqnarray}
  t_3 &=& (\mathcal{B}+\mathcal{B}') \; \langle S_1 |\vec \sigma \cdot \vec p \,| S_2 \rangle \;
  \delta_{MM'}  \nonumber\\[2mm]
   &=& (\mathcal{B}+\mathcal{B}') \; \sum_\mu (-1)^\mu \;
   \langle S_1 |\sigma_{-\mu}| S_2 \rangle \; p_\mu\; \delta_{MM'},
\end{eqnarray}
and now combining to $jm$ we have
\begin{eqnarray}
 t_3 &\to& \sum_{S_1} \, (\mathcal{B}+\mathcal{B}') \;
 \mathcal{C}(\frac{1}{2}\, \frac{1}{2}\, j; \, S_1, -S_2, m)\;
 (-1)^{1/2+S_2}\; \nonumber \\[2mm]
 &&\times \sqrt{3}\;\sum_\mu (-1)^\mu \;
 \mathcal{C}(\frac{1}{2}\, 1\, \frac{1}{2}; \, S_2, -\mu, S_1)\;
 p_\mu\; \delta_{MM'},
\end{eqnarray}
which implies $m=S_1-S_2$ and $\mu= S_2-S_1=-m$.
Permuting the order of the arguments in the second CGC,
\begin{equation}
 \mathcal{C}(\frac{1}{2}\, 1\, \frac{1}{2}; \, S_2, S_1-S_2, S_1)
 =\sqrt{\frac{2}{3}}\, (-1)^{1/2-S_2}\;
 \mathcal{C}(\frac{1}{2}\, \frac{1}{2}\, 1; \, S_1, -S_2, S_1-S_2),
\end{equation}
and then
\begin{equation}
\sum_{S_1} \mathcal{C}(\frac{1}{2}\, \frac{1}{2}\, 1; \, S_1, -S_2, S_1-S_2)\;\,
 \mathcal{C}(\frac{1}{2}\, \frac{1}{2}\, j; \, S_1, -S_2, S_1-S_2)=\delta_{1j},
\end{equation}
where we have fixed $S_1-S_2=m$.
Hence we have
\begin{equation}\label{eq:Appt3:1}
 t_3=-\sqrt{2} (\mathcal{B}+\mathcal{B}') \; (-1)^{-m}\; p_{-m} \;\delta_{MM'} \; \delta_{j1},
\end{equation}
which can be rewritten in terms of the polarization vector
$\vec \epsilon$ of the $j=1$ $D_s^*$ state as
\begin{equation}\label{eq:Appt3:2}
 t_3=-\sqrt{2} (\mathcal{B}+\mathcal{B}') \;  \vec \epsilon \cdot \vec p \;\, \delta_{MM'} \; \delta_{j1}\;,
\end{equation}
since for a vector polarization $\epsilon_m$ in spherical basis
one gets the expression of Eq.~\eqref{eq:Appt3:1}.

\subsection{Term $\boldsymbol{t_4}$}
\vspace{-0.8cm}
\begin{equation*}
  t_4 = (-1+ \mathcal{B}\mathcal{B}'\, \vec{p}^{\,2})\; \langle S_1 | \sigma^i | S_2 \rangle \;
           \langle M' |\sigma^i | M \rangle,
\end{equation*}
which can be written in spherical basis as
\begin{eqnarray*}
  t_4 &=& (-1+ \mathcal{B}\mathcal{B}'\, \vec{p}^{\,2})\; \sum_\mu (-1)^\mu \;
  \langle S_1 | \sigma_{-\mu} | S_2 \rangle \;
  \langle M' |\sigma_{\mu} | M \rangle\\[2.5mm]
   &=& (-1+ \mathcal{B}\mathcal{B}'\, \vec{p}^{\,2})\; \sum_\mu  (-1)^\mu \; \sqrt{3} \;
   \mathcal{C}(\frac{1}{2}\, 1\, \frac{1}{2}; \, S_2, -\mu, S_1)\; \sqrt{3} \;
   \mathcal{C}(\frac{1}{2}\, 1\, \frac{1}{2}; \, M, \mu, M').
\end{eqnarray*}
Combining spins to produce the $jm$ state we have
\begin{eqnarray}
 t_4 &\to& (-1+ \mathcal{B}\mathcal{B}'\, \vec{p}^{\,2})\; \sum_{S_1}\sum_\mu  (-1)^\mu \; (-1)^{1/2+S_2}\;
 \mathcal{C}(\frac{1}{2}\, \frac{1}{2}\, j; \, S_1, -S_2, m)\;\nonumber \\[2mm]
 &&\times \sqrt{3}\; \mathcal{C}(\frac{1}{2}\, 1\, \frac{1}{2}; \, S_2, -\mu, S_1)\;
 \langle M' |\sigma_\mu|M \rangle,
 \end{eqnarray}
which implies $S_1-S_2=m$, $S_2-\mu =S_1$ and hence $\mu =-m$.
Then we can permute arguments in the second CGC and find
\begin{equation*}
  \mathcal{C}(\frac{1}{2}\, 1\, \frac{1}{2};
  \, S_2, -\mu, S_1)= (-1)^{1/2-S_2} \sqrt{\frac{2}{3}}\;
  \mathcal{C}(\frac{1}{2}\, \frac{1}{2}\, 1; \, S_1, -S_2, m),
\end{equation*}
and keeping $m$ fixed
\begin{equation*}
  \sum_{S_1}\mathcal{C}(\frac{1}{2}\, \frac{1}{2}\, 1; \, S_1, m-S_1, m)\;
  \mathcal{C}(\frac{1}{2}\, \frac{1}{2}\, j; \, S_1, m-S_1, m)=\delta_{j1}.
\end{equation*}
Hence we get
\begin{equation}\label{eq:Appt4:1}
  t_4 = -(-1+ \mathcal{B}\mathcal{B}'\, \vec{p}^{\,2})\; \sqrt{2} \; \delta_{j1}\, (-1)^{-m}\;
   \langle M' |\sigma_{-m}|M \rangle,
\end{equation}
which can be rewritten in terms of the $D_s^{*-}$ polarization $\vec \epsilon$, as
\begin{equation}\label{eq:Appt4:2}
  t_4 = -(-1+ \mathcal{B}\mathcal{B}'\, \vec{p}^{\,2})\; \sqrt{2} \;
   \langle M' |\vec\sigma \cdot \vec \epsilon \,|M \rangle \; \delta_{j1}.
\end{equation}
For later use in meson decay we can write it from Eq.~\eqref{eq:Appt4:1} as
\begin{equation}\label{eq:Appt4:3}
  t_4 = -(-1+ \mathcal{B}\mathcal{B}'\, \vec{p}^{\,2})\; \sqrt{2} \; \delta_{j1}\, (-1)^{-m}\; \sqrt{3}\;
   \mathcal{C}(\frac{1}{2}\, 1\, \frac{1}{2}; \, M, -m, M').
\end{equation}

\subsection{Term $\boldsymbol{t_5}$}
\begin{equation*}
  t_5 = -2\mathcal{B}\mathcal{B}' \; \langle S_1 |\vec \sigma \cdot \vec p \,| S_2 \rangle \;
          \langle M' |\vec \sigma \cdot \vec p \,| M \rangle,
\end{equation*}
We can use the results of $t_3$ and immediately write
\begin{equation}
 t_5 = 2 \sqrt{2}\, \mathcal{B}\mathcal{B}' \; \delta_{j1} \; \vec \epsilon \cdot \vec p \;
          \langle M' |\vec \sigma \cdot \vec p \,| M \rangle,
\end{equation}
but for later use in meson decay we can write it as
\begin{eqnarray}\label{eq:Appt5:1}
  t_5 &=& 2 \sqrt{2}\, \mathcal{B}\mathcal{B}' \; \delta_{j1} \; (-1)^{-m}\, p_{-m} \sum_\mu (-1)^\mu \; p_\mu \;
          \sqrt{3}\; \mathcal{C}(\frac{1}{2}\, 1\, \frac{1}{2}; \, M, -\mu, M') \nonumber\\[2mm]
      &=& 2 \sqrt{6}\, \mathcal{B}\mathcal{B}' \; \delta_{j1} \; (-1)^{-m}\, (-1)^{M-M'} \; p_{-m} \;\,  p_{M-M'}
          \; \, \mathcal{C}(\frac{1}{2}\, 1\, \frac{1}{2}; \, M, M'-M, M'),
 \end{eqnarray}
with $m$ the polarization of the vector meson.

\subsection{Term $\boldsymbol{t_6}$}
\begin{equation*}
  t_6 = i(\mathcal{B}-\mathcal{B}')\; \epsilon_{ijk} \; p^j \;\langle S_1 | \sigma^i | S_2 \rangle \;
           \langle M' |\sigma^k | M \rangle,
\end{equation*}
which can be written in spherical basis as
\begin{equation}\label{eq:Appt6:1}
 t_6 = (\mathcal{B}-\mathcal{B}')\; (-\sqrt{2}) \; \sum_{\mu\nu}\mathcal{C}(1\, 1\, 1; \, \mu, \nu, \mu+\nu)\;
       \langle S_1 | \sigma_\mu | S_2 \rangle \;
       \langle M' |\sigma_\nu | M \rangle \; (-1)^{\mu+\nu} \; p_{-\mu -\nu},
\end{equation}
as one can see explicitly by writing $\sigma^i$ in terms of $\sigma_\mu$ (see also Ref.~\cite{rose}).

Combining to $jm$, following the steps of $t_3$, we have
\begin{eqnarray}
&& \sum_{S_1} \; \mathcal{C}(\frac{1}{2}\, \frac{1}{2}\, j; \, S_1, -S_2, m)\; (-1)^{1/2+S_2}\;
  \langle S_1 | \sigma_\mu | S_2 \rangle \nonumber\\[2mm]
&=& \sum_{S_1} \; \mathcal{C}(\frac{1}{2}\, \frac{1}{2}\, j; \, S_1, -S_2, m)\; (-1)^{1/2+S_2}\;
    \sqrt{3}\; \mathcal{C}(\frac{1}{2}\, 1\, \frac{1}{2}; \, S_2, \mu, S_1),
\end{eqnarray}
which implies $S_1 -S_2=m$ and $S_2 +\mu =S_1$, hence $\mu =m$.
Permuting the arguments of the second CGC, we have
\begin{equation*}
  \mathcal{C}(\frac{1}{2}\, 1\, \frac{1}{2}; \, S_2, \mu, S_1) =(-1)^{1/2-S_2}\;
  \sqrt{\frac{2}{3}}\;\, \mathcal{C}(\frac{1}{2}\, \frac{1}{2}\, 1; \, S_1, -S_2, \mu),
\end{equation*}
and then keeping $m$ fixed
\begin{equation}
  \sum_{S_1} \; \mathcal{C}(\frac{1}{2}\, \frac{1}{2}\, 1; \, S_1, m-S_1, m)\;
  \mathcal{C}(\frac{1}{2}\, \frac{1}{2}\, j; \, S_1, m-S_1, m)\, =\delta_{j1},
\end{equation}
so
\begin{equation}\label{eq:Appt6:2}
  t_6 \to (\mathcal{B}-\mathcal{B}')\; 2 \, \delta_{j1}\; \sum_{m\nu} \,
  \mathcal{C}(1\, 1\, 1; \, m, \nu)\; (-1)^{m+\nu}\; p_{-m-\nu}\;
  \sqrt{3}\; \mathcal{C}(\frac{1}{2}\, 1\, \frac{1}{2}; \, M, \nu, M'),
\end{equation}
which implies $\nu = M'-M$,
or equivalently
\begin{equation}\label{eq:Appt6:3}
  t_6 = -\sqrt{2}\,(\mathcal{B}-\mathcal{B}')\;  \delta_{j1}\; (-\sqrt{2}) \; \sum_{m\nu} \,
  \mathcal{C}(1\, 1\, 1; \, m, \nu)\; (-1)^{m+\nu}\; p_{-m-\nu}\;
  \langle M' | \sigma_\nu |M \rangle,
\end{equation}
which can be written in terms of the polarization vector $\vec \epsilon$
by analogy to Eq.~\eqref{eq:Appt6:1} as
\begin{equation}\label{eq:Appt6:4}
  t_6 = -i \sqrt{2}\,(\mathcal{B}-\mathcal{B}')\;  \delta_{j1}\; (\vec \epsilon \times \vec p \,) \cdot
    \langle M' | \vec \sigma |M \rangle.
\end{equation}
This term does not interfere with any of the other terms and one easily finds that
\begin{equation}\label{eq:Appt6:5}
  \overline{\sum} \sum_{\rm pol} |t_6|^2 = 4\,(\mathcal{B}-\mathcal{B}')^2\;  {\vec p}^{\, 2}.
\end{equation}

\subsection{$\boldsymbol{\overline{\sum} \sum |t|^2}$ with all terms}
Next we perform the sum and average of $|t|^2$
which will appear in the decay width of the $\Lambda_b$ state.
We have the two cases:
\begin{enumerate}[A)]
\setlength{\itemsep}{1.5pt}
\setlength{\parsep}{0.5pt}
\setlength{\parskip}{0.2pt}
\item $j=0$. We have contribution from $t_1, t_2$,
\begin{eqnarray}\label{eq:Appj0:1}
  \frac{1}{2} \sum_{M, M'} |t|^2
  &=& \frac{1}{2} \sum_{M, M'} \left| \mathcal{A}\mathcal{A}'\;
   \big[ -\sqrt{2}\, (1+ \mathcal{B}\mathcal{B}' \, \vec{p}^{\,2})\;  \delta_{MM'}
   +\sqrt{2} (\mathcal{B}+\mathcal{B}') \;
   \langle M' |\vec \sigma \cdot \vec p \,| M\rangle \big] \right|^2 \nonumber \\[2mm]
  &=& 2(\mathcal{A}\mathcal{A}')^2 \big[ (1+ \mathcal{B}\mathcal{B}' \, \vec{p}^{\,2}\,)^2 +  (\mathcal{B}+\mathcal{B}')^2 \; \vec{p}^{\,2}   \big].
\end{eqnarray}

\item $j=1$.  We get contribution from $t_3, t_4, t_5, t_6$,
\begin{eqnarray}\label{eq:Appj1:1}
  t &=& \mathcal{A}\mathcal{A}' \big[
  -\sqrt{2} (\mathcal{B}+\mathcal{B}') \;  \delta_{MM'} \;  \vec \epsilon \cdot \vec p
  -\sqrt{2} \;(-1+ \mathcal{B}\mathcal{B}'\, \vec{p}^{\,2})\;
   \langle M' |\vec\sigma \cdot \vec \epsilon \,|M \rangle  \nonumber\\[2mm]
&& +\, 2 \sqrt{2}\; \mathcal{B}\mathcal{B}' \;  \vec \epsilon \cdot \vec p \;
          \langle M' |\vec \sigma \cdot \vec p \,| M \rangle
   -i \sqrt{2}\,(\mathcal{B}-\mathcal{B}')\;  (\vec \epsilon \times \vec p \,) \cdot
    \langle M' | \vec \sigma |M \rangle \big].
\end{eqnarray}
$t_3$ and $t_4$ interfere with themselves, and there is no further interference.
We then get
\begin{eqnarray}\label{eq:Appj1:2}
  \overline{\sum} \sum |t|^2 &=& \frac{1}{2} \;
  \sum_{M,M'} \sum_{\epsilon \;{\rm pol}}|t|^2
  \nonumber\\[2mm]
   &=& (\mathcal{A}\mathcal{A}')^2 \big[\, 2(\mathcal{B}+\mathcal{B}')^2 \,\vec{p}^{\,2} + 6(-1+ \mathcal{B}\mathcal{B}'\, \vec{p}^{\,2})^2
       -8(-1+ \mathcal{B}\mathcal{B}'\, \vec{p}^{\,2})\mathcal{B}\mathcal{B}'\; \vec{p}^{\,2} \nonumber\\
   && ~~~~~~~~~~    +8(\mathcal{B}\mathcal{B}')^2\, \vec{p}^{\,4}
       +4 (\mathcal{B}-\mathcal{B}')^2 \,\vec{p}^{\,2} \,\big] \nonumber\\[2mm]
   &=& (\mathcal{A}\mathcal{A}')^2 \big[\, 6+6(\mathcal{B}^2+\mathcal{B}'^2)\,\vec{p}^{\,2}
       -8\mathcal{B}\mathcal{B}'\, \vec{p}^{\,2} +6(\mathcal{B}\mathcal{B}')^2 \,\vec{p}^{\,4} \, \big].
\end{eqnarray}
\end{enumerate}

We can see that, up to the $p$ terms the strength from $D_s^{*-}$ production
is three times bigger than for $D_s^-$ production.

\section{External emission in $\boldsymbol{B}$ decays}
\label{App:B}

We evaluate the matrix elements for the case of
\begin{equation*}
  \bar B^0 \to D_s^- D^+,\, D_s^- D^{*+}, \,D_s^{*-} D^+,\, D_s^{*-} D^{*+}.
\end{equation*}
In this case in addition to coupling the $\bar c s$ pair from the $W$ vertex to $jm$,
we must couple the quarks forming the $\bar B^0$ to $|00\rangle$
and the final $c\bar d$ pair to $j' m'$.

We have the diagram of Fig.~\ref{Fig:4} and we take the terms evaluated in the former section.

\subsection{Term $\boldsymbol{t_1}$}
We project over spin zero for the $\bar B^0$ and $j' m'$ for the final $c\bar d$ state.
Since the $b\bar d $ state couples to zero spin,
the third component of the $\bar d$ spin must be opposite to the one of the $b$ quark, $M$,
hence $\bar d$ has third component $-M$.
The phase $(-1)^{1/2+M}$ from particle-hole conjugation appears twice and can be ignored.
Furthermore, we will fix $m'$, which is $M'-M$ and sum over the other spin components.
Then we have, taking $t_1$ from Eq.~\eqref{eq:App:t1:2} and projecting over spin,
\begin{eqnarray}\label{eq:AppB:t1}
 t_1  &\to&  \sum_M  \, \mathcal{C}(\frac{1}{2}\, \frac{1}{2}\, 0; \, M, -M, 0) \;
             \mathcal{C}(\frac{1}{2}\, \frac{1}{2}\, j'; \, M', -M, m') \;
             (-\sqrt{2})\; \delta_{MM'}\;
             \delta_{j0}\; (1+ \mathcal{B}\mathcal{B}'\, \vec{p}^{\,2})\nonumber\\[2mm]
   &=&  -\sqrt{2}\; (1+ \mathcal{B}\mathcal{B}'\, \vec{p}^{\,2})\;
   \delta_{j0}\; \delta_{j'0}\; \delta_{m'0}.
\end{eqnarray}

\subsection{Term $\boldsymbol{t_2}$}
We take $t_2$ from Eq.~\eqref{eq:App:t2:2} and project over spins. We obtain
\begin{eqnarray}\label{eq:AppB:t2}
 t_2  &\to& \sqrt{2} (\mathcal{B}+\mathcal{B}') \; \delta_{j \, 0} \; \sum_{M'} (-1)^{M-M'} \sqrt{3}\;
       \mathcal{C}(\frac{1}{2}\, 1\, \frac{1}{2}; \, M, M'-M, M') \nonumber\\[2mm]
      && \times p_{M-M'} \; \mathcal{C}(\frac{1}{2}\, \frac{1}{2}\, 0; \, M, -M, 0) \;
       \mathcal{C}(\frac{1}{2}\, \frac{1}{2}\, j'; \, M', -M, m'),
\end{eqnarray}
which fixes $M'-M$ to $m'$.
On the other hand,
\begin{eqnarray}\label{eq:AppB:CGC2}
\mathcal{C}(\frac{1}{2}\, \frac{1}{2}\, 0; \, M, -M, 0) &=&
(-1)^{1/2-M}\; \frac{1}{\sqrt{2}}\;
\mathcal{C}(\frac{1}{2}\, 0\, \frac{1}{2}; \, M, 0, M)
\nonumber\\[2mm]
      &=& (-1)^{1/2-M}\; \frac{1}{\sqrt{2}},
\end{eqnarray}
\begin{equation}\label{eq:AppB:CGC3}
  \mathcal{C}(\frac{1}{2}\, 1\, \frac{1}{2}; \, M, M'-M, M')
  =(-1)^{1/2-M}\; \sqrt{\frac{2}{3}} \;
  \mathcal{C}(\frac{1}{2}\, \frac{1}{2}\, 1; \, M', -M, M'-M).
\end{equation}
Fixing $M'-M$ we have
\begin{equation}\label{eq:AppB:CGC4}
  \sum_{M'} \mathcal{C}(\frac{1}{2}\, \frac{1}{2}\, 1; \, M', -M, M'-M) \;
   \mathcal{C}(\frac{1}{2}\,
  \frac{1}{2}\, j'; \, M', -M, M'-M) =\delta_{j'1}.
\end{equation}
The state $j'm'$ is $(1, M'-M)$, which means
we have polarization of the vector as $\epsilon'_{M'-M}$.
The combination $(-1)^{-m'}\, p_{-m'}$ is the contribution
of $\vec \epsilon{\,'} \cdot \vec p$
for a vector with polarization $m'$.
Hence, the term is written as
\begin{equation}\label{eq:AppB:t2}
  t_2 = \sqrt{2}\, (\mathcal{B}+\mathcal{B}')\; \delta_{j0}\; \delta_{j'1}\;
  \vec \epsilon\,' \cdot \vec p.
\end{equation}

\subsection{Term $\boldsymbol{t_3}$}
Taking the term $t_3$ from Eq.~\eqref{eq:Appt3:2}
and proceeding as in term $t_1$, we obtain
\begin{equation}\label{eq:AppB:t3}
 t_3=-\sqrt{2} (\mathcal{B}+\mathcal{B}') \; \delta_{j1}\; \delta_{j'0} \;
 \vec \epsilon \cdot \vec p.
\end{equation}

\subsection{Term $\boldsymbol{t_4}$}
We start from $t_4$ of Eq.~\eqref{eq:Appt4:3} and project over spins. We have
\begin{eqnarray}
 t_4 &\to& -(-1+ \mathcal{B}\mathcal{B}'\, \vec{p}^{\,2})\; \sqrt{2} \;
 \delta_{j1}\, \sum_{M'}\, (-1)^{-m}\; \sqrt{3}\;
           \mathcal{C}(\frac{1}{2}\, 1\,
           \frac{1}{2}; \, M, -m, M')\nonumber \\[2mm]
        && \cdot   \mathcal{C}(\frac{1}{2}\, \frac{1}{2}\, 0; \, M, -M, 0)\;
           \mathcal{C}(\frac{1}{2}\, \frac{1}{2}\, j'; \, M', -M, m'),
 \end{eqnarray}
which fixes $m$ to $M-M'$ and $m'$ to $M'-M$, hence $m=-m'$.
Using Eqs.~\eqref{eq:AppB:CGC3},\eqref{eq:AppB:CGC2},
\begin{equation}
  \mathcal{C}(\frac{1}{2}\, 1\, \frac{1}{2}; \, M, -m, M')
  =(-1)^{1/2-M}\; \sqrt{\frac{2}{3}} \;
  \mathcal{C}(\frac{1}{2}\, \frac{1}{2}\, 1; \, M', -M, -m),
\end{equation}
\begin{equation}
  \mathcal{C}(\frac{1}{2}\,
  \frac{1}{2}\, 0; \, M, -M, 0) =(-1)^{1/2-M}\; \frac{1}{\sqrt{2}},
\end{equation}
we have, fixing $m'=-m=M'-M$,
\begin{equation}
  \sum_{M'}\, \mathcal{C}(\frac{1}{2}\, \frac{1}{2}\, 1; \, M', -M, m')\;
  \mathcal{C}(\frac{1}{2}\, \frac{1}{2}\, j'; \, M', -M, m')=\delta_{j'1}.
\end{equation}
Then
\begin{equation}\label{eq:AppB:t4}
  t_4 \to -(-1+ \mathcal{B}\mathcal{B}'\, \vec{p}^{\,2})\; \sqrt{2} \; \delta_{j1}\;
  \delta_{j'1}\; (-1)^{-m}\; \delta_{m, -m'}.
\end{equation}

Furthermore, now we combine $(j,m)$, $(j',-m)$ to
give spin $00$ of the $\bar B^0$ and then we get $(j=j'=1)$
\begin{eqnarray}\label{eq:AppB:CGC5}
   && \sum_{m'}\,  \mathcal{C}(j'\, j\, 0; \, m', -m',0)\; (-1)^{m'}
        \nonumber\\[2mm]
   &=& \sum_{m'}\, (-1)^{m'}\; (-1)^{j'-m'} \; \frac{1}{\sqrt{3}} \;
       \mathcal{C}(j'\, 0\, j; \, m', 0,m') \; =-\frac{3}{\sqrt{3}}.
%    \nonumber\\[2mm]
%   &=& -\frac{3}{\sqrt{3}}.
\end{eqnarray}
Hence, altogether we find
\begin{equation}\label{eq:AppB:t4:2}
  t_4=3\, \sqrt{\frac{2}{3}} \; \delta_{j1}\;
  \delta_{j'1}\; (-1+ \mathcal{B}\mathcal{B}'\, \vec{p}^{\,2}).
\end{equation}

In order to get the interference with $t_5$,
it is convenient to write it with the explicit polarization vector of $j$ and $j'$.
For this we start from Eq.~\eqref{eq:AppB:t4} and
realize that $(-1)^{-m}\,\delta{m,-m'}$ corresponds to
the product of the vectors $\vec \epsilon \cdot \vec \epsilon \,'$
written in spherical basis. Hence we can write
\begin{equation}\label{eq:AppB:t4:3}
  t_4=-\sqrt{2}  \; (-1+ \mathcal{B}\mathcal{B}'\, \vec{p}^{\,2}) \; \delta_{j1}\;
  \delta_{j'1}\; \vec \epsilon \cdot \vec \epsilon \,'.
\end{equation}
Note that $2 \sum \epsilon_i \, \epsilon'_i \; \epsilon_j \, \epsilon'_j = 2 \delta_{ij} \delta_{ij}=2\delta_{ii}=6$, which is the same result that we get when we take $|t_4|^2$ from Eq.~\eqref{eq:AppB:t4:2}
(there the polarizations have been already combined in the amplitude
to have $j,j'$ combined to zero spin).

\subsection{Term $\boldsymbol{t_5}$}
We start from the term $t_5$ of Eq.~\eqref{eq:Appt5:1} and project over spins.
\begin{eqnarray}\label{eq:AppB:t5:1}
  t_5 &\to& 2 \sqrt{2}\, \mathcal{B}\mathcal{B}' \; \delta_{j1} \; \sum_{M,M'}(-1)^{-m}\, (-1)^{M-M'}\;
              p_{-m}  \; p_{M-M'} \nonumber\\[2mm]
        &&  \times \sqrt{3}\; \mathcal{C}(\frac{1}{2}\, 1\, \frac{1}{2}; \, M, M'-M, M')
          \mathcal{C}(\frac{1}{2}\, \frac{1}{2}\, 0; \, M, -M, 0)\;
          \mathcal{C}(\frac{1}{2}\, \frac{1}{2}\, j'; \, M', -M, m').
 \end{eqnarray}
Once again, fixing $m'$ which is $M'-M$ and proceeding as done for the former term,
\begin{eqnarray}\label{eq:AppB:t5:CGC1}
   && \sum_{M'} \mathcal{C}(\frac{1}{2}\, 1\, \frac{1}{2}; \, M, M'-M, M')\;
                 \mathcal{C}(\frac{1}{2}\, \frac{1}{2}\, 0; \, M, -M, 0)\;
                 \mathcal{C}(\frac{1}{2}\, \frac{1}{2}\, j'; \, M', -M, m')
    \nonumber\\[2mm]
   &=& \sum_{M'} (-1)^{1/2-M}\; \frac{1}{\sqrt{2}}\; (-1)^{1/2-M} \sqrt{\frac{2}{3}}\;
                 \mathcal{C}(\frac{1}{2}\, \frac{1}{2}\, 1; \, M', -M, m')
                 \mathcal{C}(\frac{1}{2}\, \frac{1}{2}\, j'; \, M', -M, m')\nonumber\\
   &=& \frac{1}{\sqrt{3}} \delta_{j'1},
\end{eqnarray}
and we get
\begin{equation}\label{eq:AppB:t5:2}
  t_5 \to 2 \sqrt{2}\, \mathcal{B}\mathcal{B}' \; \delta_{j1} \; \delta_{j'1}\; (-1)^{-m}\, (-1)^{M-M'}\;
              p_{-m}  \; p_{M-M'}.
\end{equation}
We cannot now combine $j,j'$ to give zero,
because we have two vectors $p_{-m} \; p_{M-M'}$,
which can combine to orbital angular momentum $L=0$ or $L=2$,
and it is $(j \otimes L) \otimes j'$ that must couple to spin zero.
Because of this, it is better to write Eq.~\eqref{eq:AppB:t5:2}
in terms of the polarization vectors,
which is quite simple because $(-1)^{-m}\, p_{-m}$ is
$\vec \epsilon \cdot \vec p$ in spherical basis and
$(-1)^{M-M'} \, p_{M-M'}$ is $\vec \epsilon\, ' \cdot \vec p$.
Thus
\begin{equation}\label{eq:AppB:t5:3}
  t_5 \to 2 \sqrt{2}\, \mathcal{B}\mathcal{B}' \; \delta_{j1} \; \delta_{j'1}\;
  (\vec \epsilon \cdot \vec p\,)\; (\vec \epsilon\, ' \cdot \vec p\,).
\end{equation}

The decomposition of $\epsilon_i\, p_i \,\epsilon'_j \,p_j$ into $s$ and $d$-waves
\begin{equation*}
   p_i \,p_j \to (p_i \,p_j - \frac{1}{3} \vec p^{\,2}\, \delta_{ij})
   + \frac{1}{3} \vec p^{\,2}\, \delta_{ij}
\end{equation*}
shows that the $s$-wave, $\frac{1}{3} \vec p^{\,2}\, \delta_{ij}$,
has the same structure $\vec \epsilon \cdot \vec \epsilon\,'$ as for $t_4$
in Eq.~\eqref{eq:AppB:t4:3} and thus interferes, while the $d$-wave term will sum incoherently.

\subsection{Term $\boldsymbol{t_6}$}
We start from the term $t_6$ from Eq.~\eqref{eq:Appt6:2} and project over spins. Then
\begin{eqnarray}\label{eq:AppB:t6:2}
  t_6 &\to& 2(\mathcal{B}-\mathcal{B}')\;  \delta_{j1}\; \sum_{m\nu} \sum_{M'}\,
  \mathcal{C}(1\, 1\, 1; \, m, \nu)\; (-1)^{m+\nu}\; p_{-m-\nu}\;
  \sqrt{3}\; \mathcal{C}(\frac{1}{2}\, 1\, \frac{1}{2}; \, M, \nu, M') \nonumber\\[2mm]
   && \times \mathcal{C}(\frac{1}{2}\, \frac{1}{2}\, 0; \, M, -M, 0) \;
             \mathcal{C}(\frac{1}{2}\, \frac{1}{2}\, j'; \, M', -M, M'-M),
\end{eqnarray}
which fixes $\nu=M'-M$.
Keeping $M'-M$ fixed, we use Eq.~\eqref{eq:AppB:t5:CGC1} and obtain
\begin{equation}\label{eq:AppB:t6:3}
  t_6 \to 2(\mathcal{B}-\mathcal{B}')\;  \delta_{j1}\;  \delta_{j'1}\; \sum_{m,\nu} \,
  \mathcal{C}(1\, 1\, 1; \, m, \nu)\; (-1)^{m+\nu} \; p_{-m-\nu},
\end{equation}
which by virtue of Eq.~\eqref{eq:Appt6:1} can be recast as
\begin{equation}\label{eq:AppB:t6:4}
  t_6 = i \sqrt{2}\,(\mathcal{B}-\mathcal{B}')\;  \delta_{j1}\;
  \delta_{j'1}\; (\vec \epsilon \times \vec \epsilon\,')\cdot \vec p\,.
\end{equation}

\subsection{$\boldsymbol{\overline{\sum} \sum |t|^2}$ for all combinations}
\begin{enumerate}[A)]
\setlength{\itemsep}{1.5pt}
\setlength{\parsep}{0.5pt}
\setlength{\parskip}{0.2pt}
\item $j=0, \,j'=0$:

Only the term $t_1$ contributes and we find from Eq.~\eqref{eq:AppB:t1}
\begin{equation}\label{eq:AppB:A}
  (\mathcal{A}\mathcal{A}')^2 \; \overline{\sum} \sum |t|^2 = 2 (1+\mathcal{B}\mathcal{B}' \vec p^{\,2})^2 \; (\mathcal{A}\mathcal{A}')^2.
\end{equation}

\item $j=0, \,j'=1$:

Only the term $t_2$ contributes and we find
\begin{equation}\label{eq:AppB:B}
  (\mathcal{A}\mathcal{A}')^2 \; \overline{\sum} \sum |t|^2 = 2 (\mathcal{B}+\mathcal{B}')^2 \vec p^{\,2} \; (\mathcal{A}\mathcal{A}')^2.
\end{equation}

\item $j=1, \, j'=0$:

Only the term $t_3$ contributes and we find
\begin{equation}\label{eq:AppB:C}
  (\mathcal{A}\mathcal{A}')^2 \; \overline{\sum} \sum |t|^2 = 2 (\mathcal{B}+\mathcal{B}')^2 \vec p^{\,2} \; (\mathcal{A}\mathcal{A}')^2.
\end{equation}
It is interesting to see that the case of $j=1, j'=0$ gives
the same contribution as that of $j=0, j'=1$,
which is corroborated by experiment.

\item $j=1, \, j'=1$:

Here we have a contribution from $t_4, t_5, t_6$.
As mentioned before, $t_4$ and $t_5$ interfere partially but they do not interfere with $t_6$.
\begin{eqnarray}\label{eq:AppB:D}
&& (\mathcal{A}\mathcal{A}')^2 \; \overline{\sum} \sum |t_4+ t_5+ t_6|^2   \nonumber\\[3mm]
   &=&  (\mathcal{A}\mathcal{A}')^2\; \overline{\sum} \sum
  \left|
   -\sqrt{2}  \; (-1+ \mathcal{B}\mathcal{B}'\, \vec{p}^{\,2}) \;  \vec \epsilon \cdot \vec \epsilon\,'
   + 2 \sqrt{2}\, \mathcal{B}\mathcal{B}' \;  (\vec \epsilon \cdot \vec p\,)\;
   (\vec \epsilon\, ' \cdot \vec p\,) \right.\nonumber\\
   &&~~~~~~~~~~\left. +i \sqrt{2}\,(\mathcal{B}-\mathcal{B}')\;  (\vec \epsilon \times \vec \epsilon\,')\cdot \vec p
    \, \right|^2  \nonumber\\[2mm]
  &=&  (\mathcal{A}\mathcal{A}')^2\; \big[ 6\,(-1+ \mathcal{B}\mathcal{B}'\, \vec{p}^{\,2})^2
       -8\,\mathcal{B}\mathcal{B}'\; (-1+ \mathcal{B}\mathcal{B}'\, \vec{p}^{\,2})\,\vec{p}^{\,2} \nonumber\\
  &&~~~~~~~~~~+8\,(\mathcal{B}\mathcal{B}')^2 \; \vec{p}^{\,4}
  +4\, (\mathcal{B}-\mathcal{B}')^2 \; \vec{p}^{\,2} \,\big]  \nonumber\\[2mm]
  &=&  (\mathcal{A}\mathcal{A}')^2\; \big[\, 6+ 4 \,\mathcal{B}^2 \, \vec{p}^{\,2}
  + 4\, \mathcal{B}'^2 \, \vec{p}^{\,2} -12\,\mathcal{B}\mathcal{B}'\,\vec{p}^{\,2}
  +6\, (\mathcal{B}\mathcal{B}')^2 \; \vec{p}^{\,4} \, \big].
\end{eqnarray}
\end{enumerate}

\section{Matrix element for internal emission for baryon decay}
\label{App:C}

Using Eq.~\eqref{eq:App:sigma}, we can work out the terms $t_i$ in Eq.~\eqref{eq:ti26},
and get a more simplified expression,
where $t_6, t_7, t_8$ are incorporated in $t'_3, t'_2, t'_4, t'_5$ and $t'_9$,
\begin{equation}
  t=\mathcal{A}\mathcal{A}'\; \big[\, t'_1+ t'_2+ t'_3+ t'_4 +t'_5 +t'_9\, \big],
\end{equation}
with
\begin{eqnarray}\label{eq:AppC:t:1}
  t'_1 &=& (1- \mathcal{B}\mathcal{B}'\, \vec{p}^{\,2})\; \langle M' | S_2 \rangle \;
           \langle S_1 | M \rangle, \nonumber\\[2mm]
  t'_2 &=& (\mathcal{B}-\mathcal{B}') \; \langle S_1 | M \rangle \;
           \langle M' |\vec \sigma \cdot \vec p \,| S_2 \rangle, \nonumber\\[2mm]
  t'_3 &=& (\mathcal{B}-\mathcal{B}') \; \langle M' | S_2 \rangle \;
           \langle S_1 |\vec \sigma \cdot \vec p \,| M \rangle, \nonumber\\[2mm]
  t'_4 &=& 2\mathcal{B}\mathcal{B}'\; \langle M' |\vec \sigma \cdot \vec p \,| S_2 \rangle \;
          \langle S_1 |\vec \sigma \cdot \vec p \,| M \rangle, \\[2mm]
  t'_5 &=& -(1+ \mathcal{B}\mathcal{B}'\, \vec{p}^{\,2})\; \langle M' |\sigma^i | S_2 \rangle \;
           \langle S_1 | \sigma^i | M \rangle, \nonumber\\[2mm]
  t'_9 &=& i(\mathcal{B}+\mathcal{B}')\; \epsilon_{ijk} \; p^j \;\langle M' | \sigma^i | S_2 \rangle \;
           \langle S_1 |\sigma^k | M \rangle. \nonumber
\end{eqnarray}

We follow here a different approach than in the external emission and we choose as $z$ direction to
write the spin states, the direction of the momentum $\vec p$.
Hence $\vec p \equiv p \, \hat u_z$. Then
\begin{equation*}
  \langle S' |\vec \sigma \cdot \vec p \,| S \rangle = \langle S' | \sigma_z |S \rangle p
  =p\; (-1)^{1/2-S}\; \delta_{SS'}.
\end{equation*}

We proceed to evaluate the contribution of each term for $j=0, j=1$.

\subsection{Term $\boldsymbol{t'_1}$}
In order to project over $jm$, we must multiply by
$(-1)^{1/2+S_2} \, \mathcal{C}(\frac{1}{2}\, \frac{1}{2}\, j; \, S_1, -S_2, m)$
and sum over $S_1$.
\begin{equation}\label{eq:AppC:t1:1}
  t'_1 \to \sum_{S_1} \, (-1)^{1/2+S_2} \; \mathcal{C}(\frac{1}{2}\,
  \frac{1}{2}\, j; \, S_1, -S_2, m)\;
  \delta_{M'\,S_2}\; \delta_{S_1\, M}\; (1- \mathcal{B}\mathcal{B}'\, \vec{p}^{\,2}),
\end{equation}
which implies $m=M-M'$,
\begin{equation}\label{eq:AppC:t1:2}
  t'_1 \to (-1)^{1/2+M'}\; \mathcal{C}(\frac{1}{2}\, \frac{1}{2}\, j; \, M, -M', M-M')\;
           (1- \mathcal{B}\mathcal{B}'\, \vec{p}^{\,2}).
\end{equation}
We can split it for $j=0, j=1$. For $j=0, M-M'=0$, we get
\begin{eqnarray}\label{eq:AppC:t1:3}
  t'_1 \,(j=0) &\to&  (-1)^{1/2+M'}\; (-1)^{1/2-M}\; \frac{1}{\sqrt{2}} \; \delta_{MM'} \;
                      (1- \mathcal{B}\mathcal{B}'\, \vec{p}^{\,2})  \nonumber\\[2mm]
   &=& - \frac{1}{\sqrt{2}} \; \delta_{MM'} \; (1- \mathcal{B}\mathcal{B}'\, \vec{p}^{\,2}).
\end{eqnarray}
For $j=1$ we leave it explicitly as in Eq.~\eqref{eq:AppC:t1:2},
\begin{equation}\label{eq:AppC:t1:4}
  t'_1 \,(j=1)\to (-1)^{1/2+M'}\; \mathcal{C}(\frac{1}{2}\, \frac{1}{2}\, 1; \, M, -M', M-M')\;
           (1- \mathcal{B}\mathcal{B}'\, \vec{p}^{\,2}).
\end{equation}

\subsection{Term $\boldsymbol{t'_2}$}
\vspace{-0.8cm}
\begin{eqnarray}\label{eq:AppC:t2:1}
t'_2 &\to& (\mathcal{B}-\mathcal{B}') \; p\; \langle S_1 | M \rangle \; \langle M' |\sigma_z | S_2 \rangle \nonumber\\[2mm]
   &=& (\mathcal{B}-\mathcal{B}') \; p\; (-1)^{1/2-S_2}\; \delta_{S_2M'}\; \delta_{S_1M},
\end{eqnarray}
and projecting over $jm$,
\begin{eqnarray}\label{eq:AppC:t2:2}
t'_2 &\to& \sum_{S_1} (-1)^{1/2+S_2}\; \mathcal{C}(\frac{1}{2}\,
   \frac{1}{2}\, j; \, S_1, -S_2, m)\;
  (\mathcal{B}-\mathcal{B}') \; p\; (-1)^{1/2-S_2}\; \delta_{S_2M'}\; \delta_{S_1M} \nonumber\\[2mm]
   &=& -(\mathcal{B}-\mathcal{B}') \; p\; \mathcal{C}(\frac{1}{2}\, \frac{1}{2}\, j; \, M, -M', M-M').
\end{eqnarray}
Once again, for $j=0$ we obtain
\begin{equation}\label{eq:AppC:t2:3}
  t'_2 \,(j=0) \to -(\mathcal{B}-\mathcal{B}') \; p\; (-1)^{1/2-M}\; \frac{1}{\sqrt{2}}\; \delta_{MM'},
\end{equation}
for $j=1$ we leave it as in Eq.~\eqref{eq:AppC:t2:2},
\begin{equation}\label{eq:AppC:t2:4}
  t'_2 \,(j=1) \to -(\mathcal{B}-\mathcal{B}') \; p\; \mathcal{C}(\frac{1}{2}\, \frac{1}{2}\, 1; \, M, -M', M-M').
\end{equation}

\subsection{Term $\boldsymbol{t'_3}$}
\vspace{-0.8cm}
\begin{equation}
  t'_3 \to (\mathcal{B}'-\mathcal{B}) \;\delta_{S_2M'}\; (-1)^{1/2-M}\; \delta_{S_1M}.
\end{equation}
This term has the same structure as $t'_2$ but the phase is $(-1)^{1/2-M}$ instead of
$(-1)^{1/2-M'}$. Thus
\begin{equation}
  t'_3 \to -(\mathcal{B}'-\mathcal{B}) \; p\; (-1)^{M'-M}\;\mathcal{C}(\frac{1}{2}\, \frac{1}{2}\, j; \, M, -M', M-M').
\end{equation}
Then
\begin{equation}\label{eq:AppC:t3:3}
  t'_3  \,(j=0) = -(\mathcal{B}'-\mathcal{B}) \; p\; (-1)^{1/2-M}\; \frac{1}{\sqrt{2}}\; \delta_{MM'},
\end{equation}
\begin{equation}\label{eq:AppC:t3:4}
  t'_3  \,(j=1) = -(\mathcal{B}'-\mathcal{B}) \; p\; (-1)^{M'-M}\;
  \mathcal{C}(\frac{1}{2}\, \frac{1}{2}\, 1; \, M, -M', M-M').
\end{equation}

\subsection{Term $\boldsymbol{t'_4}$}
\begin{equation*}
  t'_4 \to 2\mathcal{B}\mathcal{B}' \; p^2 \; (-1)^{1/2-S_2}\; \delta_{S_2M'}\; (-1)^{1/2-M}\;  \delta_{S_1M},
\end{equation*}
which again is like $t'_2$ but with an extra phase $(-1)^{1/2-M}$. Hence
\begin{equation}\label{eq:AppC:t4:3}
  t'_4  \,(j=0) = -\frac{2}{\sqrt{2}}\; \mathcal{B}\mathcal{B}' \; \vec p^{\,2}\;  \delta_{MM'},
\end{equation}
\begin{equation}\label{eq:AppC:t4:4}
  t'_4  \,(j=1) = -2 \mathcal{B}\mathcal{B}' \; \vec p^{\,2}\;
  \mathcal{C}(\frac{1}{2}\, \frac{1}{2}\, 1; \, M, -M', M-M')\; (-1)^{1/2-M}.
\end{equation}

\subsection{Term $\boldsymbol{t'_5}$}
Let us write in spherical basis,
\begin{eqnarray}\label{eq:AppC:t5:1}
  \tilde{t}_5 &\equiv& \langle M' |\sigma^i | S_2 \rangle \;
  \langle S_1 | \sigma^i | M \rangle \nonumber\\[2.5mm]
   &=& \sum_\mu (-1)^\mu \; \langle M' |\sigma_\mu | S_2 \rangle \;
        \langle S_1 | \sigma_{-\mu} | M \rangle \nonumber\\[2mm]
   &=&\sum_\mu (-1)^\mu \; \sqrt{3}\;
        \mathcal{C}(\frac{1}{2}\, 1\, \frac{1}{2}; \, S_2, \mu, M')\;
        \sqrt{3}\; \mathcal{C}(\frac{1}{2}\, 1\, \frac{1}{2}; \, M, -\mu, S_1),
\end{eqnarray}
which implies $S_2 +\mu=M'$, $M-\mu=S_1$ and hence $S_2=M'-M+S_1$.
\begin{equation}\label{eq:AppC:t5:2}
  \tilde{t}_5 =3 \, (-1)^{M-S_1}\;
           \mathcal{C}(\frac{1}{2}\, 1\, \frac{1}{2}; \, M'-M+S_1, M-S_1, M')\;
           \mathcal{C}(\frac{1}{2}\, 1\, \frac{1}{2}; \, M, S_1-M, S_1).
\end{equation}

Projecting over spin $jm$, we have
\begin{eqnarray}\label{eq:AppC:t5:3}
 \tilde{t}_5  &\to& 3 \, \sum_{S_1}\, (-1)^{M-S_1}\; (-1)^{1/2+M'-M+S_1}\;
                   \mathcal{C}(\frac{1}{2}\, \frac{1}{2} \, j; \, S_1, M-M'-S_1, M-M')\;    \nonumber\\
   && \times \; \mathcal{C}(\frac{1}{2}\, 1\, \frac{1}{2}; \, M'-M+S_1, M-S_1, M') \;
             \mathcal{C}(\frac{1}{2}\, 1\, \frac{1}{2}; \, M, S_1-M, S_1).
\end{eqnarray}
Now $S_1$ appears in the three coefficients and we must construct a Racah coefficient from there.
We reorder the CGC as
\begin{eqnarray}
  \mathcal{C}(\frac{1}{2}\, 1\, \frac{1}{2}; \, M, S_1-M, S_1) &=& (-1)^{1/2-M}\;
      \sqrt{\frac{2}{3}} \;
      \mathcal{C}(\frac{1}{2}\, \frac{1}{2}\, 1; \, M, -S_1, M \!-\! S_1), \\[2mm]
  \mathcal{C}(\frac{1}{2} 1 \frac{1}{2};  M'\!\!-\!\!M\!+\!S_1, M\!\!-\!\!S_1, M')
  &=& (-1)^{1/2+1-1/2}
     \mathcal{C}(1\, \frac{1}{2}\, \frac{1}{2};
     \, M\!\!-\!\!S_1, M'\!\!-\!\!M\!+\!S_1, M'),  \\[2mm]
  \mathcal{C}(\frac{1}{2} \frac{1}{2}  j;  S_1, M\!\!-\!\!M'\!\!-\!\!S_1, M\!\!-\!\!M')
  &=& (-1)^{1/2+1/2-j}
      \mathcal{C}(\frac{1}{2} \frac{1}{2}  j;  -S_1, M'\!\!-\!\!M\!+\!S_1, M'\!\!-\!\!M),~~
\end{eqnarray}
and then apply Eq.~(6.5a) of Ref.~\cite{rose},
summing over $S_1$ keeping $M-M'$ fixed, and we obtain
\begin{eqnarray}\label{eq:AppC:t5:4}
 \tilde{t}_5  &\to& (-1)^{M'-M-j+1}\; 3 \,\sqrt{\frac{2}{3}}\;  [(2j+1)\cdot 3]^{1/2}
                     W(\frac{1}{2}\, \frac{1}{2} \, \frac{1}{2}\, \frac{1}{2}; 1j) \;
                   \mathcal{C}(\frac{1}{2}\, j\, \frac{1}{2}; \, M, M'-M, M')\;    \nonumber\\[3mm]
   &=& 6\, (-1)^{1/2+M'-j}\; W(\frac{1}{2}\, \frac{1}{2} \, \frac{1}{2}\, \frac{1}{2}; 1j) \;
                    \mathcal{C}(\frac{1}{2}\, \frac{1}{2}\, j; \, M, -M', M-M'),
 \end{eqnarray}
where $W(\cdots)$ is a Racah coefficient,
which can be calculated using formulas from the Appendix of Ref.~\cite{rose}, and we get
\begin{equation}
  W(\frac{1}{2}\, \frac{1}{2} \, \frac{1}{2}\, \frac{1}{2}; 1j)=
  (-1)^{-j}\; \Big[\,\frac{1}{2}- \frac{1}{3}\, j(j+1)\,\Big].
\end{equation}
Note that this term has the same structure as $t'_1$, and we get $t'_1 +t'_5$ substituting
\begin{eqnarray}
  (1-\mathcal{B}\mathcal{B}' \; \vec p^{\,2})&\to& (1-\mathcal{B}\mathcal{B}' \; \vec p^{\,2}) -3(1+\mathcal{B}\mathcal{B}' \; \vec p^{\,2}),
  ~{\rm for} ~j=0 \nonumber\\
  (1-\mathcal{B}\mathcal{B}' \; \vec p^{\,2})&\to& (1-\mathcal{B}\mathcal{B}' \; \vec p^{\,2}) +(1+\mathcal{B}\mathcal{B}' \; \vec p^{\,2}),
  ~ {\rm for} ~j=1 \nonumber
\end{eqnarray}
Hence
\begin{equation}\label{eq:AppC:t5:5}
   t'_1 +t'_5 =
    \left\{
    \begin{array}{ll}
     -(2+4\,\mathcal{B}\mathcal{B}' \; \vec p^{\,2} ) \;
              (-1)^{1/2+M'}\;
              \mathcal{C}(\frac{1}{2}\, \frac{1}{2}\, j; \, M, -M', M-M'),~ {\rm for} ~j=0; \\[3mm]
    2\, (-1)^{1/2+M'}\;
        \mathcal{C}(\frac{1}{2}\, \frac{1}{2}\, j; \, M, -M', M-M'),~ {\rm for} ~j=1. \\
    \end{array}
   \right.
\end{equation}

For $j=0$ it gets further simplified like $t'_1$ and we get
\begin{eqnarray}\label{eq:AppC:t5:6}
  (t'_1 +t'_5)\, (j=0) &=& \frac{1}{\sqrt{2}}\;
  (2+4\,\mathcal{B}\mathcal{B}' \; \vec p^{\,2} )\; \delta_{MM'}, \\[2mm]
  (t'_1 +t'_5)\, (j=1) &=& 2\, (-1)^{1/2+M'}\;
  \mathcal{C}(\frac{1}{2}\, \frac{1}{2}\, 1; \, M, -M', M-M').
\end{eqnarray}

\subsection{Term $\boldsymbol{t'_9}$}
We take $t'_9$ from Eq.~\eqref{eq:AppC:t:1},
and using Eq.~\eqref{eq:Appt6:1} we write it in spherical basis
\begin{equation*}
  t'_9 \to -\sqrt{2}\, (\mathcal{B}+\mathcal{B}')\; \sum_{\mu \nu}\,
              \mathcal{C}(1\, 1\, 1; \, \mu, \nu)\; \langle M' | \sigma_\mu | S_2 \rangle\;
              \langle S_1 |\sigma_\nu | M \rangle \; (-1)^{\mu+\nu}\; p_{-\mu-\nu}.
 \end{equation*}
We take again $\vec p$ in the $z$ direction and then $\mu+\nu=0$. Hence
\begin{eqnarray}\label{eq:AppC:t9:1}
   t'_9 &\to& -\sqrt{2}\, (\mathcal{B}+\mathcal{B}')\; p\; \sum_\mu \, \mathcal{C}(1\, 1\, 1; \, \mu, -\mu) \;
              \langle M' | \sigma_\mu | S_2 \rangle\; \langle S_1 |\sigma_{-\mu} | M \rangle
    \nonumber\\
   &=& -\sqrt{2}\, (\mathcal{B}+\mathcal{B}')\; p\; \sum_\mu \, \mathcal{C}(1\, 1\, 1; \, \mu, -\mu) \;
       \sqrt{3}\;  \mathcal{C}(\frac{1}{2}\, 1\, \frac{1}{2}; \, S_2, \mu, M')  \nonumber\\
    &&~~\times   \sqrt{3}\; \mathcal{C}(\frac{1}{2}\, 1\, \frac{1}{2}; \, M, -\mu, S_1).
\end{eqnarray}
And projecting over spin $jm$
\begin{eqnarray}\label{eq:AppC:t9:2}
   t'_9 &\to& -3\,\sqrt{2}\, (\mathcal{B}+\mathcal{B}')\; p\; \sum_{S_1} \, (-1)^{1/2+S_2}\;
                   \mathcal{C}(\frac{1}{2}\, \frac{1}{2}\, j; \, S_1, -S_2, m)\;
                    \nonumber\\
    &&       \times  \sum_\mu \, \mathcal{C}(1\, 1\, 1; \, \mu, -\mu) \;
                     \mathcal{C}(\frac{1}{2}\, 1\, \frac{1}{2}; \, S_2, \mu, M') \;
                     \mathcal{C}(\frac{1}{2}\, 1\, \frac{1}{2}; \, M, -\mu, S_1),
\end{eqnarray}
which requires $S_2+\mu =M'$, $M-\mu=S_1$. Hence $m=M-M'$ and
\begin{eqnarray}\label{eq:AppC:t9:3}
   t'_9 &\to& -3\,\sqrt{2}\, (\mathcal{B}+\mathcal{B}')\; p\; \sum_\mu \, (-1)^{1/2+M'-\mu}\;
               \mathcal{C}(1\, 1\, 1; \, \mu, -\mu) \nonumber\\[2mm]
      && \times \;
         \mathcal{C}(\frac{1}{2}\, \frac{1}{2}\, j; \, M-\mu, \mu-M', M-M') \nonumber\\[2.5mm]
     && \times \;\mathcal{C}(\frac{1}{2}\, 1\, \frac{1}{2}; \, M'-\mu, \mu, M') \;
               \mathcal{C}(\frac{1}{2}\, 1\, \frac{1}{2}; \, M, -\mu, M-\mu)
                      \nonumber\\[2.5mm]
    &=& -3\,\sqrt{2}\, (\mathcal{B}+\mathcal{B}')\; p\; \sum_\mu \, (-1)^{1/2+M'-\mu}\;
             \mathcal{C}(1\, 1\, 1; \, \mu, -\mu) \nonumber\\[2mm]
    && \times \;
         \mathcal{C}(\frac{1}{2}\, \frac{1}{2}\, j; \, M-\mu, \mu-M', M-M') \nonumber\\[2mm]
    && \times \; (-1)^{1/2-M'+\mu}\; \sqrt{\frac{2}{3}}\;
        \mathcal{C}(\frac{1}{2}\, \frac{1}{2}\, 1; \, M', \mu-M', \mu)\; (-1)^{1/2-M}\nonumber\\[2mm]
    && \times \; \sqrt{\frac{2}{3}}\; \mathcal{C}(\frac{1}{2}\, \frac{1}{2}\, 1; \, M, \mu-M, \mu).
\end{eqnarray}

This combination cannot be cast into a Racah coefficient, but it is easy to evaluate.
Indeed, since $\mathcal{C}(111;\, 000)=0$, $\mu$ can only be $+1$ or $-1$.
From the last two CGC we have: if $M=1/2$ then $\mu=+1$ and $M'=1/2$;
if $M=-1/2$ then $\mu =-1$ and $M'=-1/2$ and the two CGC are $1$.
Then we have
\begin{equation}\label{eq:AppC:t9:4}
  t'_9 \to 2\sqrt{2}\; (-1)^{1/2-M}\; (\mathcal{B}+\mathcal{B}')\; p\; \sum_\mu \,
  \mathcal{C}(1\, 1\, 1; \, \mu, -\mu)\;
           \mathcal{C}(\frac{1}{2}\, \frac{1}{2}\, j; \, M-\mu, \mu-M', M-M').
\end{equation}
Now we have
\begin{alignat*}{3}
& j=0:  & \qquad   & M=M'=1/2,  & \qquad  & t'_9=-\sqrt{2}\; (\mathcal{B}+\mathcal{B}')\; p \, ; \\[2mm]
&   & \qquad   & M=M'=-1/2,  & \qquad  & t'_9=\sqrt{2}\; (\mathcal{B}+\mathcal{B}')\; p .
\end{alignat*}
Hence
\begin{equation}\label{eq:AppC:t9:5}
  t'_9  \,(j=0) = \sqrt{2}\; (\mathcal{B}+\mathcal{B}')\; p\; \delta_{MM'}\; (-1)^{1/2+M}.
\end{equation}
\begin{alignat*}{3}
& j=1:  & \qquad   & M=M'=1/2,  & \qquad  & t'_9=\sqrt{2}\; (\mathcal{B}+\mathcal{B}')\; p \, ; \\[2mm]
&   & \qquad   & M=M'=-1/2,  & \qquad  & t'_9=\sqrt{2}\; (\mathcal{B}+\mathcal{B}')\; p .
\end{alignat*}
Hence
\begin{equation}\label{eq:AppC:t9:6}
  t'_9  \,(j=1) = \sqrt{2}\; (\mathcal{B}+\mathcal{B}')\; p\; \delta_{MM'}.
\end{equation}

We summarize in the following.
\begin{enumerate}[1)]
\setlength{\itemsep}{1.5pt}
\setlength{\parsep}{0.5pt}
\setlength{\parskip}{0.2pt}
\item $j=0$:
\begin{eqnarray}\label{eq:AppT:j0:1}
  t'_1+t'_5 &=& \frac{1}{\sqrt{2}}\; (2+4\,\mathcal{B}\mathcal{B}' \; \vec p^{\,2} )\; \delta_{MM'}, \nonumber\\[2mm]
  t'_2 &=& -\frac{1}{\sqrt{2}}\;(\mathcal{B}-\mathcal{B}') \; p\; (-1)^{1/2-M}\;  \delta_{MM'}, \nonumber\\[2mm]
  t'_3 &=&  -\frac{1}{\sqrt{2}}\; (\mathcal{B}'-\mathcal{B}) \; p\; (-1)^{1/2-M}\;  \delta_{MM'}, \\[2mm]
  t'_4 &=& -\sqrt{2}\; \mathcal{B}\mathcal{B}' \; \vec p^{\,2}\;  \delta_{MM'},\nonumber\\[2mm]
  t'_9 &=& \sqrt{2}\; (\mathcal{B}+\mathcal{B}')\; p\;  (-1)^{1/2+M}\; \delta_{MM'}. \nonumber
\end{eqnarray}
We can see that $t'_2$ and $t'_3$ cancel here and $t'_4$ adds coherently to $t'_1+t'_5$.
Also $t'_9$ does not interfere with those terms in $\overline{\sum} \sum |t|^2$.
Thus
\begin{equation}\label{eq:AppT:j0:2}
  \overline{\sum} \sum |t|^2 = 2\, (\mathcal{A}\mathcal{A}')^2 \; \Big[ (1+\mathcal{B}\mathcal{B}'\; \vec p^{\,2}\,)^2
                               + (\mathcal{B}+\mathcal{B}')^2 \; \vec p^{\,2}\, \Big].
\end{equation}

\item $j=1$:
\begin{eqnarray}\label{eq:AppT:j1:1}
  t'_1+t'_5 &=& 2\, (-1)^{1/2+M'}\;
              \mathcal{C}(\frac{1}{2}\, \frac{1}{2}\, 1; \, M, -M', M-M'), \nonumber\\[2mm]
  t'_2 &=& -(\mathcal{B}-\mathcal{B}') \; p\;
               \mathcal{C}(\frac{1}{2}\, \frac{1}{2}\, 1; \, M, -M', M-M'), \nonumber\\[2mm]
  t'_3 &=& -(\mathcal{B}'-\mathcal{B}) \; p\; (-1)^{M'-M}\;
               \mathcal{C}(\frac{1}{2}\, \frac{1}{2}\, 1; \, M, -M', M-M'), \\[2mm]
  t'_4 &=& -2\, \mathcal{B}\mathcal{B}' \; \vec p^{\,2}\; (-1)^{1/2-M} \;
            \mathcal{C}(\frac{1}{2}\, \frac{1}{2}\, 1; \, M, -M', M-M'),\nonumber\\[2mm]
  t'_9 &=& \sqrt{2}\; (\mathcal{B}+\mathcal{B}')\; p\; \delta_{MM'}. \nonumber
\end{eqnarray}

We can still work out the sum of $t'_2$ and $t'_3$, which now does not cancel.
Indeed, if $M'-M=\pm 1$, the two terms have opposite sign which is compensated by having
$\mathcal{B}-\mathcal{B}'$ in $t'_2$ or $\mathcal{B}'-\mathcal{B}$ in $t'_3$. Hence, they sum. If $M'-M=0$, the two terms cancel.
We can implement this by means of the factor
\begin{equation*}
  (-)(-1)^{1/2-M}\; 2\, \sqrt{2} \; \mathcal{C}(1\,1\, 1; \, M'-M, M-M', 0),
\end{equation*}
and we can write
\begin{equation*}
 t'_2 + t'_3 = 2\, \sqrt{2} \; (\mathcal{B}-\mathcal{B}') \; p \; (-1)^{1/2-M}\;
               \mathcal{C}(\frac{1}{2}\, \frac{1}{2}\, 1; \, M, -M', M-M')\;
               \mathcal{C}(1\,1\, 1; \, M'-M, M-M', 0).
\end{equation*}
\end{enumerate}

In order to see the interference of these terms in $\overline{\sum} \sum |t|^2$,
we will perform the sum over $M$ and $M'-M$,
and $\frac{1}{2}\, \sum_M \sum_{M'} \equiv \frac{1}{2}\, \sum_M \sum_{M'-M}$.
In most terms we get
\begin{equation}
  \frac{1}{2}\,  \sum_{M'-M}\sum_M \, \mathcal{C}(\frac{1}{2}\, \frac{1}{2}\, 1; \, M, -M', M-M')^2
  =\frac{1}{2}\, \sum_{M'-M} 1 = \frac{3}{2}.
\end{equation}
However, because of the $\mathcal{C}(1\,1\, 1; \, M'-M, M-M', 0)$ coefficient,
$t'_2 + t'_3$ does not interfere with any term.
Similarly $t'_9$ does not interfere with any term,
but there is an interference between $t'_1 + t'_5$ and $t'_4$.
This is because $t'_4$ comes from a term of type
$(\vec \sigma \cdot \vec p\,)(\vec \sigma \cdot \vec p\,) $
and $p_i p_j$ can combine to $s$-wave $\frac{1}{3} \, \vec p^{\,2}\; \delta_{ij}$
or $d$-wave $(p_i p_j -\frac{1}{3} \, \delta_{ij})$.
The part of $s$-wave interferes with $t'_1 + t'_5$
and we get it in the following way,
\begin{eqnarray}\label{eq:AppT:sum:1}
  \frac{1}{2}\; 2  \sum_M \sum_{M'} (t'_1 + t'_5)\, t'_4
&\to&
    -4\, \mathcal{B}\mathcal{B}' \; \vec p^{\,2} \sum_{M'-M}\sum_M \, (-1)^{1+M'-M} \;
    \mathcal{C}(\frac{1}{2}\, \frac{1}{2}\, 1; \, M, -M', M-M')^2  \nonumber\\[2mm]
   &=&  4\, \mathcal{B}\mathcal{B}' \; \vec p^{\,2} \sum_{M'-M}\, (-1)^{M'-M} \cdot 1
   \;\;= 4\, \mathcal{B}\mathcal{B}' \; \vec p^{\,2} \, (-1-1+1) \nonumber\\[2mm]
   &=& -4\, \mathcal{B}\mathcal{B}' \; \vec p^{\,2}.
\end{eqnarray}
Similarly,
\begin{eqnarray}\label{eq:AppT:sum:2}
  \frac{1}{2}  \sum_M \sum_{M'} (t'_2 + t'_3)^2
&\to&
    \frac{1}{2}\, \sum_{M'-M}\sum_M \, 8\, (\mathcal{B}-\mathcal{B}')^2\; \vec p^{\,2}\;
    \mathcal{C}(\frac{1}{2}\, \frac{1}{2}\, 1; \, M, -M', M-M')^2 \nonumber\\[2mm]
   &&  \times \;\mathcal{C}(1\,1\, 1; \, M'-M, M-M', 0)^2  \nonumber\\[2mm]
   &=&  \frac{1}{2}\; 8\, (\mathcal{B}-\mathcal{B}')^2\; \vec p^{\,2} \sum_{M'-M}
        \mathcal{C}(1\,1\, 1; \, M'-M, M-M', 0)^2  \nonumber\\[2mm]
   &=& 4\, (\mathcal{B}-\mathcal{B}')^2 \; \vec p^{\,2},
\end{eqnarray}
and
\begin{equation}\label{eq:AppT:sum:3}
  \overline{\sum_M} \sum_{M'} 2\; (\mathcal{B}+\mathcal{B}')^2 \; \vec p^{\,2} \; \delta_{MM'}
  = 2\,(\mathcal{B}+\mathcal{B}')^2 \; \vec p^{\,2}.
\end{equation}
Thus, altogether
\begin{eqnarray}\label{eq:AppT:sum:4}
 &&\overline{\sum} \sum |t|^2 \nonumber\\[2mm]
 &=&  (\mathcal{A}\mathcal{A}')^2\; \Big[ \frac{3}{2} \times 4 + 4\, (\mathcal{B}-\mathcal{B}')^2 \; \vec p^{\,2}
                 +\frac{3}{2} \times 4\, (\mathcal{B}\mathcal{B}')^2 \; \vec p^{\,4}
                 -4\, \mathcal{B}\mathcal{B}' \; \vec p^{\,2}
                 + 2\, (\mathcal{B}+\mathcal{B}')^2 \; \vec p^{\,2} \, \Big]  \nonumber\\[2mm]
 &=&  (\mathcal{A}\mathcal{A}')^2\; \Big[ 6+ 6\, \mathcal{B}^2\; \vec p^{\,2} + 6\, \mathcal{B}'^2\; \vec p^{\,2}
                 -8\, \mathcal{B}\mathcal{B}'\; \vec p^{\,2} +6\, (\mathcal{B}\mathcal{B}')^2\; \vec p^{\,4} \, \Big].
\end{eqnarray}
This result is the same obtained in Eq.~\eqref{eq:Appj1:2} for external emission.

\section{Internal emission for $\boldsymbol{B}$ decays}
\label{App:D}

We look at the diagram of Fig.~\ref{Fig:7} and take the terms $t'_i$ of
Eqs.~\eqref{eq:AppT:j0:1} and \eqref{eq:AppT:j1:1}
projecting over spin zero to $b\bar d$ quarks and
to $j'm'$ the $s,d$ quarks.
The phase from particle-hole conjugation corresponding to
the $\bar d$ state appears twice and can be ignored.
We discuss each term in detail.

\subsection{$\boldsymbol{j=0}$ case}

We have two terms contributing,
$t'_1+t'_5+t'_4 = \sqrt{2} (1+\mathcal{B}\mathcal{B}'\,\vec p^{\,2}\, ) \delta_{MM'}$
and $t'_9= \sqrt{2}\, (\mathcal{B}+\mathcal{B}')\, p\, (-1)^{1/2+M}\, \delta_{MM'}$.
Projecting over spin zero for the $B$ meson and spin $j'm'$ for $\bar K^0$,
and fixing $M'-M=m'$, we obtain
\begin{eqnarray}\label{eq:AppD:j0:1}
 t'_1+t'_5+t'_4 &\to&  \sum_M \sqrt{2}\; (1+\mathcal{B}\mathcal{B}'\;\vec p^{\,2}\,)\; \delta_{MM'}\;
                   \mathcal{C}(\frac{1}{2}\, \frac{1}{2}\, 0; \, M, -M, 0)  \nonumber\\[2mm]
                 && \times \;
                   \mathcal{C}(\frac{1}{2}\, \frac{1}{2}\, j'; \, M', -M, M'-M) \nonumber\\[2mm]
 &=&  \sqrt{2}\; (1+\mathcal{B}\mathcal{B}'\;\vec p^{\,2}\,)\; \delta_{j'0}\; \delta_{m'0}\, ;
\end{eqnarray}
\begin{eqnarray}\label{eq:AppD:j0:2}
 t'_9 &\to&  \sum_M \sqrt{2}\; (\mathcal{B}+\mathcal{B}')\; p\; (-1)^{1/2+M}\; \delta_{MM'}\;
                    \mathcal{C}(\frac{1}{2}\, \frac{1}{2}\, 0; \, M, -M, 0)  \nonumber\\[2mm]
                 && \times \;
                 \mathcal{C}(\frac{1}{2}\, \frac{1}{2}\, j'; \, M', -M, M'-M) \nonumber\\[2mm]
 &=& \sum_M \sqrt{2} \; (\mathcal{B}+\mathcal{B}')\; p\; (-1)^{1/2+M}\; (-1)^{1/2-M}\; \frac{1}{\sqrt{2}}\;
               \mathcal{C}(\frac{1}{2}\, \frac{1}{2}\, j'; \, M, -M, 0)\nonumber\\[2mm]
 &=& - \sqrt{2}\;  (\mathcal{B}+\mathcal{B}')\; p\;  \delta_{j'1}\; \delta_{m'0}\, .
\end{eqnarray}

\subsection{$\boldsymbol{j=1}$ case}

We study the different terms of Eq.\eqref{eq:AppT:j1:1}
and project them over spin zero for the $B$ meson and $j'm'$ for the $\bar K^0$.

In this case we fix $m=-m'=M-M'$, which is one degree of freedom to
sum in $\overline{\sum} \sum |t|^2$.
The fact that $\vec p$ is chosen in the $z$ direction forces $m=-m'$ here.

\begin{eqnarray}\label{eq:AppD:j1:1}
 t'_1+t'_5 &\to&  \sum_{M'} 2\; (-1)^{1/2+M'} \;
                \mathcal{C}(\frac{1}{2}\, \frac{1}{2}\, 1; \, M, -M', M-M')\;
                  \mathcal{C}(\frac{1}{2}\, \frac{1}{2}\, 0; \, M, -M, 0) \nonumber\\[2mm]
 && \times \;     \mathcal{C}(\frac{1}{2}\, \frac{1}{2}\, j'; \, M', -M, M'-M) \nonumber\\[2mm]
 &=& \sum_{M'} 2\; (-1)^{1/2+M'} \; \frac{1}{\sqrt{2}}\;  (-1)^{1/2-M} \;
                  \mathcal{C}(\frac{1}{2}\, \frac{1}{2}\, 1; \, M, -M', M-M') \nonumber\\[2mm]
 && \times \;     \mathcal{C}(\frac{1}{2}\, \frac{1}{2}\, j'; \, M', -M, M'-M) \nonumber\\[2mm]
 &=& -\sum_{M'}\sqrt{2}\;  (-1)^{M'-M} \;
                  \mathcal{C}(\frac{1}{2}\, \frac{1}{2}\, 1; \, M', -M, M'-M)\;
                  \mathcal{C}(\frac{1}{2}\, \frac{1}{2}\, j'; \, M', -M, M'-M) \nonumber\\[2mm]
 &=& -\sqrt{2}\; (-1)^{M'-M} \; \delta_{j'1}.
\end{eqnarray}

Recall that we will sum over $M'-M$ in $\overline{\sum} \sum |t|^2$.

\begin{eqnarray}\label{eq:AppD:j1:2}
 t'_2+t'_3 &\to&  \sum_{M'} 2\, \sqrt{2}\; (\mathcal{B}-\mathcal{B}')\; p\; (-1)^{1/2-M} \;
                           \mathcal{C}(\frac{1}{2}\, \frac{1}{2}\, 1; \, M, -M', M-M')
                            \nonumber\\[2mm]
  && \times \;              \mathcal{C}(1\, 1\, 1; \, M'-M, M-M', 0) \;
                            \mathcal{C}(\frac{1}{2}\, \frac{1}{2}\, 0; \, M, -M, 0)\;
                           \mathcal{C}(\frac{1}{2}\, \frac{1}{2}\, j'; \, M', -M, M'-M)\nonumber\\[2mm]
 &=& \sum_{M'} 2\, \sqrt{2}\; (\mathcal{B}-\mathcal{B}')\; p\; (-1)^{1/2-M} \;
                            \mathcal{C}(\frac{1}{2}\, \frac{1}{2}\, 1; \, M, -M', M-M')\nonumber\\[2mm]
 && \times \;  \mathcal{C}(1\, 1\, 1; \, M'-M, M-M', 0)\; (-1)^{1/2-M} \; \frac{1}{\sqrt{2}} \;
                            \mathcal{C}(\frac{1}{2}\, \frac{1}{2}\, j'; \, M', -M, M'-M) \nonumber\\[2mm]
 &=& 2\, (\mathcal{B}-\mathcal{B}')\; p\; \mathcal{C}(1\, 1\, 1; \, M'-M, M-M', 0) \nonumber\\[2mm]
 && \times \;         \sum_{M'}
                      \mathcal{C}(\frac{1}{2}\, \frac{1}{2}\, 1; \, M', -M, M'-M)\;
                      \mathcal{C}(\frac{1}{2}\, \frac{1}{2}\, j'; \, M', -M, M'-M)\nonumber\\[2mm]
 &=& 2\, (\mathcal{B}-\mathcal{B}')\; p\; \mathcal{C}(1\, 1\, 1; \, M'-M, M-M', 0) \; \delta_{j'1}.
\end{eqnarray}

\begin{eqnarray}\label{eq:AppD:j1:3}
 t'_4 &\to&  \sum_{M} (-2)\; \mathcal{B}\mathcal{B}'\; \vec p^{\,2}\;
             \mathcal{C}(\frac{1}{2}\, \frac{1}{2}\, 1; \, M, -M', M-M')\; (-1)^{1/2-M}  \nonumber\\[2mm]
 && \times \; \mathcal{C}(\frac{1}{2}\, \frac{1}{2}\, 0; \, M, -M, 0)\;
               \mathcal{C}(\frac{1}{2}\, \frac{1}{2}\, j'; \, M', -M, M'-M) \nonumber\\[2mm]
&=& -2\;  \mathcal{B}\mathcal{B}'\; \vec p^{\,2}\; \sum_{M}
                \mathcal{C}(\frac{1}{2}\, \frac{1}{2}\, 1; \, M, -M', M-M')\; (-1)^{1/2-M}  \nonumber\\[2mm]
&& \times \; \frac{1}{\sqrt{2}}\; (-1)^{1/2-M}\;
                 \mathcal{C}(\frac{1}{2}\, \frac{1}{2}\, j'; \, M', -M, M'-M) \nonumber\\[2mm]
&=& -\sqrt{2}\;  \mathcal{B}\mathcal{B}'\; \vec p^{\,2}\; \sum_{M}
                 \mathcal{C}(\frac{1}{2}\, \frac{1}{2}\, 1; \, M, -M', M-M')\;
                 \mathcal{C}(\frac{1}{2}\, \frac{1}{2}\, j'; \, M, -M', M-M') \nonumber\\[2mm]
&=& -\sqrt{2}\;  \mathcal{B}\mathcal{B}'\; \vec p^{\,2}\; \delta_{j'1},~~~{\rm for ~any}~m'.
\end{eqnarray}

\begin{eqnarray}\label{eq:AppD:j1:4}
 t'_9 &\to&  \sum_{M} \sqrt{2}\; (\mathcal{B}+\mathcal{B}')\; p\; \delta_{MM'}\;
                 \mathcal{C}(\frac{1}{2}\, \frac{1}{2}\, 0; \, M, -M, 0)\;
                 \mathcal{C}(\frac{1}{2}\, \frac{1}{2}\, j'; \, M', -M, M'-M) \nonumber\\[2mm]
 &=& \sqrt{2}\; (\mathcal{B}+\mathcal{B}')\; p\; \delta_{j'0}\; \delta_{m'0}.
\end{eqnarray}

We summarize now,
\begin{equation}\label{eq:AppD:j1:5}
  t =
    \left\{
    \begin{array}{ll}
     \sqrt{2}\; \mathcal{A}\mathcal{A}' \; (1+\mathcal{B}\mathcal{B}' \; \vec{p}^{\, 2}) \delta_{m'0}, & {\rm for~} j=0, \; j'=0; \\[2mm]
    -\sqrt{2}\; \mathcal{A}\mathcal{A}' \; (\mathcal{B}+\mathcal{B}')\; p\; \delta_{m'0},   & {\rm for~} j=0, \; j'=1; \\[2mm]
    \sqrt{2}\; \mathcal{A}\mathcal{A}' \; (\mathcal{B}+\mathcal{B}')\; p\; \delta_{m'0}, & {\rm for~} j=1,\; j'=0; \\[2mm]
    \mathcal{A}\mathcal{A}'\;  \big[ -\sqrt{2}\; (-1)^{M'-M} -\sqrt{2}\; \mathcal{B}\mathcal{B}' \; \vec{p}^{\, 2}  & \\[2mm]
                 \qquad ~
                 +2\, (\mathcal{B}-\mathcal{B}')\; p\;
                 \mathcal{C}(1\, 1\, 1; \, M'-M, M-M', 0)\, \big],~ & {\rm for~} j=1,\; j'=1. \\
    \end{array}
   \right.
\end{equation}
For $j=1,\, j'=1$, we have three terms coming respectively from $t'_1+t'_5$, $t'_4$ and $t'_2+t'_3$.
The last term does not interfere with the other two in $\overline{\sum}\sum |t|^2$,
but the first two terms interfere and we have
\begin{eqnarray}\label{eq:AppD:j1:6}
 &&  \sum_{M'-M} \big[ -\sqrt{2}\;
      (-1)^{M'-M} -\sqrt{2}\; \mathcal{B}\mathcal{B}' \; \vec{p}^{\, 2} \,\big]^2 \nonumber\\[2mm]
 &=& 2 \sum_{M'-M} \big[ 1+ (\mathcal{B}\mathcal{B}')^2\; \vec{p}^{\; 4}
                +2\, (-1)^{M'-M}\; \mathcal{B}\mathcal{B}' \; \vec{p}^{\, 2} \, \big] \nonumber\\[2mm]
 &=& 2 \, \big[ 3+ 3\, (\mathcal{B}\mathcal{B}')^2\; \vec{p}^{\; 4} -2\, \mathcal{B}\mathcal{B}' \; \vec{p}^{\, 2}\, \big],
\end{eqnarray}
\begin{eqnarray}\label{eq:AppD:j1:7}
 &&  \sum_{M-M'}4\,(\mathcal{B}-\mathcal{B}')^2\; \vec{p}^{\, 2} \;
     \mathcal{C}(1\, 1\, 1; \, M'-M, M-M', 0)^2 \nonumber\\[2mm]
 &=& 4\, (\mathcal{B}-\mathcal{B}')^2\; \vec{p}^{\, 2}.
\end{eqnarray}
Hence finally we get
\begin{equation}\label{eq:AppD:j1:8}
  \overline{\sum}\sum |t|^2 =
    \left\{
    \begin{array}{ll}
     2\,(\mathcal{A}\mathcal{A}')^2 \; (1+\mathcal{B}\mathcal{B}' \, \vec{p}^{\, 2})^2,   & {\rm for~} j=0,\; j'=0; \\[2mm]
    2\, (\mathcal{A}\mathcal{A}')^2 \; (\mathcal{B}+\mathcal{B}')^2 \; \vec{p}^{\, 2},    & {\rm for~} j=0,\; j'=1; \\[2mm]
    2\, (\mathcal{A}\mathcal{A}')^2 \; (\mathcal{B}+\mathcal{B}')^2 \, \vec{p}^{\, 2},    & {\rm for~} j=1,\; j'=0; \\[2mm]
    (\mathcal{A}\mathcal{A}')^2 \big[ 6+4\,\mathcal{B}^2\,\vec{p}^{\, 2} +4\, \mathcal{B}'^2 \;\vec{p}^{\, 2}
               -12\,\mathcal{B}\mathcal{B}' \;\vec{p}^{\, 2}
               +6\, (\mathcal{B}\mathcal{B}')^2 \; \vec{p}^{\, 4}\, \big], & {\rm for~} j=1,\; j'=1. \\
    \end{array}
   \right.
\end{equation}

It is remarkable that we have obtained the same results
that we obtained for external emission in Section \ref{subsec:form2}
(see the summary in Eqs.~\eqref{eq:A}-\eqref{eq:D}),
even if the topology is different
and the original matrix elements also different.

\end{appendix}


\begin{thebibliography}{99}
%\cite{cai}
\bibitem{cai0}
 M.~Antonelli {\it et al.},
  %``Flavor Physics in the Quark Sector,''
  Phys.\ Rept.\  {\bf 494}, 197 (2010).
%  doi:10.1016/j.physrep.2010.05.003
%  [arXiv:0907.5386 [hep-ph]].
  %%CITATION = doi:10.1016/j.physrep.2010.05.003;%%
  %290 citations counted in INSPIRE as of 23 May 2018

%\cite{thomas}
\bibitem{thomas}
  T.~E.~Browder and K.~Honscheid,
  %``$B$ mesons,''
  Prog.\ Part.\ Nucl.\ Phys.\  {\bf 35}, 81 (1995).
%  doi:10.1016/0146-6410(95)00042-H
%  [hep-ph/9503414].
  %%CITATION = doi:10.1016/0146-6410(95)00042-H;%%
  %108 citations counted in INSPIRE as of 23 May 2018

%\cite{Buchalla}
\bibitem{Buchalla}
  G.~Buchalla, A.~J.~Buras and M.~E.~Lautenbacher,
  %``Weak decays beyond leading logarithms,''
  Rev.\ Mod.\ Phys.\  {\bf 68}, 1125 (1996).
%  doi:10.1103/RevModPhys.68.1125
%  [hep-ph/9512380].
  %%CITATION = doi:10.1103/RevModPhys.68.1125;%%
  %2268 citations counted in INSPIRE as of 23 May 2018

%1
%\cite{IJMPE52}
\bibitem{IJMPE52}
  B.~El-Bennich, A.~Furman, R.~Kaminski, L.~Lesniak, B.~Loiseau and B.~Moussallam,
  %``CP violation and kaon-pion interactions in B ---> K pi+ pi- decays,''
  Phys.\ Rev.\ D {\bf 79}, 094005 (2009)
  Erratum: [Phys.\ Rev.\ D {\bf 83}, 039903 (2011)].
%  doi:10.1103/PhysRevD.83.039903, 10.1103/PhysRevD.79.094005
%  [arXiv:0902.3645 [hep-ph]].
  %%CITATION = doi:10.1103/PhysRevD.83.039903, 10.1103/PhysRevD.79.094005;%%
  %65 citations counted in INSPIRE as of 24 Feb 2018

%2
%\cite{IJMPE53}
\bibitem{IJMPE53}
  O.~Leitner, J.-P.~Dedonder, B.~Loiseau and R.~Kaminski,
  %``K* resonance effects on direct CP violation in B -> pi pi K,''
  Phys.\ Rev.\ D {\bf 81}, 094033 (2010)
  Erratum: [Phys.\ Rev.\ D {\bf 82}, 119906 (2010)].
%  doi:10.1103/PhysRevD.81.094033, 10.1103/PhysRevD.82.119906
%  [arXiv:1001.5403 [hep-ph]].
  %%CITATION = doi:10.1103/PhysRevD.81.094033, 10.1103/PhysRevD.82.119906;%%
  %15 citations counted in INSPIRE as of 24 Feb 2018

%3
%\cite{IJMPE61}
\bibitem{IJMPE61}
  J.~W.~Li, D.~S.~Du and C.~D.~Lu,
  %``Determination of $f_0-\sigma$ mixing angle through $B_s^0 \to J/\Psi~f_0(980)(\sigma)$ decays,''
  Eur.\ Phys.\ J.\ C {\bf 72}, 2229 (2012).
%  doi:10.1140/epjc/s10052-012-2229-1
%  [arXiv:1212.5987 [hep-ph]].
  %%CITATION = doi:10.1140/epjc/s10052-012-2229-1;%%
  %13 citations counted in INSPIRE as of 24 Feb 2018

%4
%\cite{IJMPE62}
\bibitem{IJMPE62}
 W.~Ochs,
  %``The Status of Glueballs,''
  J.\ Phys.\ G {\bf 40}, 043001 (2013).
%  doi:10.1088/0954-3899/40/4/043001
%  [arXiv:1301.5183 [hep-ph]].
  %%CITATION = doi:10.1088/0954-3899/40/4/043001;%%
  %123 citations counted in INSPIRE as of 24 Feb 2018

%5
%\cite{IJMPE63}
\bibitem{IJMPE63}
X.~W.~Kang, B.~Kubis, C.~Hanhart and U.~G.~Mei\ss ner,
  %``$B_{l4}$ decays and the extraction of $|V_{ub}|$,''
  Phys.\ Rev.\ D {\bf 89}, 053015 (2014).
%  doi:10.1103/PhysRevD.89.053015
%  [arXiv:1312.1193 [hep-ph]].
  %%CITATION = doi:10.1103/PhysRevD.89.053015;%%
  %81 citations counted in INSPIRE as of 24 Feb 2018

%6
%\cite{IJMPE72}
\bibitem{IJMPE72}
 B.~El-Bennich, O.~Leitner, J.-P.~Dedonder and B.~Loiseau,
  %``The Scalar Meson f0(980) in Heavy-Meson Decays,''
  Phys.\ Rev.\ D {\bf 79}, 076004 (2009).
%  doi:10.1103/PhysRevD.79.076004
%  [arXiv:0810.5771 [hep-ph]].
  %%CITATION = doi:10.1103/PhysRevD.79.076004;%%
  %39 citations counted in INSPIRE as of 24 Feb 2018

%7
%\cite{IJMPE76}
\bibitem{IJMPE76}
M.~Sayahi and H.~Mehraban,
  %``Final state interaction in B$^{0} \to J/psi$ $\pi^{+} \pi^{-}$ decay,''
  Phys.\ Scripta {\bf 88}, 035101 (2013).
%  doi:10.1088/0031-8949/88/03/035101
  %%CITATION = doi:10.1088/0031-8949/88/03/035101;%%
  %5 citations counted in INSPIRE as of 24 Feb 2018

%8
%\cite{IJMPE78}
\bibitem{IJMPE78}
H.~W.~Ke, X.~Q.~Li and Z.~T.~Wei,
  %``Whether new data on D(s) ---> f(0)(980) e+ nu(e) can be understood if f(0)(980) consists of only the conventional q anti-q structure,''
  Phys.\ Rev.\ D {\bf 80}, 074030 (2009).
%  doi:10.1103/PhysRevD.80.074030
%  [arXiv:0907.5465 [hep-ph]].
  %%CITATION = doi:10.1103/PhysRevD.80.074030;%%
  %29 citations counted in INSPIRE as of 24 Feb 2018

%9
%\cite{IJMPE79}
\bibitem{IJMPE79}
N.~N.~Achasov and A.~V.~Kiselev,
  %``Light scalars in semi-leptonic decays of heavy quarkonia,''
  Phys.\ Rev.\ D {\bf 86}, 114010 (2012).
%  doi:10.1103/PhysRevD.86.114010
%  [arXiv:1206.5500 [hep-ph]].
  %%CITATION = doi:10.1103/PhysRevD.86.114010;%%
  %15 citations counted in INSPIRE as of 24 Feb 2018

%10
%\cite{IJMPE81}
\bibitem{IJMPE81}
A.~H.~Fariborz, R.~Jora, J.~Schechter and M.~N.~Shahid,
  %``Probing pseudoscalar and scalar mesons in semileptonic decays of $D_s^+$, $^D+$ and $D^0$,''
  Int.\ J.\ Mod.\ Phys.\ A {\bf 30}, 1550012 (2015).
%  doi:10.1142/S0217751X15500128
%  [arXiv:1407.7176 [hep-ph]].
  %%CITATION = doi:10.1142/S0217751X15500128;%%
  %4 citations counted in INSPIRE as of 24 Feb 2018

%11
%\cite{IJMPE83}
\bibitem{IJMPE83}
Y.~J.~Shi and W.~Wang,
  %``Chiral Dynamics and S-wave contributions in Semileptonic $D_s/B_s$ decays into $\pi^+\pi^-$,''
  Phys.\ Rev.\ D {\bf 92}, 074038 (2015).
%  doi:10.1103/PhysRevD.92.074038
%  [arXiv:1507.07692 [hep-ph]].
  %%CITATION = doi:10.1103/PhysRevD.92.074038;%%
  %8 citations counted in INSPIRE as of 24 Feb 2018

%12
%\cite{IJMPE84}
\bibitem{IJMPE84}
U.~G.~Mei\ss ner and W.~Wang,
  %``Generalized Heavy-to-Light Form Factors in Light-Cone Sum Rules,''
  Phys.\ Lett.\ B {\bf 730}, 336 (2014).
%  doi:10.1016/j.physletb.2014.02.009
%  [arXiv:1312.3087 [hep-ph]].
  %%CITATION = doi:10.1016/j.physletb.2014.02.009;%%
  %35 citations counted in INSPIRE as of 24 Feb 2018


%13
%\cite{IJMPE89}
\bibitem{IJMPE89}
 J.~T.~Daub, C.~Hanhart and B.~Kubis,
  %``A model-independent analysis of final-state interactions in $ {\overline{B}}_{d/s}^0\to J/\psi \pi \pi $,''
  JHEP {\bf 1602}, 009 (2016).
%  doi:10.1007/JHEP02(2016)009
%  [arXiv:1508.06841 [hep-ph]].
  %%CITATION = doi:10.1007/JHEP02(2016)009;%%
  %34 citations counted in INSPIRE as of 24 Feb 2018

%14
%\cite{IJMPE47}
\bibitem{IJMPE47}
W.~H.~Liang and E.~Oset,
  %``$B^0$ and $B^0_s$ decays into $J/\psi$ $f_0(980)$ and $J/\psi$ $f_0(500)$ and the nature of the scalar resonances,''
  Phys.\ Lett.\ B {\bf 737}, 70 (2014).
%  doi:10.1016/j.physletb.2014.08.030
%  [arXiv:1406.7228 [hep-ph]].
  %%CITATION = doi:10.1016/j.physletb.2014.08.030;%%
  %61 citations counted in INSPIRE as of 24 Feb 2018

%\cite{Zhao:2018zcb}
\bibitem{Zhao:2018zcb}
  Z.~X.~Zhao,
  %``Weak Decays of Singly Heavy Baryons in Light-Front Approach,''
  arXiv:1803.02292 [hep-ph].
  %%CITATION = ARXIV:1803.02292;%%

%15
%\cite{review2016}
\bibitem{review2016}
E.~Oset {\it et al.},
  %``Weak decays of heavy hadrons into dynamically generated resonances,''
  Int.\ J.\ Mod.\ Phys.\ E {\bf 25}, 1630001 (2016).
%  doi:10.1142/S0218301316300010
%  [arXiv:1601.03972 [hep-ph]].
  %%CITATION = doi:10.1142/S0218301316300010;%%
  %33 citations counted in INSPIRE as of 24 Feb 2018

%16
%\cite{IJMPE111}
\bibitem{IJMPE111}
 H.~Muramatsu {\it et al.} [CLEO Collaboration],
  %``Dalitz analysis of D0 ---> K0(S) pi+ pi-,''
  Phys.\ Rev.\ Lett.\  {\bf 89}, 251802 (2002); 
  Erratum: [Phys.\ Rev.\ Lett.\  {\bf 90}, 059901 (2003)].
%  doi:10.1103/PhysRevLett.89.251802
%  [hep-ex/0207067].
  %%CITATION = doi:10.1103/PhysRevLett.89.251802;%%
  %184 citations counted in INSPIRE as of 24 Feb 2018

%17
%\cite{IJMPE112}
\bibitem{IJMPE112}
L.~L.~Chau,
  %``Quark Mixing in Weak Interactions,''
  Phys.\ Rept.\  {\bf 95}, 1 (1983).
%  doi:10.1016/0370-1573(83)90043-1
  %%CITATION = doi:10.1016/0370-1573(83)90043-1;%%
  %587 citations counted in INSPIRE as of 24 Feb 2018

%18
%\cite{IJMPE113}
\bibitem{IJMPE113}
L.~L.~Chau and H.~Y.~Cheng,
  %``Analysis of Exclusive Two-Body Decays of Charm Mesons Using the Quark Diagram Scheme,''
  Phys.\ Rev.\ D {\bf 36}, 137 (1987);
  Addendum: [Phys.\ Rev.\ D {\bf 39}, 2788 (1989)].
%  doi:10.1103/PhysRevD.39.2788, 10.1103/PhysRevD.36.137
  %%CITATION = doi:10.1103/PhysRevD.39.2788, 10.1103/PhysRevD.36.137;%%
  %203 citations counted in INSPIRE as of 24 Feb 2018

%19
%\cite{IJMPE114}
\bibitem{IJMPE114}
H.~Y.~Cheng and C.~W.~Chiang,
  %``Hadronic D decays involving even-parity light mesons,''
  Phys.\ Rev.\ D {\bf 81}, 074031 (2010).
%  doi:10.1103/PhysRevD.81.074031
%  [arXiv:1002.2466 [hep-ph]].
  %%CITATION = doi:10.1103/PhysRevD.81.074031;%%
  %27 citations counted in INSPIRE as of 24 Feb 2018

%20
%\cite{uniex}
\bibitem{uniex}
S. Bifani, in LHCb Seminar at CERN (April 18th 2017),

https://indico.cern.ch/event/580620/.

%21
%\cite{jorge}
\bibitem{jorge}
  L.~S.~Geng, B.~Grinstein, S.~J\" ager, J.~Martin Camalich, X.~L.~Ren and R.~X.~Shi,
  %``Towards the discovery of new physics with lepton-universality ratios of $b\to s\ell\ell$ decays,''
  Phys.\ Rev.\ D {\bf 96}, 093006 (2017).
%  doi:10.1103/PhysRevD.96.093006
%  [arXiv:1704.05446 [hep-ph]].
  %%CITATION = doi:10.1103/PhysRevD.96.093006;%%
  %75 citations counted in INSPIRE as of 25 Feb 2018

%22
%\cite{private}
\bibitem{private}
J.~Martin Camalich, private communication.

%23
%\cite{pdg}
\bibitem{pdg}
C. Patrignani {\it et al.} (Particle Data Group), Chin.\ Phys.\ C {\bf 40}, 100001 (2016) and 2017 update.

%#########################################################  New

%\cite{Beneke}
\bibitem{Beneke}
  M.~Beneke and M.~Neubert,
  %``QCD factorization for B ---> PP and B ---> PV decays,''
  Nucl.\ Phys.\ B {\bf 675}, 333 (2003).
%  doi:10.1016/j.nuclphysb.2003.09.026
%  [hep-ph/0308039].
  %%CITATION = doi:10.1016/j.nuclphysb.2003.09.026;%%
  %879 citations counted in INSPIRE as of 17 May 2018

%\cite{Beneke2}
\bibitem{Beneke2}
  M.~Beneke, G.~Buchalla, M.~Neubert and C.~T.~Sachrajda,
  %``QCD factorization for exclusive, nonleptonic B meson decays: General arguments and the case of heavy light final states,''
  Nucl.\ Phys.\ B {\bf 591}, 313 (2000).
%  doi:10.1016/S0550-3213(00)00559-9
%  [hep-ph/0006124].
  %%CITATION = doi:10.1016/S0550-3213(00)00559-9;%%
  %1140 citations counted in INSPIRE as of 17 May 2018

%\cite{Rosner}
\bibitem{Rosner}
  M.~Gronau, O.~F.~Hernandez, D.~London and J.~L.~Rosner,
  %``Decays of B mesons to two light pseudoscalars,''
  Phys.\ Rev.\ D {\bf 50}, 4529 (1994).
%  doi:10.1103/PhysRevD.50.4529
%  [hep-ph/9404283].
  %%CITATION = doi:10.1103/PhysRevD.50.4529;%%
  %396 citations counted in INSPIRE as of 17 May 2018

%\cite{Neubert3}
\bibitem{Neubert3}
  M.~Neubert and B.~Stech,
  %``Nonleptonic weak decays of B mesons,''
  Adv.\ Ser.\ Direct.\ High Energy Phys.\  {\bf 15}, 294 (1998).
%  doi:10.1142/9789812812667_0004
%  [hep-ph/9705292].
  %%CITATION = doi:10.1142/9789812812667_0004;%%
  %281 citations counted in INSPIRE as of 17 May 2018


%\cite{AliLu2}
\bibitem{AliLu2}
  A.~Ali, G.~Kramer and C.~D.~Lu,
  %``Experimental tests of factorization in charmless nonleptonic two-body B decays,''
  Phys.\ Rev.\ D {\bf 58}, 094009 (1998).
%  doi:10.1103/PhysRevD.58.094009
%  [hep-ph/9804363].
  %%CITATION = doi:10.1103/PhysRevD.58.094009;%%
  %536 citations counted in INSPIRE as of 17 May 2018

%\cite{Keum}
\bibitem{Keum}
  Y.~Y.~Keum and H.~N.~Li,
  %``Nonleptonic charmless B decays: Factorization versus perturbative QCD,''
  Phys.\ Rev.\ D {\bf 63}, 074006 (2001).
 % doi:10.1103/PhysRevD.63.074006
%  [hep-ph/0006001].
  %%CITATION = doi:10.1103/PhysRevD.63.074006;%%
  %206 citations counted in INSPIRE as of 17 May 2018

%\cite{Du}
\bibitem{Du}
  D.~S.~Du, H.~J.~Gong, J.~F.~Sun, D.~S.~Yang and G.~H.~Zhu,
  %``Phenomenological analysis of charmless decays B ---> PV with QCD factorization,''
  Phys.\ Rev.\ D {\bf 65}, 094025 (2002);
  Erratum: [Phys.\ Rev.\ D {\bf 66}, 079904 (2002)].
%  doi:10.1103/PhysRevD.65.094025, 10.1103/PhysRevD.66.079904
%  [hep-ph/0201253].
  %%CITATION = doi:10.1103/PhysRevD.65.094025, 10.1103/PhysRevD.66.079904;%%
  %104 citations counted in INSPIRE as of 17 May 2018

%\cite{Chay}
\bibitem{Chay}
  J.~Chay and C.~Kim,
  %``Nonleptonic B decays into two light mesons in soft collinear effective theory,''
  Nucl.\ Phys.\ B {\bf 680}, 302 (2004).
%  doi:10.1016/j.nuclphysb.2003.12.027
%  [hep-ph/0301262].
  %%CITATION = doi:10.1016/j.nuclphysb.2003.12.027;%%
  %80 citations counted in INSPIRE as of 17 May 2018

%\cite{cdLu}
\bibitem{cdLu}
  R.~H.~Li, C.~D.~Lu and H.~Zou,
  %``The B(B(s)) ---> D(s) P, D(s) V, D*(s) P and D*(s) V decays in the perturbative QCD approach,''
  Phys.\ Rev.\ D {\bf 78}, 014018 (2008).
%  doi:10.1103/PhysRevD.78.014018
%  [arXiv:0803.1073 [hep-ph]].
  %%CITATION = doi:10.1103/PhysRevD.78.014018;%%
  %61 citations counted in INSPIRE as of 17 May 2018

%\cite{Grinstein}
\bibitem{Grinstein}
  B.~Grinstein and R.~F.~Lebed,
  %``SU(3) decomposition of two-body B decay amplitudes,''
  Phys.\ Rev.\ D {\bf 53}, 6344 (1996).
%  doi:10.1103/PhysRevD.53.6344
%  [hep-ph/9602218].
  %%CITATION = doi:10.1103/PhysRevD.53.6344;%%
  %61 citations counted in INSPIRE as of 17 May 2018

%\cite{cai}
\bibitem{cai}
  X.~Liu, Z.~J.~Xiao and C.~D.~Lu,
  %``The Pure annihilation type B(c) ---> M(2) M(3) decays in the perturbative QCD approach,''
  Phys.\ Rev.\ D {\bf 81}, 014022 (2010).
%  doi:10.1103/PhysRevD.81.014022
%  [arXiv:0912.1163 [hep-ph]].
  %%CITATION = doi:10.1103/PhysRevD.81.014022;%%
  %39 citations counted in INSPIRE as of 17 May 2018

%\cite{weiwang}
\bibitem{weiwang}
  W.~Wang and C.~D.~Lu,
  %``Distinguishing two kinds of scalar mesons from heavy meson decays,''
  Phys.\ Rev.\ D {\bf 82}, 034016 (2010).
%  doi:10.1103/PhysRevD.82.034016
%  [arXiv:0910.0613 [hep-ph]].
  %%CITATION = doi:10.1103/PhysRevD.82.034016;%%
  %18 citations counted in INSPIRE as of 17 May 2018

%\cite{molina}
\bibitem{molina}
  R.~Molina, M.~D{\" o}ring and E.~Oset,
  %``Determination of the compositeness of resonances from decays: the case of the $B^0_s\to J/\psi f_1(1285)$,''
  Phys.\ Rev.\ D {\bf 93}, 114004 (2016).
%  doi:10.1103/PhysRevD.93.114004
%  [arXiv:1604.02574 [hep-ph]].
  %%CITATION = doi:10.1103/PhysRevD.93.114004;%%
  %1 citations counted in INSPIRE as of 17 May 2018

%\cite{weilu}
\bibitem{weilu}
  W.~Wang, Y.~M.~Wang, D.~S.~Yang and C.~D.~Lu,
  %``Charmless Two-body B(B(s)) ---> VP decays In Soft-Collinear-Effective-Theory,''
  Phys.\ Rev.\ D {\bf 78}, 034011 (2008).
%  doi:10.1103/PhysRevD.78.034011
%  [arXiv:0801.3123 [hep-ph]].
  %%CITATION = doi:10.1103/PhysRevD.78.034011;%%
  %54 citations counted in INSPIRE as of 17 May 2018

%\cite{alilu}
\bibitem{alilu}
  A.~Ali, G.~Kramer, Y.~Li, C.~D.~Lu, Y.~L.~Shen, W.~Wang and Y.~M.~Wang,
  %``Charmless non-leptonic $B_s$ decays to $PP$, $PV$ and $VV$ final states in the pQCD approach,''
  Phys.\ Rev.\ D {\bf 76}, 074018 (2007).
%  doi:10.1103/PhysRevD.76.074018
%  [hep-ph/0703162 [HEP-PH]].
  %%CITATION = doi:10.1103/PhysRevD.76.074018;%%
  %212 citations counted in INSPIRE as of 28 May 2018


%\cite{Luwei}
\bibitem{Luwei}
  C.~D.~Lu, Y.~L.~Shen and W.~Wang,
  %``Final state interaction in B ---> KK decays,''
  Phys.\ Rev.\ D {\bf 73}, 034005 (2006).
%  doi:10.1103/PhysRevD.73.034005
%  [hep-ph/0511255].
  %%CITATION = doi:10.1103/PhysRevD.73.034005;%%
  %38 citations counted in INSPIRE as of 17 May 2018

%\cite{kuo}
\bibitem{kuo}
  H.~Y.~Cheng, C.~W.~Chiang and A.~L.~Kuo,
  %``Global analysis of two-body D$\to$VP decays within the framework of flavor symmetry,''
  Phys.\ Rev.\ D {\bf 93}, 114010 (2016).
%  doi:10.1103/PhysRevD.93.114010
%  [arXiv:1604.03761 [hep-ph]].
  %%CITATION = doi:10.1103/PhysRevD.93.114010;%%
  %11 citations counted in INSPIRE as of 17 May 2018

%\cite{Gronau}
\bibitem{Gronau}
  B.~Bhattacharya, M.~Gronau and J.~L.~Rosner,
  %``CP asymmetries in singly-Cabibbo-suppressed $D$ decays to two pseudoscalar mesons,''
  Phys.\ Rev.\ D {\bf 85}, 054014 (2012); Erratum: [Phys.\ Rev.\ D {\bf 85}, 079901 (2012)].
%  doi:10.1103/PhysRevD.85.079901, 10.1103/PhysRevD.85.054014
%  [arXiv:1201.2351 [hep-ph]].
  %%CITATION = doi:10.1103/PhysRevD.85.079901, 10.1103/PhysRevD.85.054014;%%
  %126 citations counted in INSPIRE as of 17 May 2018

%\cite{Fazio}
\bibitem{Fazio}
  P.~Colangelo, F.~De Fazio and W.~Wang,
  %``Nonleptonic $B_s$ to charmonium decays: analyses in pursuit of determining the weak phase $\beta_s$,''
  Phys.\ Rev.\ D {\bf 83}, 094027 (2011).
%  doi:10.1103/PhysRevD.83.094027
%  [arXiv:1009.4612 [hep-ph]].
  %%CITATION = doi:10.1103/PhysRevD.83.094027;%%
  %39 citations counted in INSPIRE as of 17 May 2018

\iffalse
%\cite{caiplus}
\bibitem{caiplus}
  A.~Ali, G.~Kramer, Y.~Li, C.~D.~Lu, Y.~L.~Shen, W.~Wang and Y.~M.~Wang,
  %``Charmless non-leptonic $B_s$ decays to $PP$, $PV$ and $VV$ final states in the pQCD approach,''
  Phys.\ Rev.\ D {\bf 76}, 074018 (2007).
%  doi:10.1103/PhysRevD.76.074018
%  [hep-ph/0703162 [HEP-PH]].
  %%CITATION = doi:10.1103/PhysRevD.76.074018;%%
  %212 citations counted in INSPIRE as of 17 May 2018
\fi

%\cite{Sanda}
\bibitem{Sanda}
  R.~H.~Li, X.~X.~Wang, A.~I.~Sanda and C.~D.~Lu,
  %``Decays of $B$ meson to two charmed mesons,''
  Phys.\ Rev.\ D {\bf 81}, 034006 (2010).
%  doi:10.1103/PhysRevD.81.034006
%  [arXiv:0910.1424 [hep-ph]].
  %%CITATION = doi:10.1103/PhysRevD.81.034006;%%
  %25 citations counted in INSPIRE as of 17 May 2018

%\cite{Manohar3}
\bibitem{Manohar3}
  E.~E.~Jenkins, M.~E.~Luke, A.~V.~Manohar and M.~J.~Savage,
  %``Semileptonic B(c) decay and heavy quark spin symmetry,''
  Nucl.\ Phys.\ B {\bf 390}, 463 (1993).
%  doi:10.1016/0550-3213(93)90464-Z
%  [hep-ph/9204238].
  %%CITATION = doi:10.1016/0550-3213(93)90464-Z;%%
  %69 citations counted in INSPIRE as of 17 May 2018

%\cite{Mannel}
\bibitem{Mannel}
  G.~Kramer, T.~Mannel and W.~F.~Palmer,
  %``Angular correlations in the decays B ---> V V using heavy quark symmetry,''
  Z.\ Phys.\ C {\bf 55}, 497 (1992).
%  doi:10.1007/BF01565112
  %%CITATION = doi:10.1007/BF01565112;%%
  %48 citations counted in INSPIRE as of 17 May 2018


%24
%\cite{Neubert}
\bibitem{Neubert}
M.~Neubert,
  %``Heavy quark symmetry,''
  Phys.\ Rept.\  {\bf 245}, 259 (1994).
%  doi:10.1016/0370-1573(94)90091-4
%  [hep-ph/9306320].
  %%CITATION = doi:10.1016/0370-1573(94)90091-4;%%
  %1421 citations counted in INSPIRE as of 25 Feb 2018

%25
%\cite{manohar}
\bibitem{manohar}
%Heavy Quark Physics, Cambridge Monographs on Particle Physis, Nuclear Physics and
A.~V.~Manohar and M.~B.~Wise,
  {\it Heavy quark physics},
  Camb.\ Monogr.\ Part.\ Phys.\ Nucl.\ Phys.\ Cosmol.\  {\bf 10}, 1-191 (2000).
  %%CITATION = CMPCE,10,1;%%
  %386 citations counted in INSPIRE as of 25 Feb 2018

%26
%\cite{itzy}
\bibitem{itzy}
C.~Itzykson and J.~B.~Zuber, {\it Quantum Field Theory}, Mecraw-Hill, 1980.

%27
%\cite{mandl}
\bibitem{mandl}
F.~Mandl and G.~Shaw, {\it Quantum Field Theory}, John Wiley and Sons, 1984.

%28
%\cite{xieliang}
\bibitem{xieliang}
 J.~J.~Xie, W.~H.~Liang and E.~Oset,
  %``Hidden charm pentaquark and $\Lambda(1405)$ in the $\Lambda^0_b \to \eta_c K^- p (\pi \Sigma)$ reaction,''
  Phys.\ Lett.\ B {\bf 777}, 447 (2018).
%  doi:10.1016/j.physletb.2017.12.064
%  [arXiv:1711.01710 [hep-ph]].
  %%CITATION = doi:10.1016/j.physletb.2017.12.064;%%
  %1 citations counted in INSPIRE as of 25 Feb 2018


%\cite{Dillig}
\bibitem{Dillig}
  R.~D.~Bent, P.~W.~F.~Alons and M.~Dillig,
  %``Energy dependence of the He-3 (p, pi+) He-4 and H-3 (n, pi-) He-4 reactions,''
  Nucl.\ Phys.\ A {\bf 511}, 541 (1990).
%  doi:10.1016/0375-9474(90)90109-Y
  %%CITATION = doi:10.1016/0375-9474(90)90109-Y;%%
  %8 citations counted in INSPIRE as of 17 May 2018

%\cite{Vijande}
\bibitem{Vijande}
  J.~Vijande, F.~Fernandez and A.~Valcarce,
  %``Constituent quark model study of the meson spectra,''
  J.\ Phys.\ G {\bf 31}, 481 (2005).
%  doi:10.1088/0954-3899/31/5/017
%  [hep-ph/0411299].
  %%CITATION = doi:10.1088/0954-3899/31/5/017;%%
  %205 citations counted in INSPIRE as of 17 May 2018

%\cite{david}
\bibitem{david}
  P.~G.~Ortega, D.~R.~Entem and F.~Fernández,
  %``Hadronic molecules in the open charm and open bottom baryon spectrum,''
  Phys.\ Rev.\ D {\bf 90}, 114013 (2014).
%  doi:10.1103/PhysRevD.90.114013
  %%CITATION = doi:10.1103/PhysRevD.90.114013;%%
  %3 citations counted in INSPIRE as of 17 May 2018

%\cite{Ortega}
\bibitem{Ortega}
  P.~G.~Ortega, D.~R.~Entem and F.~Fernandez,
  %``Quark model description of the \Lambda_c(2940)^+ as a molecular D^*N state and the possible existence of the \Lambda_b(6248),''
  Phys.\ Lett.\ B {\bf 718}, 1381 (2013).
%  doi:10.1016/j.physletb.2012.12.025
%  [arXiv:1210.2633 [hep-ph]].
  %%CITATION = doi:10.1016/j.physletb.2012.12.025;%%
  %37 citations counted in INSPIRE as of 17 May 2018

%\cite{bassel1}
\bibitem{bassel1}
 M.~Wirbel, B.~Stech and M.~Bauer,
  %``Exclusive Semileptonic Decays of Heavy Mesons,''
  Z.\ Phys.\ C {\bf 29}, 637 (1985).
%  doi:10.1007/BF01560299
  %%CITATION = doi:10.1007/BF01560299;%%
  %1698 citations counted in INSPIRE as of 28 May 2018

%\cite{bassel2}
\bibitem{bassel2}
 M.~Bauer, B.~Stech and M.~Wirbel,
  %``Exclusive Nonleptonic Decays of D, D(s), and B Mesons,''
  Z.\ Phys.\ C {\bf 34}, 103 (1987).
%  doi:10.1007/BF01561122
  %%CITATION = doi:10.1007/BF01561122;%%
  %1795 citations counted in INSPIRE as of 28 May 2018

%\cite{bassel3}
\bibitem{bassel3}
M.~Bauer and M.~Wirbel,
  %``Form-Factor Effects in Exclusive d and B Decays,''
  Z.\ Phys.\ C {\bf 42}, 671 (1989).
%  doi:10.1007/BF01557675
  %%CITATION = doi:10.1007/BF01557675;%%
  %330 citations counted in INSPIRE as of 28 May 2018

%\cite{neubert2}
\bibitem{neubert2}
   M.~Neubert, V.~Rieckert, Q.P.~Xu and B.~Stech in {\it Heavy Flavours},
   Edited by A.J.~Buras and H.~Lindner (World Scientific, Singapore, 1992).
   
\iffalse
%\cite{klaus}
\bibitem{klaus}
  T.~E.~Browder and K.~Honscheid,
  %``$B$ mesons,''
  Prog.\ Part.\ Nucl.\ Phys.\  {\bf 35}, 81 (1995).
%  doi:10.1016/0146-6410(95)00042-H
%  [hep-ph/9503414].
  %%CITATION = doi:10.1016/0146-6410(95)00042-H;%%
  %108 citations counted in INSPIRE as of 17 May 2018
\fi

%\cite{Browder}
\bibitem{Browder}
  T.~E.~Browder, K.~Honscheid and D.~Pedrini,
  %``Nonleptonic decays and lifetimes of $b$ quark and $c$ quark hadrons,''
  Ann.\ Rev.\ Nucl.\ Part.\ Sci.\  {\bf 46}, 395 (1996).
%  doi:10.1146/annurev.nucl.46.1.395
%  [hep-ph/9606354].
  %%CITATION = doi:10.1146/annurev.nucl.46.1.395;%%
  %114 citations counted in INSPIRE as of 17 May 2018

%29
%\cite{Bramon}
\bibitem{Bramon}
  A.~Bramon, A.~Grau and G.~Pancheri,
  %``Intermediate vector meson contributions to V0 ---> P0 P0 gamma decays,''
  Phys.\ Lett.\ B {\bf 283}, 416 (1992).
%  doi:10.1016/0370-2693(92)90041-2
  %%CITATION = doi:10.1016/0370-2693(92)90041-2;%%
  %98 citations counted in INSPIRE as of 26 Feb 2018


%\cite{glauber}
\bibitem{glauber}
R.J.~Glauber, {\it High-energy collision theory}, Lectures of Theoretical Physics, Vol. I
(Wiley Interscience, New York, 1959).

%\cite{oset}
\bibitem{oset}
  R.~Guardiola and E.~Oset,
  %``Short Range Correlations in High-Energy Hadron Collision,''
  Nucl.\ Phys.\ A {\bf 234}, 458 (1974).
%  doi:10.1016/0375-9474(74)90575-2
  %%CITATION = doi:10.1016/0375-9474(74)90575-2;%%
  %10 citations counted in INSPIRE as of 17 May 2018

%31
%\cite{rose}
\bibitem{ramos}
 C.~E.~Jimenez-Tejero, A.~Ramos and I.~Vidana,
  %``Dynamically generated open charmed baryons beyond the zero range approximation,''
  Phys.\ Rev.\ C {\bf 80}, 055206 (2009).
%  doi:10.1103/PhysRevC.80.055206
%  [arXiv:0907.5316 [hep-ph]].
  %%CITATION = doi:10.1103/PhysRevC.80.055206;%%
  %57 citations counted in INSPIRE as of 19 Mar 2018

%32
%\cite{feijoo}
\bibitem{feijoo}
 G.~Montana, A.~Feijoo and A.~Ramos,
  %``A meson-baryon molecular interpretation for some $\Omega_c$ excited baryons,''
  Eur.\ Phys.\ J.\ A {\bf 54}, 64 (2018).
%  doi:10.1140/epja/i2018-12498-1
%  [arXiv:1709.08737 [hep-ph]].
  %%CITATION = doi:10.1140/epja/i2018-12498-1;%%
  %15 citations counted in INSPIRE as of 23 May 2018

%33
%\cite{omegac}
\bibitem{omegac}
 V.~R.~Debastiani, J.~M.~Dias, W.~H.~Liang and E.~Oset,
  %``Molecular $\Omega_c$ states generated from coupled meson-baryon channels,''
  Phys.\ Rev.\ D {\bf 97}, 094035 (2018).
%  doi:10.1103/PhysRevD.97.094035
%  [arXiv:1710.04231 [hep-ph]].
  %%CITATION = doi:10.1103/PhysRevD.97.094035;%%
  %18 citations counted in INSPIRE as of 21 Jun 2018


%30
%\cite{rose}
\bibitem{rose}
M.~E.~Rose, {\it Elementary Theory of Angular Momentum}, John Wiley and Sons, 1957.




























\end{thebibliography}
  \end{document}